\documentclass[reprint,superscriptaddress,amsmath,aps,showpacs,pra,nofootinbib,longbibliography]{revtex4-2}
\usepackage{graphicx}
\usepackage{float}
\usepackage{color}
\usepackage{physics}
\usepackage{amsfonts}
\usepackage{hyperref}
\usepackage{color}
\usepackage{xcolor}
\usepackage{dsfont} 
\usepackage{amsmath}
\usepackage{romannum}
\usepackage{textcomp}
\usepackage[T1]{fontenc}
\hypersetup{
     colorlinks   = true,
     citecolor    = blue
}
\def\degree{{}^{\circ}}
\def\eye{\mathbb{I}}

\graphicspath{{figures_main_comp/}}

\makeatletter
\newcommand{\stoptoc}{%
  \let\oldaddcontentsline\addcontentsline
  \renewcommand{\addcontentsline}[3]{}%
}

\newcommand{\resumetoc}{%
  \let\addcontentsline\oldaddcontentsline
}
\makeatother

\begin{document}

\stoptoc

\title{Straintronics and twistronics in bilayer graphene}
\author{Federico Escudero}
\thanks{These authors contributed equally to this paper.}
\affiliation{IMDEA Nanoscience, Faraday 9, 28049 Madrid, Spain}
\author{Dong Wang}
\thanks{These authors contributed equally to this paper.}
\affiliation{Key Laboratory of Artificial Micro- and Nano-structures of Ministry of Education and School of Physics and Technology, Wuhan University, Wuhan 430072, China}
\author{Pierre A. Pantale\'on}
\affiliation{IMDEA Nanoscience, Faraday 9, 28049 Madrid, Spain}
\author{Shengjun Yuan}
\affiliation{Key Laboratory of Artificial Micro- and Nano-structures of Ministry of Education 
and School of Physics and Technology, Wuhan University, Wuhan 430072, China}
\affiliation{School of Artificial Intelligence, Wuhan University, Wuhan 430072, China}
\affiliation{Wuhan Institute of Quantum Technology, Wuhan, 430206, China}
\author{Francisco Guinea}
\affiliation{IMDEA Nanoscience, Faraday 9, 28049 Madrid, Spain}
\affiliation{Donostia International Physics Center, Paseo Manuel de Lardiz\'{a}bal 4, San Sebasti\'an 20018, Spain}
\author{Zhen Zhan}
\email{zhenzhanh@gmail.com}
\affiliation{IMDEA Nanoscience, Faraday 9, 28049 Madrid, Spain}
\date{\today}

\begin{abstract}
The interplay of twist and strain in bilayer graphene enables the formation of moiré patterns and narrow bands that host correlated and topological phases. While magic-angle twisted bilayer graphene has been widely studied, strain provides an additional and realistic control knob for band engineering. In this work, we first generate a global method to construct commensurate supercells for arbitrary twist and heterostrain. Then, using atomistic tight-binding and strain-extended continuum models to study the commensurate structures, we identify configurations that minimize the bandwidth beyond the magic angle. The results reveal a strong dependence of band narrowing and topology on strain type, magnitude, direction and lattice relaxation. Particularly, shear strain produces a stronger distortion than uniaxial strain. Including electron-electron interactions through a self-consistent Hartree potential shows that strain broadens the bare bands while reducing electrostatic renormalization. Strain also drives topological transitions as the narrow and remote bands hybridize, establishing twisted and strained bilayer graphene as a tunable platform for flat-band and topological phenomena.
\end{abstract}

\maketitle
\clearpage
\pagenumbering{arabic}
\setcounter{page}{1}
\section{Introduction}
The discovery of correlated phases and unconventional superconductivity in twisted bilayer graphene (TBG) has attracted significant attention over the last few years \cite{andrei2021marvels,cao_correlated_2018, cao_unconventional_2018, kerelsky_maximized_2019, yankowitz2019tuning, lu2019superconductors, saito2020independent, zondiner2020cascade, xu2020correlated, choi2021correlation, cao2021nematicity}. These phenomena are intrinsically connected to the emergence of very flat bands due to the moiré potential \cite{lopes2007graphene, trambly2010localization, suarez2010flat, bistritzer_moire_2011}, induced by the lattice mismatch created by the twist or strain \cite{andrei2020graphene,moon2013optical, koshino2015interlayer, san2012non,bi_designing_2019, escudero_designing_2024}. The quenching of the kinetic energy in the flat bands promotes the appearance of the observed electronic correlations. Any approach to understand the nature of the correlated phases in TBG must then start from a solid understanding of the nature and origin of the flat bands, supported by accurate modelling methods. 

With only a relative twist, the flat bands are well-known to arise at an infinite set of \textit{magic angles}, the largest one being around $1^{\circ}$~\cite{lopes2007graphene, bistritzer_moire_2011, tarnopolsky2019origin}. However, in practice the samples usually have at least some residual strain~\cite{kapfer2023programming, hou2024strain}, defects~\cite{alden2013strain, lu2013twisting, schmucker2015raman} or even twist angle disorders~\cite{wilson2020disorder, schapers2022raman, beechem2014rotational}, typically arising during their fabrication~\cite{gadelha2021twisted}. The presence of strains, in particular, can significantly modify the geometrical and electronic properties of the system since the moiré acts as a magnifying glass~\cite{huder2018electronic, bi_designing_2019, kazmierczak_strain_2021, Mesple2021Heterostrain, mannai2021twistronics, sinner_strain-induced_2023, wang2023unusual, escudero_designing_2024}. Although this may be seen as undesirable, one can conversely use strains as an additional tune parameter in the system~\cite{jiang2017visualizing,mao2020evidence,kazmierczak_strain_2021, kapfer2023programming, pena2023moire}. Compared to the twist, which only rotates the layers, strains distort the layers and thus can lead to a plethora of moiré patterns with potentially rich properties~\cite{kogl_moire_2023, escudero_designing_2024}. This has motivated recent experimental advances in developing new techniques to induce and manipulate the strain in moiré heterostructures~\cite{pena2023moire,hasan2025strain,gao2021heterostrain, sequeira2024manipulating}, opening a path to straintronics and twistronics, whereby the electronic properties can be tuned by the combined interplay between twist and strain. Among different types of strain, uniaxial heterostrain
is the most common type, and has been observed in many experimental samples \cite{huder2018electronic,kerelsky_maximized_2019,xie2019spectroscopic,Mesple2021Heterostrain,wang2023unusual}. Consequently, most theoretical works focus on the effects of the uniaxial heterostrain. Recently, under
some new well-developed techniques, shear heterostrain
is introduced to manipulate the moiré patterns \cite{yu2024twist,ouyang2025,carrasco2025twistraintronics}.  

A natural question, then, is how does the presence of arbitrary strain and twist modifies the narrow bands in twisted bilayer graphene. An initial step involves computing the electronic spectra of the system and study its evolution under different kinds of relevant strain. Since narrow bands generally arise only under small lattice mismatches~\cite{lopes2007graphene, bistritzer_moire_2011}, at which the supercells can contain tens of thousands of atoms, previous studies have mostly employed extensions of effective continuum models under strain~\cite{bi_designing_2019, san2012non,escudero_designing_2024}, or topological heavy fermion models \cite{herzog2025topological,herzog2025kekule}. Yet, despite its importance, there is no comprehensive study of the electronic properties under strain by means of more realistic approaches, such as atomistic models. The challenge for that relies not only on the huge size of the supercell, but also on the fact that under twist and strain the system is in general incommensurate, and it is not clear at which configurations there can actually be a commensurate solution \cite{zhu2026twisted}. The commensurate and incommensurate structures may in fact have distinct ground-state properties \cite{gonccalves2024incommensurability}.

In this work, we present a comprehensive analysis of the tight-binding (TB) properties of twisted and strained bilayer graphene (TSBG), focusing on the optimal conditions for the emergence of narrow bands. Two relevant types of heterostrain are considered: uniaxial and shear (in the following, without specifying, strain refers to heterostrain). We first tackle the problem of obtaining commensurate structures under both twist and strain. 
We find that by generally adding a small biaxial strain, one can always find particular twist and strain values at which the system is commensurate. Using these commensurate structures, we then obtain the band structure and the density of states (DOS) by using a full atomistic TB model. We find that: (\romannum{1}) albeit the strain tends to increase the bandwidth at the (nonstrain) magic angle~\cite{bi_designing_2019, escudero_designing_2024}, there are yet other twist angles at which the bandwidth can be minimum, resulting in a shift of magic angle with strain; (\romannum{2}) the emergence of narrow bands depends critically on the strain direction; (\romannum{3}) the shear strain produces a stronger distortion of the geometry and electronic properties of TSBG; (\romannum{4}) the gap between the narrow and remote bands (induced by the lattice relaxation) is mainly determined by the strain-dependent bandwidth of the narrow bands.

Our atomistic results are then compared with that of the strain-extended continuum model~\cite{pereira2009tight, vozmediano2010gauge, amorim2016novel, naumis2017electronic}, which introduces two main modifications: (\romannum{1}) the change in the moir\'e vectors by which electrons in the different layers are coupled through the moir\'e potential~\cite{kogl_moire_2023, escudero_designing_2024}; (\romannum{2}) the introduction of strain-induced fields~\cite{suzuura2002phonons, manes2007symmetry}. We show that with just a few suitable parameter choices, the continuum model yields results in excellent agreement with the TB ones. In particular, we find that the strain-induced gauge potential, accounting for the change in the hopping energies within the Dirac approximation~\cite{suzuura2002phonons}, plays a key role in capturing the electronic behavior in TSBG. Using the strain-extended continuum model, we extend our analysis of the bandwidth evolution under twist and strain. We find that the twist angle at which the bandwidth is minimum sensitively depends on the strain direction. Yet, we remarkably see that the minimum bandwidth of the narrow bands (at the optimal twist angle) increases practically linearly with the strain magnitude.

We further consider the effect of electrostatic interactions, as accounted by the Hartree potential~\cite{guinea2018electrostatic, cea2019electronic,goodwin2020hartree}. 
Due to the increase of bandwidth under strain, the Hartree effect is weaker than that in the only twisted configurations. Consequently, as the strain increases, there is a competition between the increase of bandwidth of the bare bands, and the decrease of the Hartree potential. We show the synergy between these effects can lead to bandwidths under twist and strain that are actually comparable, if not smaller, than those with only twist angles. As any comprehensive account of correlated phases must consider such strong renormalization of the spectra by electrostatic interactions~\cite{cea2021coulomb, phong2021band,parker2021strain,wagner2022global,hu2023symmetric}, we conclude that TSBG has the potential to be a  platform for new and rich correlated phenomena.

Finally, we analyze the influence of strain on the topology of the narrow bands around the magic angle. By introducing a small mass term that breaks the inversion symmetry, we compute the valley Chern number $C$ of the narrow bands for different strain magnitudes and directions. Due to the strain effect in both increasing the narrow bandwidth and reducing the remote bandgap, there are topological ($C=\pm1$) to trivial ($C=0$) transitions as the strain increases, with a non-trivial dependence on the strain direction. We show that these topological transitions take place when the narrow bands close their gap with the remote bands. In the noninteracting case, we find that the sum of Chern number in top and bottom narrow band is always zero, i.e., they are both topological or both trivial. However, upon taking into account the electrostatic interactions we find asymmetrical topological transitions, whereby one narrow band can be topological while the other is trivial. We associate this behavior to the asymmetrical renormalization of the narrow bands due to the Hartree potential, which implies that they close their gap with the remote bands at different strain magnitudes. 

The paper is organized as follow: In Sec. \ref{TB method}, we introduce a general formalism for moiré commensurability with any twist and strain, and discuss the Dirac point shifts due to the moiré deformation. In Sec. \ref{TB results}, we calculate the electronic structures by using the TB and continuum methods, including the uniaxial strain, shear strain and lattice relaxations. These two models yield results in excellent agreement. Then, we investigate the narrow band modulation by both twist and strain. In Sec. \ref{interaction_part}, we study the effect of the electronic interactions. The band topology with strain is investigated in Sec. \ref{topo}. Our conclusions follow in Sec. \ref{sec:Conclusions}.

\section{Moiré commensurability with twist and strain}

\label{TB method}

In order to employ the TB method to obtain the band structure, we first require a commensurate structure. Although for only twist configurations the set of commensurate twist angles is well known, there is no close expression for the set of twist and strain that give commensurate structures. In what follows we will explicitly describe a global method to construct a commensurate structure of TSBG for a given twist angle and strain, taking into account different types of strain observed in experiments.

\subsection{Structural analysis of the moiré patterns}

\subsubsection{Commensurate structures with twist}

\begin{figure}[t!]
    \centering
    \includegraphics[width=\linewidth]{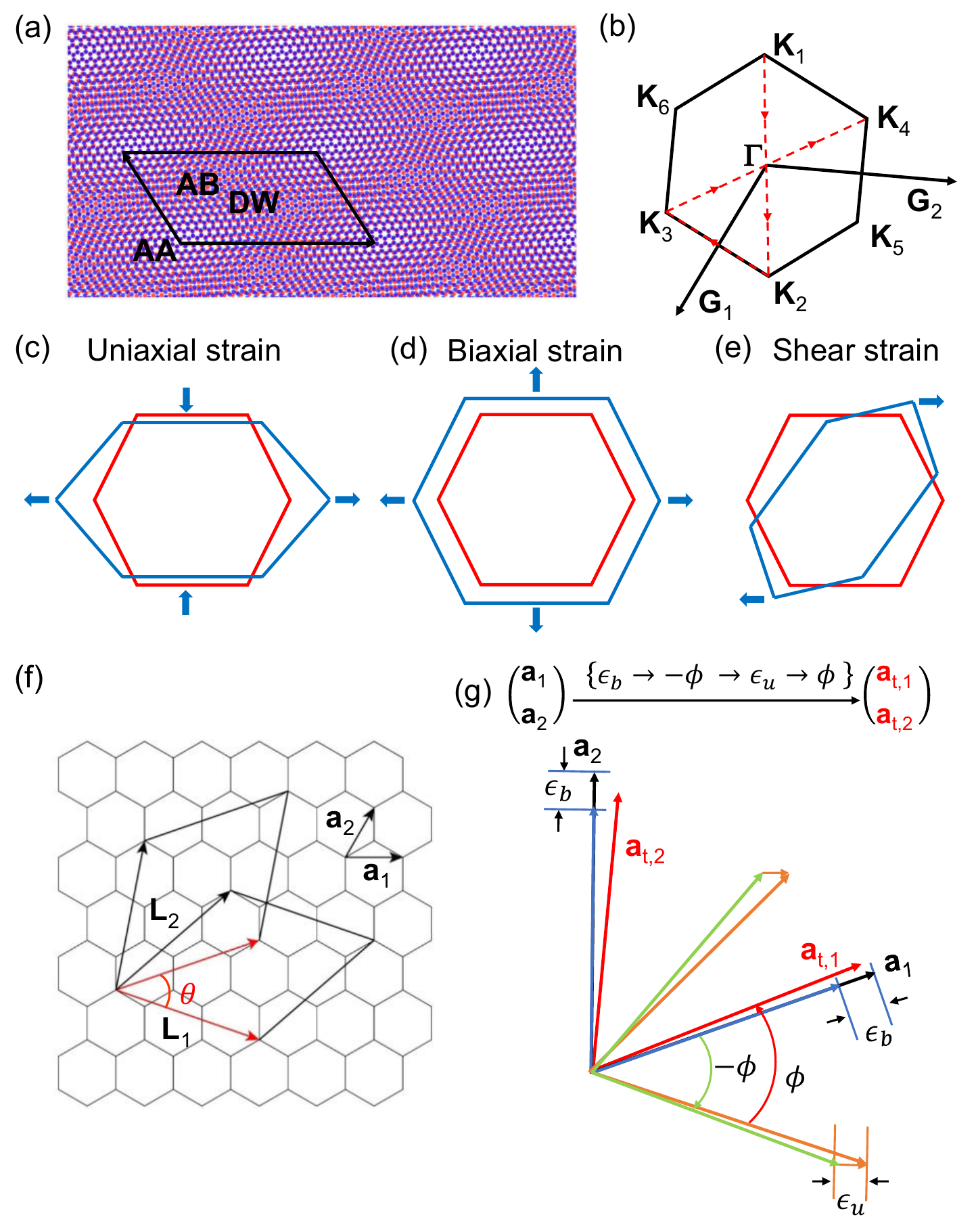}
    \caption{(a) Moiré pattern of bilayer graphene with a twist angle $\theta=3.89\degree$ and uniaxial strain $\epsilon_{u}=3.7\%$. The atoms in the top and bottom layers are plotted with blue and red dots, respectively. The AA, AB and DW stacking regions are labeled. Due to the strain the AA regions are elliptical. (b) Illustration of the moir\'e Brillouin zone, with the reciprocal lattice vectors labeled by $\mathbf{G}_1$ and $\mathbf{G}_2$. We label six corner $K$ points. The red dashed line is the momentum path for the band structure plots.  (c), (d), (e) Schematically show the uniaxial, biaxial and shear strains, respectively. Only the top layer (blue line) is deformed. The undeformed bottom layer is plot with red lines. (f) Schematic construction of the moir\'e cell with the lattice vectors of the two graphene layers. (g) Schematic construction of the moir\'e cell in three steps by following the Eq. (\ref{strainvec}) with $\mathcal{E} \to \mathcal{E}_{b}+\mathcal{E}_{u/s}$: (1) an isotropic rescaling corresponding to $\epsilon_{b}$; (2) an anisotropic rescaling corresponding to $\epsilon_{u}$ in a direction given by $\phi$; (3) a rotation by an angle $\theta$. We only plot the first two steps in (g).}
    \label{fig_structure}
\end{figure}

We consider two graphene layers rotated in the plane by an angle $\theta$, with the rotation origin at the AA site, and assume the constructed supercell is commensurate with only one moir\'e pattern~\cite{Mele2010Commensuration}. Each moir\'e pattern contains three different high-symmetry stackings, namely AA, AB and domain wall (DW) stackings, similar to the strained configuration shown in Fig. \ref{fig_structure}(a). For the non-strain case, the rotation angle for the commensurate condition is \cite{lopes2007graphene,trambly2012numerical}
\begin{equation}
    \cos \theta=\frac{3i^2+3i+1/2}{3i^2+3i+1},\quad i=0,1,2\ldots
    \label{no_strain_commensurate_condition}
\end{equation}
where $i$ is an integer. The commensurate supercell vectors are
\begin{align}
\label{twist_vec}
\mathbf{L}_{1} & =i\mathbf{a}_{1}+\left(i+1\right)\mathbf{a}_{2},\nonumber\\
\mathbf{L}_{2} & =-\left(i+1\right)\mathbf{a}_{1}+\left(2i+1\right)\mathbf{a}_{2},
\end{align}
where $\mathbf{a}_1=a(1,0)$ and $\mathbf{a}_2=a(1/2,\sqrt{3}/2)$ are the lattice vectors of monolayer graphene, with $a\simeq 2.46\text{\AA}$ being the lattice constant.
Therefore, a commensurate supercell with twist angle $\theta$ and vectors $\mathbf{L}_{1,2}$ is exclusively identified by the integer $i$. 

When strain is introduced into the system, Eqs. (\ref{no_strain_commensurate_condition}) and (\ref{twist_vec}) become invalid, and a general formalism is needed to determine the twist, strain and moir\'e vectors of the commensurate structure.   

\subsubsection{Uniaxial, shear and biaxial strains}

We introduce three important strains that are reported by experiments. Assuming that the 2D system has a homogeneous
(position-independent) strain. Then, an arbitrary two-dimensional strain tensor is given by
\begin{equation}
   \mathcal{E}=
    \begin{pmatrix}
        \epsilon_{xx} & \epsilon_{xy} \\
         \epsilon_{xy}& \epsilon_{yy}
    \end{pmatrix},
    \label{eq_strain_type}
\end{equation}
where $\epsilon_{ij}=\left(\partial u_{i}/\partial x_{j}+\partial u_{j}/\partial x_{i}\right)/2$, with $\mathbf{u}$ the displacement vector that accounts for the deformation induced by stress \cite{landau2012theory, gurtin2010mechanics}. The three relevant types of strain that we shall consider are uniaxial, shear and biaxial (Figs. \ref{fig_structure}(c)-(e)). Their corresponding strain tensors read \cite{pereira2009tight, bi_designing_2019, kogl_moire_2023, escudero_designing_2024}
\begin{align}
\mathcal{E}_{u} & =R_{\phi}\left(\begin{array}{cc}
\epsilon & 0\\
0 & -\nu\epsilon
\end{array}\right)R_{-\phi} & \left(\mathrm{uniaxial}\right),\\
\mathcal{E}_{s} & =R_{\phi}\left(\begin{array}{cc}
0 & \epsilon\\
\epsilon & 0
\end{array}\right)R_{-\phi} & \left(\mathrm{shear}\right),\\
\mathcal{E}_{b} & =\left(\begin{array}{cc}
\epsilon & 0\\
0 & \epsilon
\end{array}\right) & \left(\mathrm{biaxial}\right).
\end{align}
Here $R_{\phi}\equiv R\left(\phi\right)$ is the rotation matrix, $\epsilon$ is the strain magnitude, $\phi$ is the strain direction relative to the $x$ axis, and $\nu$ is the Poisson's ratio ($\nu\simeq0.16$ in graphene). The shear strain can be written as
\begin{equation}
\mathcal{E}_{s}=R_{\phi+\pi/4}\left(\begin{array}{cc}
\epsilon & 0\\
0 & -\epsilon
\end{array}\right)R_{-\phi-\pi/4}.
\end{equation}
Comparing with the uniaxial strain tensor we then have the relation
\begin{equation}
\mathcal{E}_{s}\left(\epsilon,\phi\right)=\mathcal{E}_{u}\left(\epsilon,\phi+\pi/4,\nu\rightarrow1\right).\label{eq:uni_shear}
\end{equation}
This means that a shear strain with magnitude $\epsilon$ and direction $\phi_{s}$ can be thought as uniaxial strain with magnitude $\epsilon$ and direction $\phi_{u}=\phi_{s}+\pi/4$, but with Poisson's ratio $\nu\rightarrow1$ (i.e., the limit case in which the lateral contraction is equal to the applied longitudinal tension). 

In this paper, we restrict ourselves to TBG with a heterostrain, which refers to relative strains between two layers. In general, the heterostrain can be unintentional or intentional. The unintentional heterostrain is commonly generated without control during the sample growth or fabrication \cite{huder2018electronic}. The intentional heterostrain can be induced externally and designed carefully by well-established strain techniques, for instance, the substrate out-of-plane bending \cite{gao2021heterostrain}, process-induced strain \cite{pena2023moire}, and sliding-based strain \cite{sequeira2024manipulating,kapfer2023programming,carrasco2025twistraintronics} (see e.g. Ref. \cite{escudero2025geometrical} for more details about these strain techniques). In our model, we will specifically assume that the top layer is strained and the bottom layer is rotated. Note that our definition of heterostrain differs from that in Refs.~\cite{bi_designing_2019,escudero_designing_2024}, where the two layers are strained in opposite directions with equal magnitude, i.e. a symmetric configuration. Instead, our definition is closer to the experimental cases in Ref.~\cite{huder2018electronic,kerelsky_maximized_2019}, where the concept of heterostrain was first introduced. These two types of heterostrain show slightly different modification of the geometrical and electronic properties of the moiré systems. However, these two definitions induce practically the same perturbation effects to the TBG around the first magic angle and are indistinguishable in both theory \cite{escudero2025} and experiments \cite{yu2024twist,carrasco2025twistraintronics}. 

\subsubsection{Moiré geometry with twist and strain}
An application of strain in the top layer, and a rotation in the bottom layer, transform their lattice vectors as
\begin{align}
\mathbf{a}_{t,i} & =\left(\mathbb{I}+\mathcal{E}\right)\mathbf{a}_{i},\nonumber\\
\mathbf{a}_{b,i} & =R_{\theta}\mathbf{a}_{i},
\label{strainvec}
\end{align}
where $\eye$ is the $2\times2$ identity matrix. The reciprocal vectors follow as
\begin{align}
\mathbf{b}_{t,i} & =\left(\mathbb{I}+\mathcal{E}\right)^{-1}\mathbf{b}_{i},\nonumber\\
\mathbf{b}_{b,i} & =R_{\theta}\mathbf{b}_{i},
\label{reciprocal_vector}
\end{align}
where $\mathbf{b}_{i}$ are the reciprocal lattice vectors of the honeycomb lattice. The reciprocal moiré vectors $\mathbf{G}_{i}$ (see Fig. \ref{fig_structure}(b)) can then be calculated by taking the difference between the deformed lattice vectors in each layer~\cite{artaud_universal_2016,kogl_moire_2023, escudero_designing_2024,koshino2015interlayer}
\begin{equation}
\mathbf{G}_{i}=\mathbf{b}_{t,i}-\mathbf{b}_{b,i}.
\label{vec_g}
\end{equation}
The real space moiré vectors $\mathbf{L}_{i}$ are determined by the relation $\mathbf{L}_{i}\cdot\mathbf{G}_{j}=2\pi\delta_{ij}$. In principle, the moiré vectors $\mathbf{G}_{i}$ and $\mathbf{L}_{i}$ define the moiré structure \cite{kogl_moire_2023,bi_designing_2019}. However, in practice, the definition of $\mathbf L_i$ does not guarantee commensurability in the supercell. 

\subsubsection{Commensurate structures with twist and strain}

In this part, we develop a general geometrical formalism for commensurate structures in TSBG. The analysis of the commensurate supercell can be performed by expressing the moir\'e lattice vectors as function of those of the two graphene layers, as shown in Fig. \ref{fig_structure}(f)
\begin{equation}
    \begin{pmatrix}
        \mathbf{L}_{1}\\\mathbf{L}_{2}
    \end{pmatrix}
    =
    \begin{pmatrix}
        i&j\\k&l
    \end{pmatrix}
    \begin{pmatrix}
        \mathbf{a}_{t,1}\\\mathbf{a}_{t,2}
    \end{pmatrix}
    =\begin{pmatrix}
        m&n\\q&r
    \end{pmatrix}
    \begin{pmatrix}
        \mathbf{a}_{b,1}\\\mathbf{a}_{b,2}
    \end{pmatrix}.
    \label{general_form_comensurate_condition}
\end{equation}
Then the top and bottom graphene lattice can be related by a Park-Madden transformation matrix
\begin{equation}
    \begin{aligned}
    \begin{pmatrix}
        \mathbf{a}_{t,1}\\\mathbf{a}_{t,2}
    \end{pmatrix}
    &=
    \frac{1}{il-jk}
    \begin{pmatrix}
        lm-jq&ln-jr\\-km+iq&-kn+ir
    \end{pmatrix}
    \begin{pmatrix}
        \mathbf{a}_{b,1}\\\mathbf{a}_{b,2}
    \end{pmatrix} \\
    &=
    \begin{pmatrix}
        a&b\\c&d
    \end{pmatrix}
    \begin{pmatrix}
        \mathbf{a}_{b,1}\\\mathbf{a}_{b,2}
    \end{pmatrix}        
    \end{aligned}.
    \label{matrix_1}
\end{equation}
For an arbitrary transformation matrix composed of contributions of twist and strain, there are four variables ($a, b, c, d$) corresponding to a pair of eight integers ($i,j,k,l,m,n,q,r$) that define the Park-Madden matrix. The set of eight integers can be determined experimentally through atomically-resolved microscopy \cite{huder2018electronic,artaud_universal_2016}. 

The analysis of the commensurate supercell also can be expressed as a function of the elementary geometrical deformations in Eq.~(\ref{strainvec}), which in general depends on four parameters: the twist angle and the three components of the strain tensor. These four parameters completely determine, in principle, the (2 $\times$ 2) matrix in Eq.~(\ref{matrix_1}). If, for simplicity, one assumes that the system contains only uniaxial or shear strain, then one is left with only three parameters, namely, the twist and the strain magnitude and direction. Therefore, we introduce an additional -extremely small- biaxial strain in the top layer, so that Eq.~(\ref{strainvec}) becomes $\mathbf{a}_{t,i}\rightarrow\left(\mathbb{I}+\mathcal{E}_{u/s}+\mathcal{E}_{b}\right)\mathbf{a}_{i}$ [see Fig.~\ref{fig_structure}(g)]. Then the system has four parameters ($\theta,\epsilon_{u/s},\phi,\epsilon_{b}$) that can \textit{fully} determine the set of eight integers in Eq.~\eqref{matrix_1}.
To obtain the commensurate solutions, it is convenient to rewrite $\mathbf{a}_{t,i}\simeq\left(\mathbb{I}+\mathcal{E}_{b}\right)\left(\mathbb{I}+\mathcal{E}_{u/s}\right)\mathbf{a}_{i}$ by taking $\mathcal{E}_{b}\mathcal{E}_{u/s}\rightarrow0$ under the limit of small deformations. The transformation matrix that links $\left(\mathbf{a}_{t,1},\mathbf{a}_{t,2}\right)$ with $\left(\mathbf{a}_{b,1},\mathbf{a}_{b,2}\right)$ can then be written as
\begin{equation}
\left(\begin{array}{c}
\mathbf{a}_{t,1}\\
\mathbf{a}_{t,2}
\end{array}\right)=P\left(1+\epsilon_{b}\right)\left(\mathbb{I}+\mathcal{E}_{u/s}\right)R_{\theta}P^{-1}\left(\begin{array}{c}
\mathbf{a}_{b,1}\\
\mathbf{a}_{b,2}
\end{array}\right),
\label{matrix_2}
\end{equation}
where 
$P=\left(\begin{array}{cc}
1 & 0\\
1/2 & \sqrt{3}/2
\end{array}\right)$ 
is the basis of the triangular lattice vectors. By combining then Eqs.~(\ref{matrix_1}) and (\ref{matrix_2}) we can relate the physical parameters ($\theta,\epsilon_{u/s},\phi,\epsilon_{b}$) to the eight integers ($i,j,k,l,m,n,q,r$) \cite{artaud_universal_2016}. 

The described procedure to obtain commensurate structures can be generalized to any arbitrary strain tensor, with the inclusion of lattice relaxation effects \cite{lai2025moire}. A detailed, step-by-step algorithm of how to obtain commensurate structures with any twist and strain, with some example solutions, can be found in the Secs. S1 and S2 of the Supplemental Materials (SM) \cite{SM}.

\subsection{Dirac point shifts from moiré geometry deformation}
\label{sec_mBZ}

When the strain is small, we can consider that the length of the two moiré vectors remains, to a first approximation, practically equal. In that case, the borders of the mBZ are given by the three points (see Fig. \ref{fig_structure}(b)) \cite{escudero_designing_2024}: 
\begin{align}
\mathbf{K}_{1} & =-\frac{(1+2\chi)\mathbf{G}_{1}-\lambda\mathbf{G}_{2}}{2(1+\chi)},\nonumber\\
\mathbf{K}_{3} & =\mathbf{K}_{1}+\mathbf{G}_{1},\nonumber\\
\mathbf{K}_{5} & =\mathbf{K}_{1}+\mathbf{G}_{1}-\lambda\mathbf{G}_{2},
\end{align} 
and their negatives. Here $\chi=\abs{\vb{G}_{1} \cdot \vb{G}_{2}}/\abs{\vb{G}_{1} \cdot \vb{G}_{1}}$ and $\lambda=\mathrm{sign}(\vb{G}_{1}\cdot \vb{G}_{2})+\delta_{0,\vb{G}_{1}\cdot \vb{G}_{2}}$. 
When the angle between $\vb{G}_1$ and $\vb{G}_2$ is $120\degree$, the moiré pattern is triangular and the points above collapse to yield a hexagonal mBZ.  

\begin{figure}
    \centering
    \includegraphics[width=0.9\linewidth]{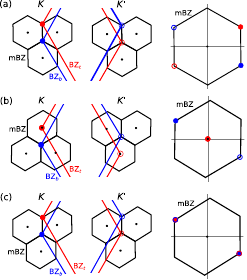}
    \caption{(a) Position of the Dirac points projected within the mBZ, for the commensurate solutions of magic angle $\theta\sim1.05^{\circ}$ without strain. (b) The same case as (a) for  uniaxial strain $\epsilon_{u}\sim0.1\%$. (c) The same case as (a) for shear strain $\epsilon_{s}\sim0.1\%$. In each case, plots on the left and middle sides show the Dirac points in the top (red) and bottom (blue) layers, and the mBZ periodically translated from the origin for the $K$ and $K^{\prime}$ valleys, respectively. Plots on the right side show the position of the Dirac points within the mBZ, when translated by the moir\'e vectors. In (b) and (c), the positions of Dirac points from the top strained layer are slightly shifted from the five possible positions in the mBZ (see the Eq. \eqref{projection}). }
    \label{fig:fold}
\end{figure}

Next, we check how the undeformed Dirac points, at the corners of the graphene BZ, are mapped into the mBZ. First we translate Eq. (\ref{general_form_comensurate_condition}) to the reciprocal space as:
\begin{align}
\left(\begin{array}{c}
\mathbf{b}_{t,1}\\
\mathbf{b}_{t,2}
\end{array}\right) & =\left(\begin{array}{cc}
i & k\\
j & l
\end{array}\right)\left(\begin{array}{c}
\mathbf{G}_{1}\\
\mathbf{G}_{2}
\end{array}\right),\nonumber\\
\left(\begin{array}{c}
\mathbf{b}_{b,1}\\
\mathbf{b}_{b,2}
\end{array}\right) & =\left(\begin{array}{cc}
m & q\\
n & r
\end{array}\right)\left(\begin{array}{c}
\mathbf{G}_{1}\\
\mathbf{G}_{2}
\end{array}\right).
\end{align}
The projection of a Dirac point at $\mathbf{K}_{0}=\mathbf{b}_{1}/3+2\mathbf{b}_{2}/3$ is then given by:
\begin{align}
\mathbf{K}_{0} & =\left(\begin{array}{cc}
\mathbf{G}_{1}, & \mathbf{G}_{2}\end{array}\right)\left(\begin{array}{c}
\frac{1}{3}i+\frac{2}{3}j\\
\frac{1}{3}k+\frac{2}{3}l
\end{array}\right)\nonumber \\
 & =\left(\begin{array}{cc}
\mathbf{G}_{1}, & \mathbf{G}_{2}\end{array}\right)\left(\begin{array}{c}
\frac{1}{3}m+\frac{2}{3}n\\
\frac{1}{3}q+\frac{2}{3}r
\end{array}\right).\label{projection}
\end{align}
For different pairs of ($i,j,k,l,m,n,q,r$), there are five different types of 
projections: $(0, 0)$, $(0, \frac{1}{3})$, 
$(\frac{1}{3}, 0)$, $(\frac{1}{3}, \frac{1}{3})$, $(\frac{1}{3}, \frac{2}{3})$. These projections will always fall into the five high-symmetry points.

The situation under strain changes because the borders of the mBZ are no longer located at $\mathbf{G}_{1}/3+2\mathbf{G}_{2}/3$ (and translations by reciprocal moiré vectors), as in the non-strain case. Consequently, the Dirac points in the mBZ are rather located at arbitrary, strain-dependent positions. Some examples of the geometrical positions of the Dirac points in TSBG are plotted in Figs. \ref{fig:fold}(b)-(c). The Dirac points are effectively away from the corners of the mBZ, and are no longer degenerate. It should be noted that our analysis here only accounts for the geometrical position of the Dirac points in each layer, i.e., the projection of the borders of their respective BZ. As we will discuss in the following sections, under strain and relaxation the Dirac points are also sightly shifted by strain-induced fields. Moreover, due to the broken symmetries under strain, the actual position of the moiré Dirac points is further influenced by the moiré potential that couples the two layers~\cite{sinner_strain-induced_2023,escudero_designing_2024, escudero2025}.

In the particular non-strain case, the commensurate condition is given by Eq. \eqref{no_strain_commensurate_condition}, and Eq. (\ref{projection}) becomes
\begin{equation}
\mathbf{K}_{0}=\left(\begin{array}{cc}
\mathbf{G}_{1}, & \mathbf{G}_{2}\end{array}\right)\left(\begin{array}{c}
2/3\\
1/3
\end{array}\right),
\label{K0_nostrain}
\end{equation}
which means that if there is no strain, the Dirac cones are always projected to the corners of the mBZ (Fig. \ref{fig:fold}(a)). As discussed in the following section, the shift of the Dirac cones and the deformation of the mBZ under strain explain why, in the TB calculations, the $K_1$, $K_2$, $K_3$, $K_4$ and $\Gamma$ points appear displaced compared to the unstrained case.

\subsection{Numerical models}

We first compute the electronic structure of commensurate TSBG by using a full atomistic TB Hamiltonian. Then, we construct a general effective continuum model to describe the TB results. A key step for comparing the TB and continuum results is identifying the Dirac points in the strained mBZ, discussed in Sec.~\ref{sec_mBZ} (see also Fig.~\ref{fig:fold}).

\subsubsection{Tight-binding model}
The TB Hamiltonian of the TSBG is generated by only considering the $p_z$ orbital of the carbon atom as \cite{trambly2012numerical}
\begin{equation}
H=\sum_i\varepsilon_i c^{\dagger}_{i} c_{i}+ \sum_{\langle i,j \rangle} t_{ij}c^{\dagger}_{i} c_{j},
\end{equation}
where $c_i$ is an annihilation operator for the $i$ state, $\varepsilon_i$ is the on-site potential, $\langle i,j \rangle$ is the sum over index with $i \neq j$, $t_{ij}$ is the hopping integral between $i$ and $j$ orbitals, which follows the Slater-Koster (SK) formalism \cite{los2005improved, kolmogorov2005registry}. As detailed above, including a very small biaxial strain guarantees commensurate solutions for arbitrary twist and uniaxial/shear strain, from which we calculate the bands. For conciseness, in the main text we quote the approximate twist and strain values of the commensurate solutions. The exact strain and twist parameters are listed in Sec.~\textcolor{red}{S2} \cite{SM}. We identify the valley character with a valley operator~\cite{ramires_electrically_2018,ramires_impurity-induced_2019}, and relax the moiré supercell with LAMMPS~\cite{plimpton1995fast}. All the TB calculations are performed in the TBPLaS simulator \cite{li2023tbplas}. Details of the TB calculations are given in Sec.~\textcolor{red}{S3} \cite{hams_fast_2000,polizzi_density-matrix-based_2009, SM}.

\subsubsection{Strain-extended continuous model}
 In the continuum model, the electronic properties of the system are accounted by the coupling of the Dirac points in each layer with an effective moir\'e-induced interlayer potential. Neglecting couplings between different valleys in each layer, the continuum model Hamiltonian for the $K$ valley takes the form~\cite{huder2018electronic,bi_designing_2019,pantaleon2022interaction}
\begin{equation}
H=\left(\begin{array}{cc}
h_{b}\left(\mathbf{k}\right)+\mathcal{S}_{b} & U^{\dagger}\left(\mathbf{r}\right)\\
U\left(\mathbf{r}\right) & h_{t}\left(\mathbf{k}\right)+\mathcal{S}_{t}
\end{array}\right),
\end{equation}
where the index $\ell=b,t$ refers to the bottom and top layers, respectively. $h_{\ell}\left(\mathbf{k}\right)$ is the Dirac Hamiltonian relative to the twisted and strained Dirac points:
\begin{equation}
h_{\ell}\left(\mathbf{k}\right)=-\hbar v\boldsymbol{\sigma}\cdot R^T_{\theta_{\ell}}\left(1+\mathcal{E}_{\ell}\right)\left(\mathbf{k}-\mathbf{K}_{\ell}\right),
\end{equation}
where $\boldsymbol{\sigma}=\left(\sigma_{x},\sigma_{y}\right)$ are the Dirac matrices and $\mathbf{K}_{\ell}=\left(1-\mathcal{E}_{\ell}\right)R_{\ell}\left(\theta_{\ell}\right)\mathbf{K}_0$, where $\mathbf{K}_0=-\left(2\mathbf{b}_{1}+\mathbf{b}_{2}\right)/3$ is the Dirac point of a honeycomb layer. The strain introduces an additional term $\mathcal{S}_{\ell}$ that includes a deformation and gauge potential~\cite{suzuura2002phonons, manes2007symmetry, vozmediano2010gauge, Oliva2013Understanding,Oliva2015Generalizing}
\begin{equation}
\mathcal{S}_{\ell}=\mathbb{I}V_{\ell}-\hbar v\boldsymbol{\sigma}\cdot R^T_{\theta_{\ell}}\left(1+\mathcal{E}_{\ell}\right)\mathbf{A}_{\ell},
\end{equation}
where
\begin{align}
V_{\ell} & =g\left(\epsilon_{xx}^{\ell}+\epsilon_{yy}^{\ell}\right),\label{eq:V_strain}\\
\mathbf{A}_{\ell} & =\frac{\sqrt{3}}{2a}\beta\left(\epsilon_{xx}^{\ell}-\epsilon_{yy}^{\ell},-2\epsilon_{xy}^{\ell}\right),\label{eq:A_strain}
\end{align}
with $g=4$ eV and $\beta=3.14$ for graphene~\cite{vozmediano2010gauge, bi_designing_2019}. The scalar potential $V_{\ell}$ shifts the Dirac points in energy, resembling the effect of a perpendicular electric field. The vector potential $\mathbf{A}_{\ell}$ shifts the Dirac points in momentum and accounts for the strain-induced change in the hopping energies within the Dirac approximation~\cite{suzuura2002phonons, manes2007symmetry}. 

The moiré-induced coupling potential $U\left(\mathbf{r}\right)$ depends on the interplay between twist and strain through its Fourier expansion in terms of the moiré vectors~\cite{moon2013optical, koshino2015interlayer, escudero_designing_2024}. At small deformations (i.e., low twist and strain) the Fourier expansion can be truncated to the first three leading order terms~\cite{escudero_designing_2024}
\begin{equation}
U\left(\mathbf{r}\right)=U_{1}+U_{2}e^{i\mathbf{G}_{1}\cdot\mathbf{r}}+U_{3}e^{i\left(\mathbf{G}_{1}+\mathbf{G}_{2}\right)\cdot\mathbf{r}},
\label{U_strain}
\end{equation}
where
\begin{equation}
U_{j}=\left(\begin{array}{cc}
u_{0} & u_{1}e^{-i\omega_{j}}\\
u_{1}e^{i\omega_{j}} & u_{0}
\end{array}\right),
\end{equation}
with $\omega_{j}=\left(j-1\right)2\pi/3$. Here $u_{0}$ and $u_{1}$ are the effective AA and AB/BA hopping amplitudes. The values of these hopping energies are expected to depend on the local lattice deformations. In general, a rigid configuration implies equal hopping energies $u_{0}=u_{1}$. This results in the remote bands always touching the flat middle bands, i.e., there is no gap between them \cite{bistritzer_moire_2011, moon2013optical}. However, relaxation effects tend to shrink and increase the interlayer distance of the energetic AA regions, compared to the most favorable AB/BA, thus leading to an effective smaller hopping $u_{0}<u_{1}$ \cite{koshino2018maximally,koshino2020effective,carr2019exact,carr2018relaxation,ezzi2024analytical,ceferino2024pseudomagnetic}. The main effect of this is to open a gap between the flat middle bands and the remote bands.

\begin{figure*}[htbp]
    \centering
    \includegraphics[width=\linewidth]{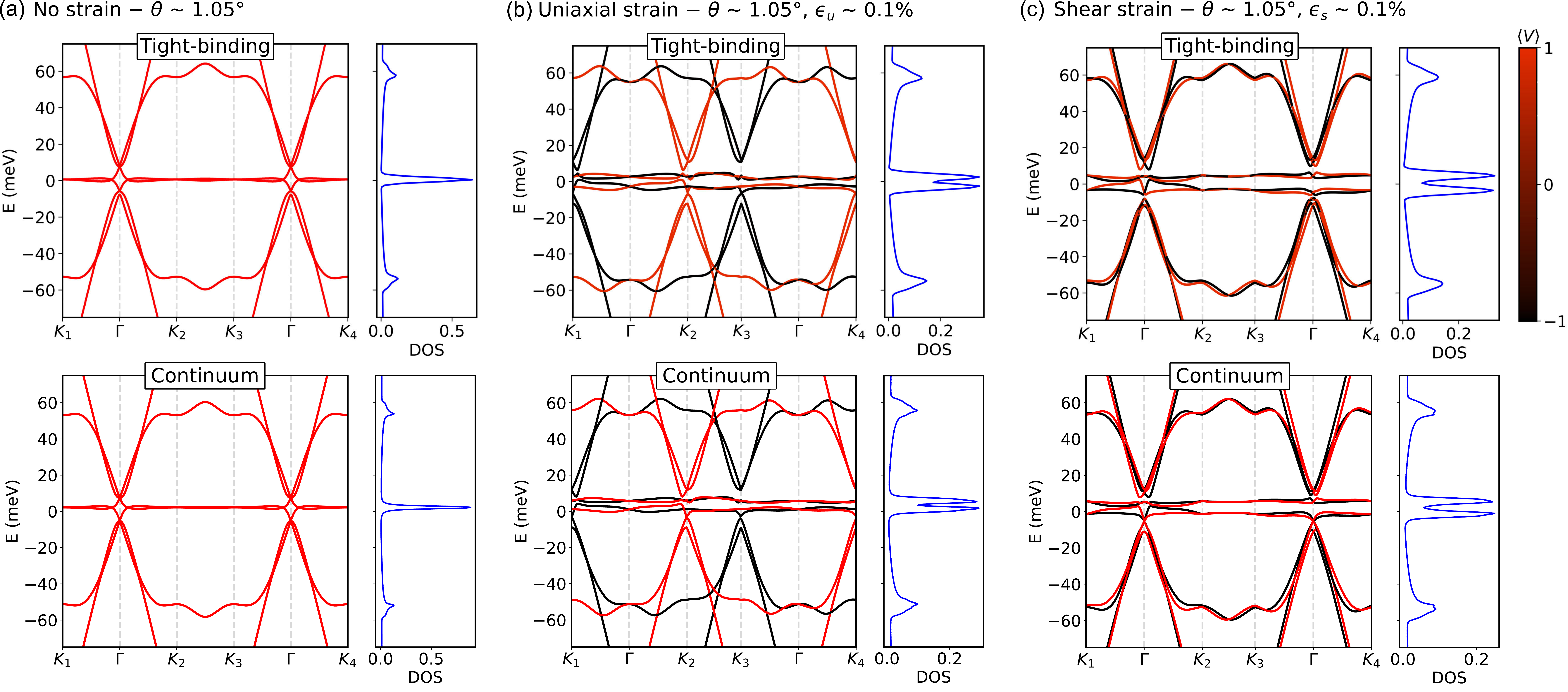}
    \caption{Band structure and DOS for the commensurate structures of TBG with $\theta=1.05\degree$, $\phi=0\degree$ and (a) no strain, (b) uniaxial strain $\epsilon_{u}=0.1\%$, (c) shear strain $\epsilon_{s}=0.1\%$, calculated by the TB (top panel) and continuum (bottom panel) models. In the TB band structures of (b) and (c), the color represents the expectation of the valley operator, with $\langle \hat{V}_z \rangle \approx 1$ if a state belongs to valley $K$ (red line) and $\langle \hat{V}_z \rangle \approx-1$ if a state belongs to valley $K'$ (black line). The momentum path is illustrated in Fig. \ref{fig_structure}(b).}
    \label{fig_uni_band}
\end{figure*}

\section{Electronic structures}
\label{TB results}

\subsection{Strain effect}

Figure~\ref{fig_uni_band} shows both the TB and continuum model results for the band structure and DOS at the first magic angle $\theta\sim1.05^\circ$, for different cases: without strain, and with uniaxial and shear strains, both with directions $\phi=0$ (see also Sec. \textcolor{red}{S4} for the cases with $\theta=1.6^\circ$ and $\theta=0.93^\circ$, and with different strain strengths \cite{SM}). The fitted continuum model parameters $u_{0}~=~u_{1}~=~0.1~\mathrm{eV}$ and $\hbar v / a = 2.15~\mathrm{eV}$ provide excellent agreement with the TB calculations. As described below, strain introduces four generic features in the electronic structures.

First, the middle narrow bands are extremely sensitive to strain. In the absence of strain, the band structure at the first magic angle exhibits characteristic flat bands near the charge neutrality point (CNP), leading to a pronounced peak in the DOS. Upon introducing strain into the system, we see that the narrow peak splits into two peaks in the DOS, with reduced magnitude~\cite{Nguyen2015Strain}. The strain broadens the width (the difference between the extreme values within one band) of the narrow bands. Moreover, the energy separation (indicated by the separation of the van Hove singularities (vHs)) between the conduction (CB) and valence (VB) bands increases with increasing strain strength \cite{SM}. In TSBG with $\theta=1.6^{\circ}$, strain induces multiple DOS peaks from both the valence and conduction bands, showing that strain can effectively generate higher-order vHs~\cite{huder2018electronic,bi_designing_2019,Mesple2021Heterostrain}. We emphasize that for each twist angle, the minimum bandwidth always appears around the small strain region (see narrow band dome in Sec. \textcolor{red}{S5} \cite{SM}).

Second, the strain breaks the $C_3$ symmetry and lifts the valley degeneracy along the high-symmetry points of the mBZ~\cite{huder2018electronic,bi_designing_2019,Long2023Electronic}. In the unstrained case, the Dirac points are located at the corners of the mBZ. The conduction and valence narrow bands are connected by two Dirac crossings in each valley, protected by the $C_{2z}T$ symmetry. When strain is introduced, the breaking of $C_3$ symmetry lifts the valley degeneracy, as confirmed by the expectation value of the valley operator at each band state (details in Sec.~\textcolor{red}{S3} \cite{SM}). In the band structure, this manifests as a separation between the red and black curves corresponding to different valleys. A similar valley splitting occurs in TBG/hBN heterostructures, where the aligned hBN substrate breaks valley degeneracy~\cite{long2022atomistic,Long2023Electronic}. This lifting of valley degeneracy may account for the experimentally observed fourfold, rather than eightfold, Landau level degeneracy near the CNP \cite{cao_unconventional_2018,yankowitz2019tuning}. Nevertheless, the band structures of the $K$ and $K'$ valleys remain related by time-reversal symmetry, and the conduction and valence narrow bands stay connected, indicating that strain alone cannot open a gap in TBG due to the preserved $C_{2z}T$ symmetry.

Third, the Dirac points are no longer located at the corner of the mBZ, but rather around the five possible projections without strain, see Eq. \eqref{projection}. This further reflects the lack of $C_3$ symmetries in the presence of strain. The geometric position of the Dirac points with $0.1\%$ strain (uniaxial and shear) are plotted in Figs. \ref{fig:fold}(b)-(c). However, the exact positions of the Dirac points, identified from the energy maps (Fig. \ref{fig_energy_map}), are found to be slightly different. As noted before, and discussed in detail in the continuum model results, there are two additional sources of corrections of the Dirac point position: (\romannum{1}) the gauge potential induced by strain (external strain and lattice relaxation); (\romannum{2}) the deformed moiré potential that couples the two graphene layers \cite{escudero2025}. Moreover, there is an energy shifting of the two Dirac points within a valley, resulting in a finite (lower) density of states at the CNP. 

Finally, we observe that the remote bands always give two additional peaks flanking the middle narrow bands, which do not change under low strain. In particular, the DOS peaks around $\pm 60$ meV from the remote bands are unchanged in the presence of strain \cite{SM}. This behavior is consistent with recent experimental results that suggest that the remote bands are insensitive to strain~\cite{yu2024twist} and their optical interband transitions can be used as a fingerprint of the twist angle~\cite{Li2024Infrared}.  

\subsection{Comparison between uniaxial and shear strain}
When comparing the effects of uniaxial and shear strain, we observe clear differences. Under shear strain, the energy separation between the vHs is larger, and the middle bands become narrower across most regions of the mBZ (see also Fig. \ref{fig_energy_map}). This can be attributed to geometric modulation: for the same strain magnitude, shear strain produces a stronger distortion of the moiré pattern than uniaxial strain, see Eq. \eqref{eq:uni_shear}. Similar trends are found for twist angles beyond the first magic angle \cite{SM}. 

In many experimental TSBG samples, the distorted moir\'e pattern is commonly interpreted under the assumption of uniaxial heterostrain. This assumption is frequently used when extracting twist and strain from STM topography. However, recent STM measurements have shown that shear strain can also appear as the dominant contribution \cite{yu2024twist,carrasco2025twistraintronics}. Particularly, the large energy separation of the narrow bands can only be explained by the effect of the shear strain \cite{yu2024twist}. Therefore, in order to properly interpret the experimental observations, it is necessary to identify the strain type.

Within the continuum model, the difference between the uniaxial and shear strain effect can be related to their corresponding scalar and gauge potentials. For uniaxial strain, the scalar potential is independent of the strain direction 
\begin{equation}
V_{u}=g\epsilon\left(1-\nu\right)\sim3.35\epsilon\,\mathrm{eV},
\end{equation}
while for shear strain it vanishes, $V_{s}=0$. On the other hand, the magnitude of the vector potential for uniaxial reads
\begin{equation}
\left|\mathbf{A}_{u}\right|=\frac{\sqrt{3}}{2a}\beta\epsilon\left(1+\nu\right)\sim12.8\epsilon\,\mathrm{nm^{-1}},
\end{equation}
which is independent of the strain direction. For shear strain one has the same expression, but with $\nu\rightarrow1$ [or, equivalently, with a strain magnitude $\epsilon_{s}\rightarrow\epsilon\left(1+\nu\right)$]; see Eq. \eqref{eq:uni_shear}.  Since $\left|\mathbf{K}\right|\sim17\,\,\mathrm{nm^{-1}}$ in graphene, the momentum space shift induced by $\mathbf{A}$ is very small at low strain magnitudes, even on the scale of the moiré BZ. However, such a small shift can still significantly alter the electronic properties. It is worth noting that the net shift of the decoupled Dirac points in each monolayer, namely $\mathbf{K}\rightarrow\left(1-\mathcal{E}\right)R_{\theta}\mathbf{K}^{0}-\mathbf{A}$, does not generally match the actual position of the moiré Dirac points within the mBZ. This is due to the effect of the moiré potential and the role of the strain in breaking the symmetries of the system \cite{escudero_designing_2024, escudero2025}. 

\begin{figure}[t]
    \centering
        \includegraphics[width=\linewidth]{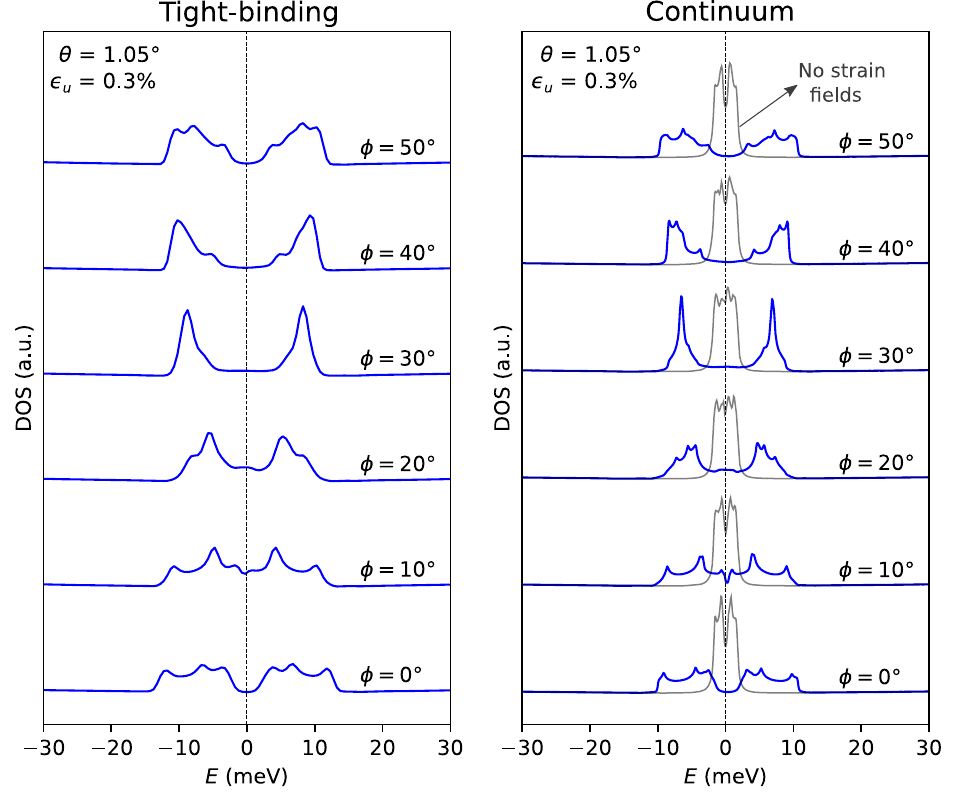}
   \caption{Evolution of DOS with uniaxial strain direction $\phi$ in TSBG with $\theta=1.05\degree$ and $\epsilon_u=0.3\%$, calculated by the TB (left side) and continuum (right side) models. We label $E=0\ \text{meV}$ with black vertical dashed lines. The curves are relatively shifted to make the plot clear.} 
    \label{fig_dos_compare}
\end{figure}

\subsection{Strain direction effect}

In unstrained TBG, the system possesses sixfold rotational symmetry, allowing the strain direction to be restricted to the range $\phi \in [0, 60^{\circ})$. Since uniaxial and shear strains are related by a $45^{\circ}$ rotation [cf. Eq.~(\ref{eq:uni_shear})], we focus on the uniaxial strain case. The evolution of the DOS as a function of strain direction is shown in Fig.~\ref{fig_dos_compare} (see also the Secs. S6 and S9 for extended results \cite{SM}). The results agree well with previous continuum model studies~\cite{bi_designing_2019}. 
We again observe an excellent agreement between the TB and continuum results. Interestingly, we see that the main features of the DOS, namely the highly sensitive splitting of the vHs as a function of the strain direction, is well captured only when the strain fields $V$ and $\mathbf{A}$ are included. Thus, although the strain effect in the continuum model Hamiltonian generally comes from both the change in the moiré potential $U\left(\mathbf{r}\right)$ (due to the geometric variation of the strained moiré vectors) and the inclusion of the strain-induced potential $\mathcal{S}_{\ell}$, the latter seem to have a greater effect on the middle narrow bands. 

As the strain direction changes, the DOS peaks exhibit strong variations in both intensity and energy, particularly for $\theta = 1.05^{\circ}$. At the magic angle, when $\phi=30^{\circ}$ only one prominent peak appears in each band, while for $\theta = 1.6^{\circ}$ a sharp peak emerges at the CNP when $\phi = 20^{\circ}$ \cite{SM}. The strain direction also shifts the position of the Dirac points within the mBZ (see Fig.~\textcolor{red}{S6} \cite{SM}). Our results reveal an important aspect often overlooked in previous studies: the strain direction plays a decisive role in determining both the bandwidth and the energy separation of the narrow bands.

\begin{figure*}
    \centering
    \includegraphics[width=\linewidth]{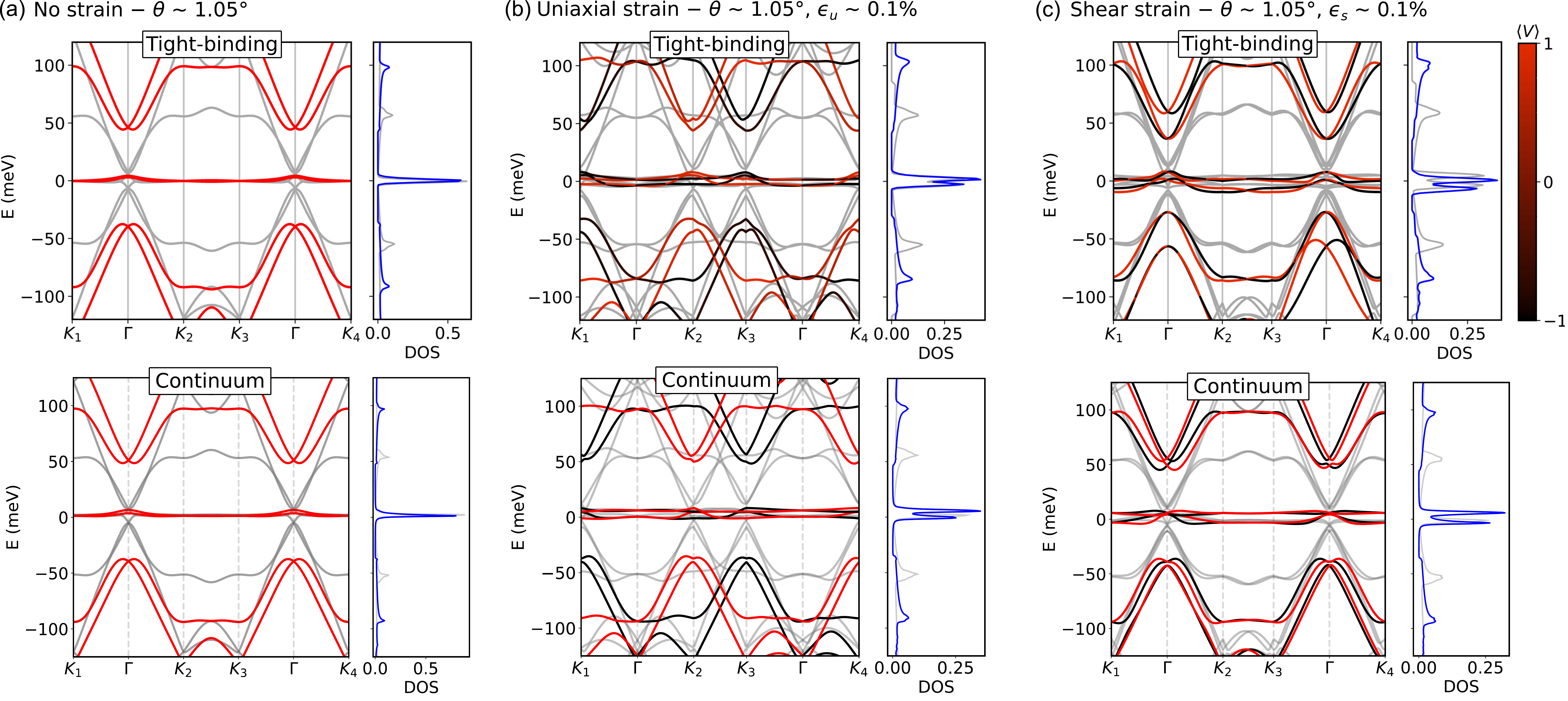}
    \caption{Band structure and DOS for relaxed TBG with $\theta=1.05\degree$ and (a) no strain, (b) uniaxial strain $\epsilon_{u}=0.1\%$, (c) shear strain $\epsilon_{s}=0.1\%$. The band structure and DOS of rigid cases are plotted with gray dots. The colors in the band structure are the same as in Figure \ref{fig_uni_band}. Note that, in the plots, the energy range in the relaxed case is almost two times larger than the energy range of the rigid case in Figure \ref{fig_uni_band}.}
    \label{fig_relax_band}
\end{figure*}

Specifically, we observe that at low twist angles the main effect of the strain on the electronic properties - in particular the strain direction - comes from the gauge potential $\mathbf{A}$. For any rotation $\phi$ of the strain tensor $\mathcal{E}\rightarrow R_{\phi}\mathcal{E}R_{-\phi}$, as considered in Fig. \ref{fig_dos_compare}, the gauge potential $\mathbf{A}$ transforms as
\begin{equation}
\mathbf{A}\left(\mathcal{E}\right)\rightarrow\mathbf{A}\left(R_{\phi}\mathcal{E}R_{-\phi}\right)=R_{-2\phi}\mathbf{A}\left(\mathcal{E}\right),
\end{equation}
that is, it simply rotates by $-2\phi$. Thus, the magnitude of the gauge potential is invariant, and only the direction of $\mathbf{A}$ changes when the strain direction is modified. The high sensitivity of the DOS to the strain direction around the magic angle, as seen in Fig. \ref{fig_dos_compare}, then reflects that it is the actual direction of the vector potential that plays the most significant role in modifying the electronic spectra. This is in line with previous studies indicating that the moiré coupling, and the emergence of flat bands, depends critically on the relative orientation between the Dirac points and the momentum transfer vectors \cite{bi_designing_2019, escudero_designing_2024, escudero2024diagrammatic}. 

In fact, in linear Dirac band systems, the gauge potential is the relevant term that significantly modifies the electronic properties \cite{bi_designing_2019}, and the scale potential only shifts the Dirac points in energy \cite{escudero_designing_2024,bi_designing_2019}. However, in parabolic band systems, the scale potential has a prominent contribution to the band structure, for example, by controlling the energy gap and flattening the band edges \cite{choi2010controlling}. In particular, in rhombohedral multilayer graphene with a quasi-one dimensional geometry, the scalar term\footnote{The relevant effect of the scalar potential is brought to our attention by discussions with Haim Beidenkopf´s group.} induces enriched correlated behaviors.   

\subsection{Lattice relaxation effect}

\begin{figure}
    \centering
    \includegraphics[width=0.9\linewidth]{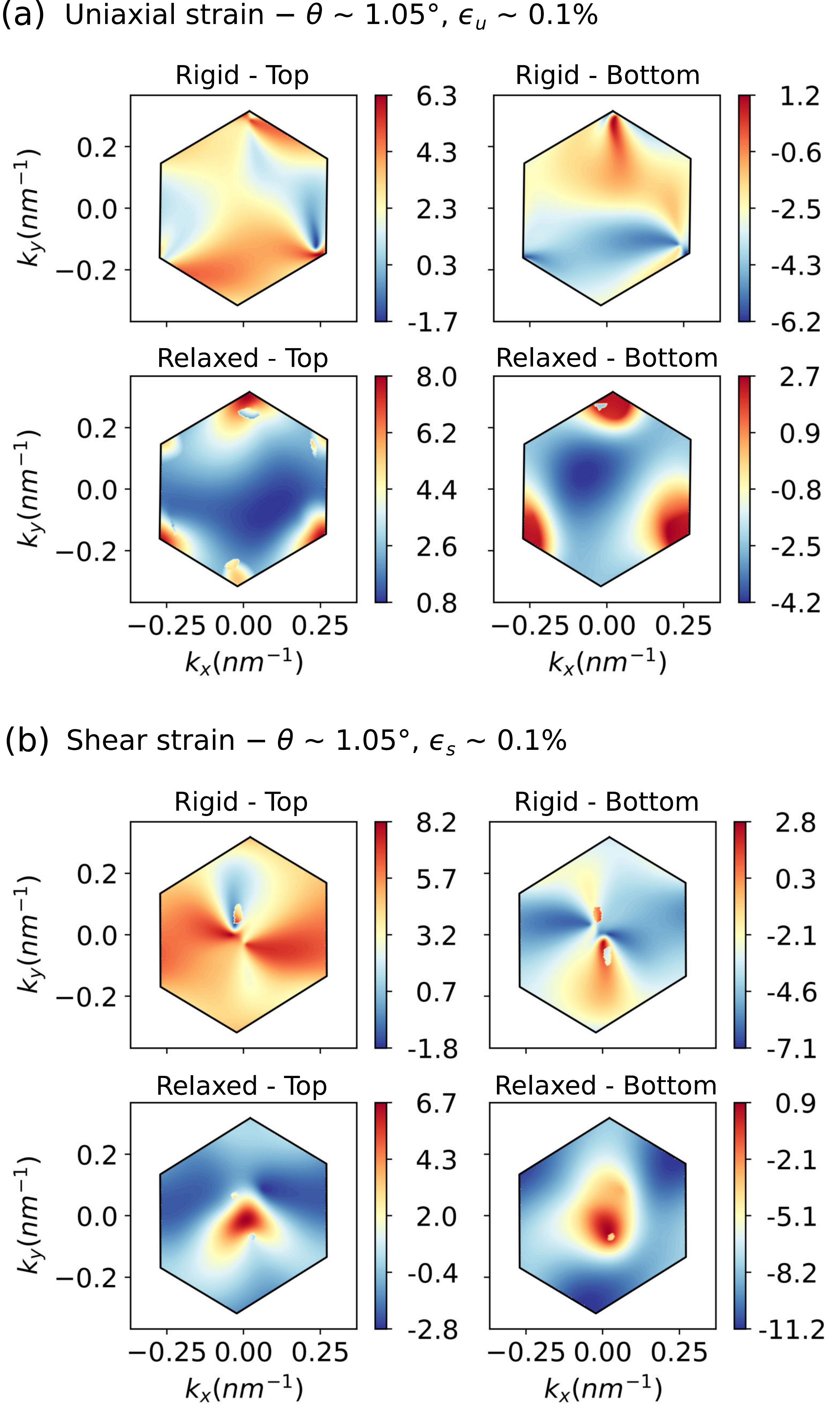}
    \caption{Comparison of energy map of the top and bottom narrow bands (tight-binding results) between relaxed and rigid  TBG with $\theta=1.05\degree$ and (a) uniaxial strain $\epsilon_{u}=0.1\%$, (b) shear strain $\epsilon_{s}=0.1\%$. The mBZ is illustrated with black line. The color represents the energy with unit meV. We only show the energy map of narrow bands from valley $K$ in the mBZ, which are identified by the valley operator.}
    \label{fig_energy_map}
\end{figure}

The relaxed geometry exhibits two main global features. First, the lattice relaxation patterns of graphene moiré structures, with and without strain, are qualitatively similar: the AA regions shrink, while the AB regions expand into triangular domains~\cite{guinea2019continuum}. Particularly, at low strain, the in-plane and out-of-plane displacements of TBG without and with strain show a high degree of consistency (see Sec.~\textcolor{red}{S7} \cite{SM}). Second, strain drives a structural transition in the DW network \cite{ouyang2025}. In nonstrain TBG, the DWs are of the shear type, characterized by a Burgers vector parallel to the DW. When strain is applied, the angle between the Burgers vector and the DW boundary changes, leading to a mixed configuration containing both shear and tensile DWs~\cite{lebedeva2020two,mesple2023giant}. The atomic displacements within the DW region differ significantly between the strained and unstrained cases. This structural discrepancy may alter the electronic states localized at the DWs, which typically lie at higher energies~\cite{nguyen2021electronic,timmel2020dirac}. 

As noted above, the simplest way to account for relaxation effects in the continuum model is to consider an unequal ratio $u_{0}<u_{1}$ between the hopping energies of AA and Bernal stacking. However, although this captures the opening of a gap between the narrow and remote bands, it still misses to capture a relaxation-induced particle-hole asymmetry \cite{koshino2020effective, kang2023pseudomagnetic}. This effect can be captured by including next-order nonlocal (momentum-dependent) corrections to the moiré potential \cite{fang2019angle, koshino2020effective, kwan2021kekule}. To leading order, the matrix elements of the nonlocal moiré potential
$U_{\mathrm{NL}}$ read \cite{fang2019angle, kwan2021kekule}
\begin{align}
\left\langle \mathbf{k},t\right|U_{\mathrm{NL}}\left|\mathbf{k}',b\right\rangle  & =\sum_{j=1}^{3}U_{\mathrm{NL,}j}\left(\mathbf{k},\mathbf{k}'\right)\delta_{\mathbf{k}'-\mathbf{k},\mathbf{G}'_{j}},
\end{align}
where $\mathbf{G}'_{1}=\mathbf{0},\mathbf{G}'_{2}=\mathbf{G}_{2},\mathbf{G}'_{3}=\mathbf{G}_{1}+\mathbf{G}_{2}$, while $U_{\mathrm{NL,}j}\left(\mathbf{k},\mathbf{k}'\right)=-\left(T_{j}p_{-}+T_{j}^{\dagger}p_{+}\right)/2$ with
\begin{align}
T_{j} & =\left(\begin{array}{cc}
\lambda_{1}e^{-i\omega_{j}} & \lambda_{2}e^{i\omega_{j}}\\
\lambda_{3} & \lambda_{1}e^{-i\omega_{j}}
\end{array}\right).
\end{align}
Here, as before, $\omega_{j}=\left(j-1\right)2\pi/3$, while $p_{\pm}=p_{x}\pm ip_{y}$, where $\mathbf{p}=\left(p_{x,}p_{y}\right)$ is the vector sum of the momenta in the top and bottom layers, relative to the positions of their Dirac points, i.e., $\mathbf{p}=\left(\mathbf{k}-\mathbf{K}_{t}\right)+\left(\mathbf{k}'-\mathbf{K}_{b}\right)$. 

Figure~\ref{fig_relax_band} presents a comparative analysis of the relaxed and rigid band structures and DOS under strain in both TB and continuum cases. To fit the TB results we consider $u_{1}=0.096\,\mathrm{eV}$, $u_{0}=0.05952\,\mathrm{eV}$ for the local moiré potential, $\hbar v/a=2.13\,\mathrm{eV}$, and $\lambda_{1}=9\,\mathrm{meV\cdot nm}$, $\lambda_{2}=18\,\mathrm{meV\cdot nm}$
and $\lambda_{3}=0$ for the nonlocal moiré potential \cite{kwan2021kekule}. The obtained results show again good agreement with the TB results with relaxation.
Improvements in the continuum model could be further obtained by accounting the local distortions of the AA, AB and DW regions, which introduce periodic pseudomagnetic magnetic fields \cite{vafek2023continuum, kang2023pseudomagnetic, kang2025analytical}.

The most notable effect remains the gap opening between the remote and narrow bands induced by relaxation. Moreover, the lattice relaxation increases the energy separation between the valence and conduction narrow bands, and broadens the width of the narrow bands, as also shown in Fig. \ref{fig_energy_map}. The lattice relaxation also introduces a pronounced electron–hole asymmetry in the TB results. The DOS peak in the conduction band is larger than that in the valence band, although both peaks exhibit nearly equal magnitudes at $\theta = 0.93^{\circ}$ and $\theta = 1.6^{\circ}$ \cite{SM}. The asymmetry between the DOS peaks associated with the narrow bands, as well as its evolution with twist angle, is consistent with recent experimental observations in which similar behavior was reported as the twist angle and strain were varied in the same device~\cite{yu2024twist}. The continuum model effectively captures the particle-hole asymmetry only when the nonlocal moiré potential is taken into account \cite{SM}. Interestingly, we find that the gap between the remote conduction and valence bands remains nearly constant under strain, while the separation between the remote and narrow bands decreases as strain increases, reflecting their broadening. 

Since the main effect of the nonlocal moiré potential is to introduce a small particle-hole asymmetry, from here on we simplify the continuum model by keeping only the local moiré potential \cite{moon2013optical, koshino2018maximally}. This approximation aligns with the motivation of the continuum model, which aims to provide a minimal model that captures the main features seen in the TB band structures \cite{lopes2007graphene, bistritzer_moire_2011}. Although our results below can be directly extended to include the effect of the nonlocal moiré potential, we expect this to only slightly change quantitatively the strain-dependence behavior, without affecting our main conclusions.

\subsection{Narrow bands with strain and twist}
\begin{figure*}[t]
    \centering
        \includegraphics[width=1\linewidth]{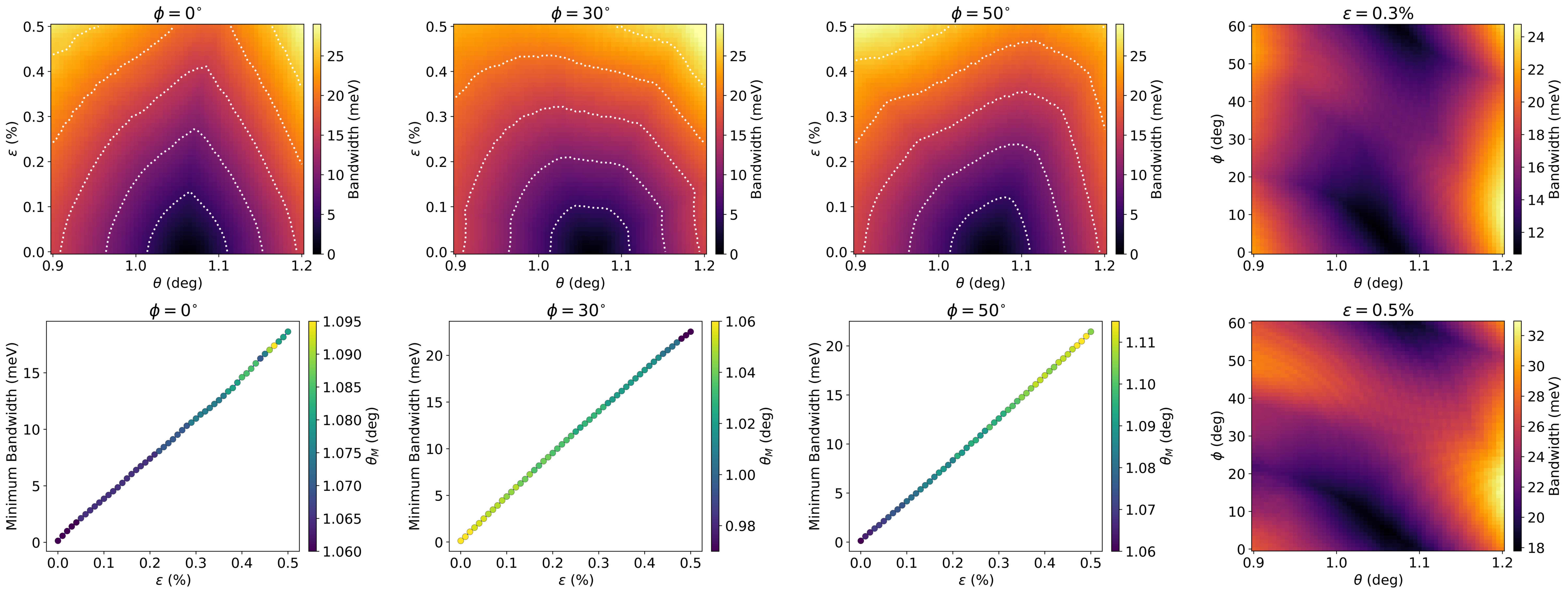}
    \caption{Numerical continuum model results for the bandwidth evolution in the top narrow band, as a function of the twist angle $\theta$ and: (i) uniaxial strain magnitude $\epsilon$ along fixed directions $\phi=0^{\circ}, 50^{\circ}, 60^{\circ}$; white dotted-lines indicate constant bandwidths from 5 to 25 meV, in steps of 5 meV. (ii) strain direction for fixed magnitudes $\epsilon=0.3\%,0.5\%$. The first three bottom panels show the minimum bandwidth at $\theta_{M}$ (right colorbar) as a function of the strain magnitude $\epsilon$, following an almost linear relation. All the results correspond to the \emph{relaxed} configuration with parameters $\hbar v/a=2.13\:\mathrm{eV}$, $u_{1}=0.096\,\mathrm{eV}$ and $u_{0}=0.05952\,\mathrm{eV}$.}\label{fig:magic_strain}
\end{figure*}

In line with the TB results, the continuum model reflects that the strain effectively increases the bandwidth of the narrow bands around the magic angle. Nevertheless, there is always a minimum bandwidth depending on the twist and strain. Thus, one can still identify potential twist and strain configurations at which electronic correlations could be maximized. Here, we shall particularly focus on the identification of the twist angle at which the bandwidth of the narrow bands is minimum.

Figure \ref{fig:magic_strain} shows the bandwidth evolution as a function of the twist angle $\theta$ and the uniaxial strain magnitude $\epsilon$, with different directions $\phi$, for the relaxed configuration of continuum model parameters. In general, we observe that the twist angle at which the bandwidth is minimum tends to shift in the presence of strain. The shift is nonuniform and depends non-trivially on the strain direction. Thus, we observe that as the strain magnitude increases, the magic angle tends to increase when $\phi=0^{\circ}$, but it tends to decrease when $\phi=30^{\circ}$. For shear strain one obtains a similar behavior as in Fig. \ref{fig:magic_strain}, only that the effect is stronger for the same strain magnitude, and the dependence with the strain direction is shifted  [cf. Eq. \eqref{eq:uni_shear}].

Remarkably, the minimum bandwidth always seems to follow a linear dependence with the strain magnitude \cite{bi_designing_2019}.  
We have checked that this behavior persist even without the strain-fields \cite{SM}. The main difference then is that the bandwidth evolution becomes almost insensitive to the strain direction, but scales linearly with the strain magnitude. Such linear dependence of the bandwidth is roughly due to the linear dependence of the moiré vector with the strain strength [see Eq. \eqref{vec_g}], which introduce the strain effect through the moiré potential given by Eq. \eqref{U_strain}.

It should be noted that under strain a minimum bandwidth does not necessarily correlate to a higher DOS. This is because with strain the narrow bands are not, in general, uniformly flat over the whole mBZ. A band at a particular twist and strain configuration can have, for instance, a larger bandwidth than at other configuration, but yet be flatter over a wider region of the mBZ. Consequently, the configuration with higher bandwidth would still have higher vHs. In this sense, a minimum bandwidth should only be considered as an indicator for the appearance of strong electronic correlations.

Although a nonzero strain tends to increase the bandwidth, it is crucial that such an increase depends on the twist angle.  Since most samples are, at least, likely to inherit some kind of random strain 
\cite{choi2021correlation, kerelsky_maximized_2019, hsieh2023domain, cazeaux2023relaxation} (e.g., due to their fabrication method), which can vary from sample to sample, our results highlight that the notion of \textit{magic angle} is intrinsically connected to the experimental conditions of the system.

\begin{figure*}[t]
    \includegraphics[width=0.85\linewidth]{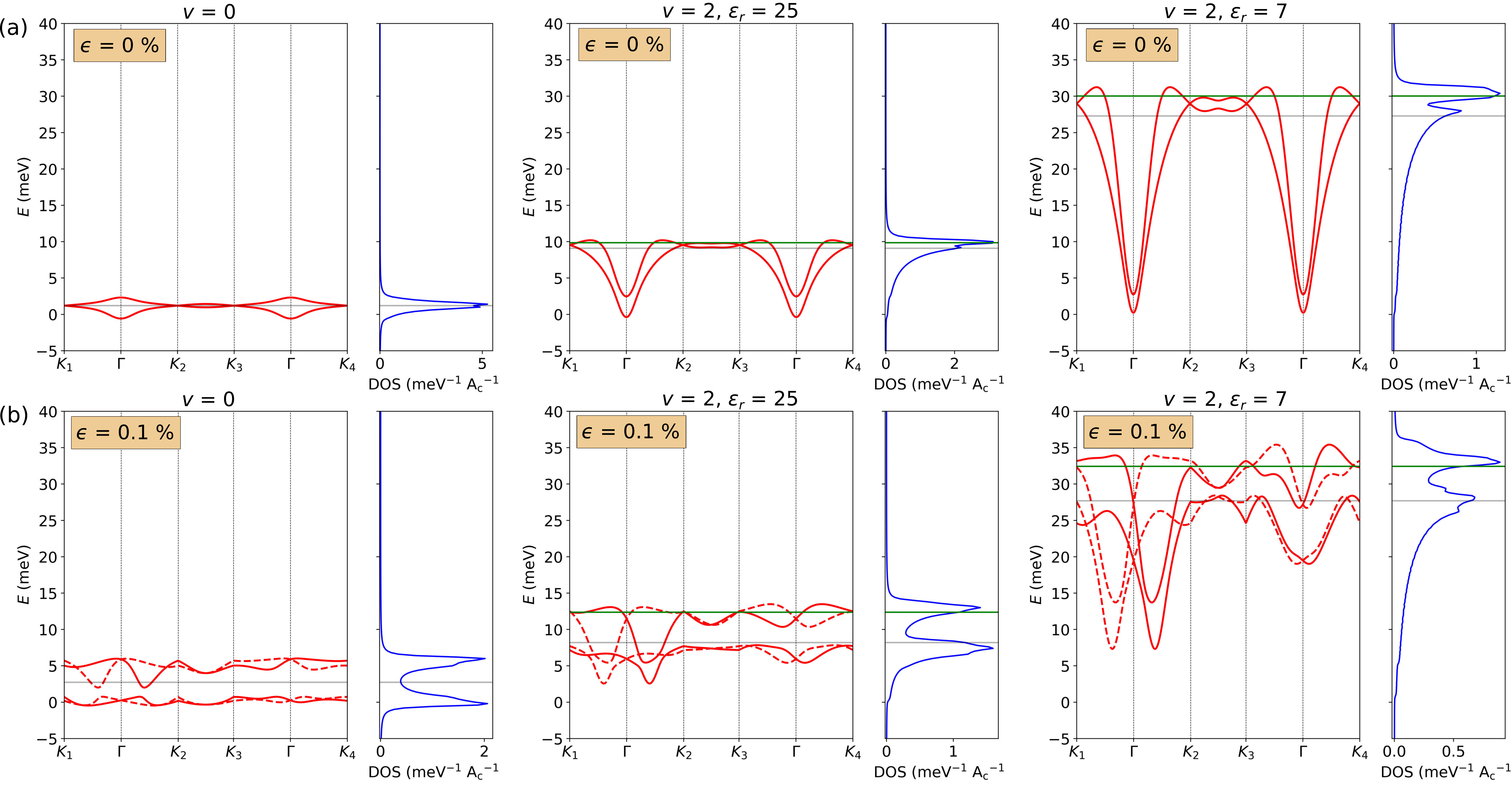}
    \caption{Evolution of the band structure and DOS as a function of the electrostatic interactions (self-consistent Hartree), from the non-interacting case at charge neutrality (filling $\nu=0$), to a filling $\nu=2$ with different dielectric constants $\varepsilon_{r}$. Panel (a) shows the results for the no strain case at $\theta=1.05^{\circ}$, while panel (b) shows the results for $\theta=1.05^{\circ}$ and uniaxial strain with magnitude $\epsilon=0.1\%$ and direction $\phi=30^{\circ}$. In all cases, the horizontal gray and green lines indicate the CNP and the Fermi level, respectively. In the band structures, the solid and dashed lines correspond to the $K$ and $K'$ valleys. All the results are for the relaxed configuration with continuum model parameters $\hbar v/a=2.13\:\mathrm{eV}$, $u_{1}=0.096\,\mathrm{eV}$ and $u_{0}=0.0592\,\mathrm{eV}$.}\label{fig:Hartree_bands}
\end{figure*}

\section{Strain and electrostatic interactions}
\label{interaction_part}

Our analysis so far has neglected the role of electron-electron interactions in the system. However, these interactions are actually crucial in the narrow band regime due to the quenching of the kinetic energy. In this section we will address, in particular, the role of the electrostatic interactions within the continuum model, as accounted by the Hartree potential~\cite{guinea2018electrostatic, cea2019electronic,goodwin2020hartree}. 
Our main interest will be the effect of electrostatic interactions on the bandwidth and charge density of the twisted and strained bilayer configurations. 

The Hartree interaction is the direct (classical) interaction of an electron with the surrounding charge density:
\begin{equation}
V_{H}\left(\mathbf{r}\right)=\int d\mathbf{r}'v_{C}\left(\mathbf{r}-\mathbf{r}'\right)\delta\rho\left(\mathbf{r}'\right),
\end{equation}
where $v_{C}\left(\mathbf{r}-\mathbf{r}'\right)$ is the bare Coulomb potential and $\delta\rho\left(\mathbf{r}'\right)$ is the electronic charge density with respect to CNP. Replacing the plane-wave expansion of the Bloch states in TSBG leads to
\begin{align}
V_{H}\left(\mathbf{r}\right) & =\sum_{\mathbf{G}\neq0}V_{H}\left(\mathbf{G}\right)e^{-i\mathbf{G}\cdot\mathbf{r}},\\
V_{H}\left(\mathbf{G}\right) & =\frac{v_{C}\left(\mathbf{G}\right)}{A_{c}}\sum_{\mathbf{k},\mathbf{G}'}\sum_{n,\eta,i}^{\prime}u_{n,\mathbf{k},\eta,i}^{*}\left(\mathbf{G}'+\mathbf{G}\right)u_{n,\mathbf{k},\eta,i}\left(\mathbf{G}'\right),
\end{align}
where $u_{n,\mathbf{k},\eta,i}\left(\mathbf{G}\right)$ are the Fourier coefficients of the band, valley/spin and layer/sublattice indices $n,\eta,i$, respectively, and $v_{C}\left(\mathbf{G}\right)$ is the Fourier transform of the bare Coulomb potential (see SM for details \cite{SM}). We consider a gated configuration of two metallic plates \cite{bernevig2021twisted}, for which $v_{C}\left(\mathbf{G}\right)=e^{2}\tanh\left(d\left|\mathbf{G}\right|\right)/2\varepsilon_{0}\varepsilon_{r}\left|\mathbf{G}\right|$, where $d$ is the distance between the two metallic plates, and $\varepsilon_{r}$ is the relative primitivity of the system. For the numerical calculations we set $d=40\,\mathrm{nm}$. Note that the $\mathbf{G}=0$ term in $V_{H}\left(\mathbf{r}\right)$ is neglected because it is canceled by the background positive charge (jellium model)~\cite{guinea2018electrostatic,rademaker2018charge}. 

\begin{figure*}[t]
    \includegraphics[width=0.8\linewidth]{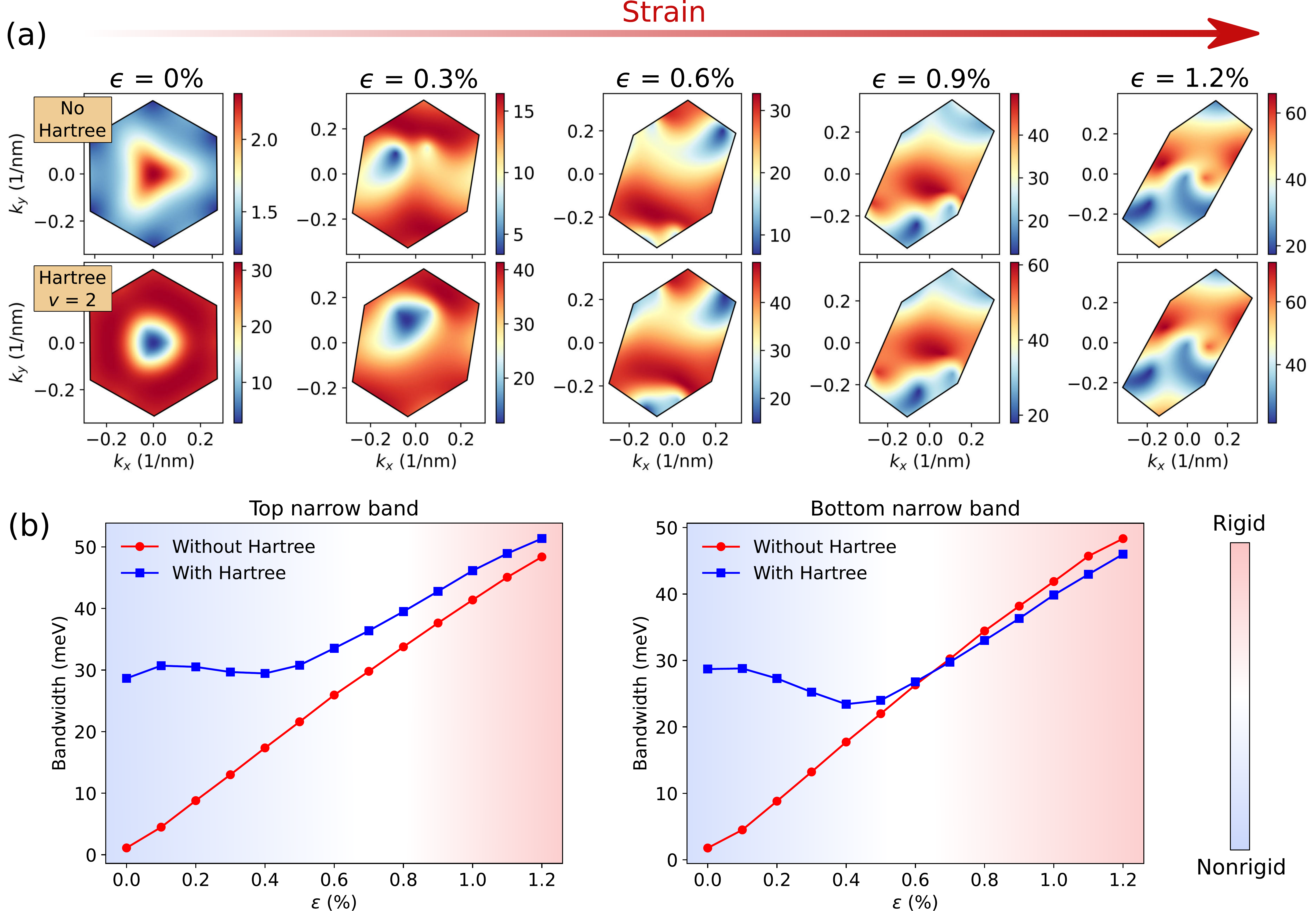}
    \caption{Evolution from a nonrigid to a rigid Hartree effect as the strain increases. Panel (a) shows the continuum model density plot of the top narrow band for $\theta=1.05^{\circ}$ and uniaxial strain with direction $\phi=20^{\circ}$ and increasing strain magnitudes $\epsilon$. The top and bottom density panels show, for each strain magnitude, the bands without Hartree and with Hartree (filling $\nu=2$ and $\varepsilon_{r}=7$), respectively. Panel (b) shows the bandwidth in the top and bottom narrow bands, with and without Hartree, as a function of the uniaxial strain magnitude $\epsilon$ with direction $\phi=20^{\circ}$; other parameters as in panel (a). The colormap indicates, schematically, the nonrigid to rigid transition as the strain increases and the bandwidths become similar. }\label{fig:Hartree_strain}
\end{figure*}

In unstrained TBG, the moiré pattern is perfectly triangular, and the largest Fourier components of the Hartree potential correspond to the first six reciprocal vectors of equal magnitude~\cite{guinea2018electrostatic,cea2019electronic}. Owing to the $\mathcal{C}_{3}$ symmetry, each reciprocal vector contributes equally to the charge density and the Hartree potential, which therefore follow the same spatial profile~\cite{cea2019electronic}. Under strain, however, the moiré pattern becomes distorted and loses its triangular symmetry~\cite{kogl_moire_2023,escudero_designing_2024}, making the contributions from different Fourier components inequivalent. As a result, the Hartree potential is no longer proportional to the charge density. In our numerical calculations, we therefore include all Fourier components $V_{H}(\mathbf{G})$ within the reciprocal moiré vectors of the continuum model.

The Hartree potential is diagonal in the valley/spin and sublattice/layer flavors, with matrix elements \cite{cea2020band}
\begin{equation}
\bra{\mathbf{k}+\mathbf{G}'-\mathbf{G},\eta',i'}\hat{V}_{H}\ket{\mathbf{k}+\mathbf{G}',\eta,i} =\delta_{\eta\eta'}\delta_{ii'}V_{H}\left(\mathbf{G}\right).
\end{equation}
Since the Hartree potential depends on the occupied Bloch states from CNP, the total Hamiltonian $\hat{H}=\hat{H}_{0}+\hat{V}_{H}$ is solved self-consistently, for different filling factors, up until convergence.

Figure \ref{fig:Hartree_bands} shows the numerical results for the evolution of the band structure and density of states, from CNP ($\nu=0$), to a filling of $\nu=+2$ (two electrons per moiré unit cell), for different dielectric constants $\varepsilon_{r}$. The results correspond to $\theta=1.05^{\circ}$ with no strain (top panels), and uniaxial strain with magnitude $\epsilon=0.1\%$ and direction $\phi=30^{\circ}$ (bottom panels). In both cases, we observe that as the dielectric constant diminish and the Hartree potential increases, the band structure is strongly reshaped~\cite{guinea2018electrostatic, cea2019electronic, goodwin2020hartree,Ezzi2024Aselfconsistent}. 

Interestingly, by comparing the nonstrain with the strain case, we observe that the interacting band structures end up having similar bandwidths, despite having quite different single-particle bandwidth. This is because the Hartree potential diminish as the bare active bands increase their bandwidth, essentially due to the decrease in the quenching of the kinetic energy. As a result, under strain there is a competition between the increase of the non-interacting bandwidth and the decrease of the Hartree potential. The synergy between both effects determines the effective bandwidth under strain and electrostatic interactions. It may actually be that under strain the net bandwidth becomes comparable, or even smaller, than the one corresponding to the no-strain scenarios. 

As the strain increases, the Hartree effect not only weakens but also evolves into a nearly uniform energy shift. In other words, it changes from a non-rigid shift in the unstrained case, to an almost rigid (and small) shift at larger strains. This behavior indicates that the charge density across different momentum points becomes increasingly uniform as strain grows. Figure~\ref{fig:Hartree_strain} illustrates this transition from non-rigid to rigid behavior, for the case of uniaxial strain. While the non-interacting and interacting bands differ noticeably at low strain, they become nearly identical at higher strain values, implying that the Hartree potential becomes effectively negligible. The strain threshold for this transition depends non-trivially on the strain direction (and, more generally, on the strain type). It also varies with the filling factor because of the asymmetric Hartree renormalization of the conduction and valence bands, depending on whether the system is electron- or hole-doped.

Besides the local Hartree potential, within mean-field the electrons also experience the nonlocal (exchange) Fock potential that accounts for correlations due to the Pauli principle \cite{cea2020band}. The Fock potential modifies quantitatively the renormalization of the bands \cite{cea2019electronic}, but its main effect is to induce polarized broken symmetry phases. Previous works have shown that under particular strain configurations, the Fock potential can stabilize different orders, such as the so-called Kramers invervalley-coherent (KIVC) order, or the incommensurate Kekulé spiral (IKS) order \cite{parker2021strain, kwan2021kekule, wagner2022global, herzog2025kekule}. These results, however, were obtained under particular combinations of twist and strain (e.g., only uniaxial strain). Although a detailed account of strain-induced broken symmetry phases is beyond the scope of this paper, our Hartree results point out that different combinations of twist and strain could potentially lead to a plethora of competing orders.

\begin{figure}[t]
    \centering
        \includegraphics[width=\linewidth]{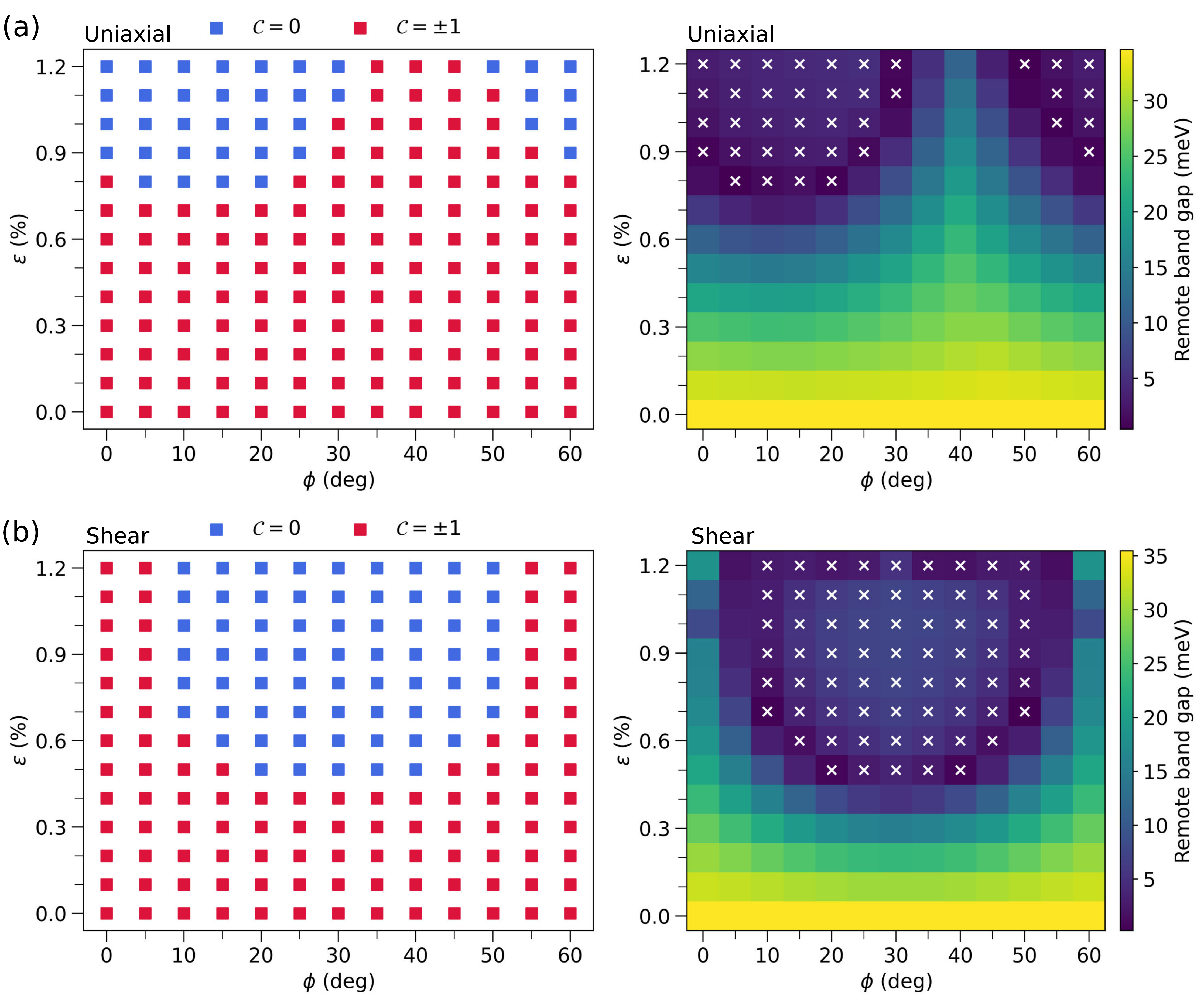}
    \caption{Strain-induced topology evolution of the narrow bands, for the non-interacting case at $\theta=1.05^{\circ}$ with (a) uniaxial heterostrain and (b) shear strain. In each case, the top left panel show the Chern number in the top and bottom narrow bands as a function of the strain direction $\phi$ and magnitude $\epsilon$. For all parameters, $C=+1\left(-1\right)$ for the top (bottom) bands. The top right panel shows the gap between the top narrow bands and the remote bands. The minimum gap, at which it closes and then re-open, corresponds to the topological transitions $\left|C\right|=1\rightarrow0$ in (a); the white crosses indicate the regions where the narrow bands are trivial ($C=0$).}\label{fig:Chern_0}
\end{figure}

\begin{figure}[t]
    \centering
        \includegraphics[width=\linewidth]{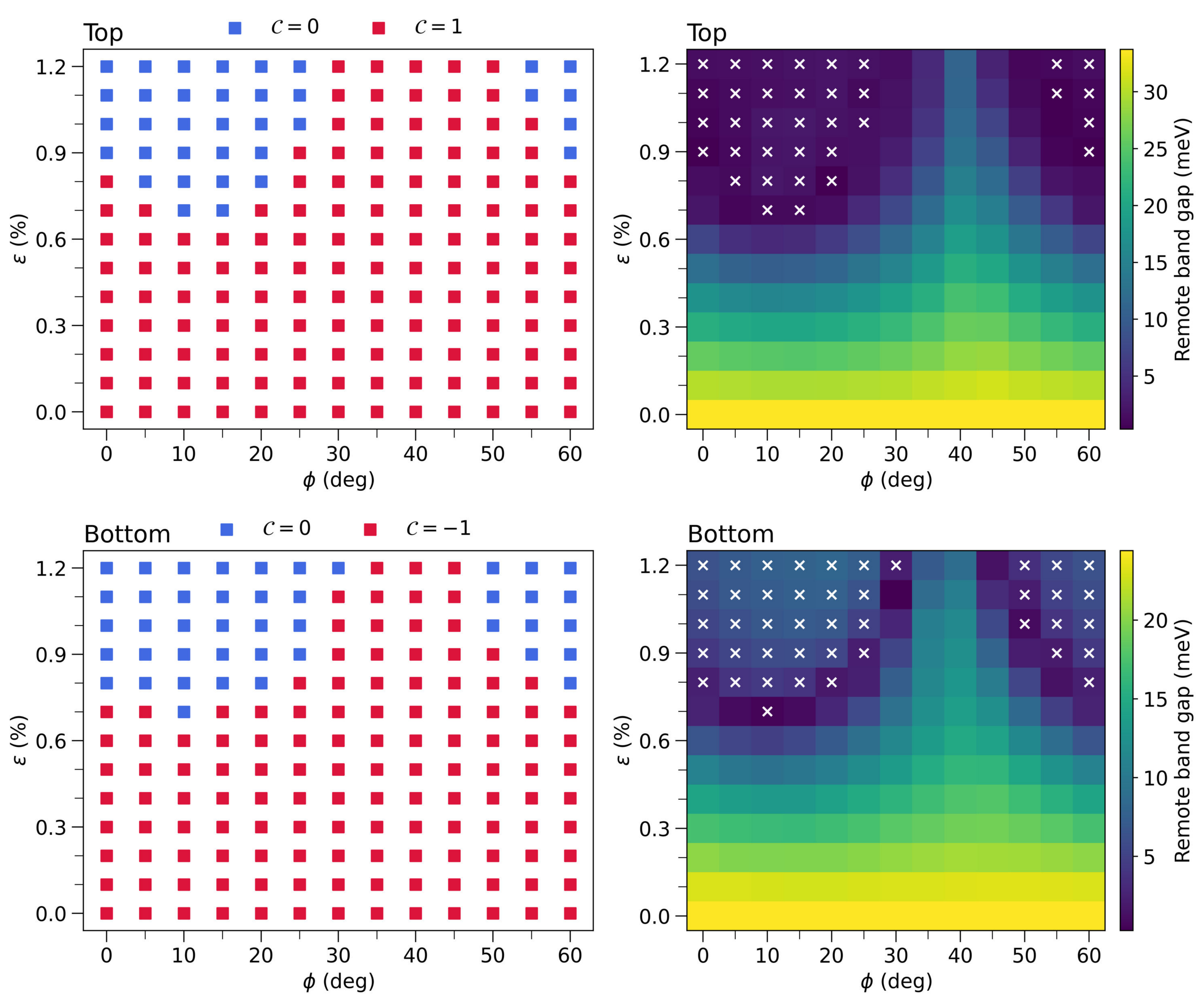}
    \caption{Strain-induced topology evolution of the top and bottom narrow bands with electrostatic interaction (self-consistent Hartree potential), at $\theta=1.05^{\circ}$ with uniaxial heterostrain, for a filling of $\nu=+2$ electrons per moiré unit cell and dielectric constant $\varepsilon_{r}=7$. The top (bottom) panel shows the Chern number in the top (bottom) narrow band, and their corresponding gap to the remote bands, as a function of the uniaxial strain direction $\phi$ and magnitude $\epsilon$; the white crosses in the remote band gap indicate where the narrow band is trivial ($C=0$). Due to the electrostatic interactions, the topological transitions $\left|C\right|=1\rightarrow0$ at which the remote band gap closes and the reopens is different for each narrow band, leading to strain configuration at which only one band topological (or trivial).}\label{fig:Chern_Hartree}
\end{figure}

\section{Band Topology with Strain}
\label{topo}

The strain-induced reshaping of the electronic properties in TSBG is expected to influence the valley-dependent topology of the band structure. Previous studies have indeed reported a rich topological phase diagram arising from the interplay between twist and strain, both in TMDs~\cite{bi_designing_2019} and in bilayer graphene~\cite{pantaleon2021tunable,pantaleon2022interaction}. However, a detailed analysis of how the topology of the narrow bands in TSBG evolves with different types of strain is still lacking.

Since strain preserves time-reversal symmetry, the Chern numbers of opposite valleys remain equal in magnitude and opposite in sign, yielding an overall zero Chern number, i.e., a topologically trivial system. Nevertheless, each valley can still host non-trivial topology. A clear signature of such valley topology, even without breaking time-reversal symmetry, is the nonlinear Hall effect~\cite{sodemann2015quantum,low2015topological}, which can be used to probe the topological character of the narrow bands~\cite{Sinha2022Berry}. In what follows, we therefore focus on the valley-resolved topology of the TSBG band structure.

We study the topology evolution of the bands by computing the Chern number
\begin{equation}
\mathcal{C}_{n}=\frac{1}{2\pi}\int\boldsymbol{\Omega}_{n}\left(\mathbf{k}\right)\cdot d\mathbf{k},
\end{equation}
where $\boldsymbol{\Omega}_{n}\left(\mathbf{k}\right)=i\left\langle \partial_{\mathbf{k}}\psi_{n\mathbf{k}}\right|\times\left|\partial_{\mathbf{k}}\psi_{n\mathbf{k}}\right\rangle $
is the Berry curvature of the $n$-band Bloch states $\psi_{n\mathbf{k}}$,
$d\mathbf{k}$ is a reciprocal-space surface vector, and the integration is over a moiré unit cell. To obtain $\mathcal{C}_{n}$ numerically we use the Fukui-Hatsugai-Suzuki method \cite{fukui2005chern}, considering different strain parameters. For comparison, we analyze separately both the noninteracting and the interacting cases with electrostatic interactions. All the results are obtained using the continuum model for the relaxed configuration. The Chern numbers are computed for the top and bottom narrow bands by introducing a small mass term $\sim m\sigma_{z}$ in the TSBG Hamiltonian that breaks the inversion symmetry and opens a gap at the Dirac points.

\begin{figure*}[t]
    \includegraphics[width=0.8\linewidth]{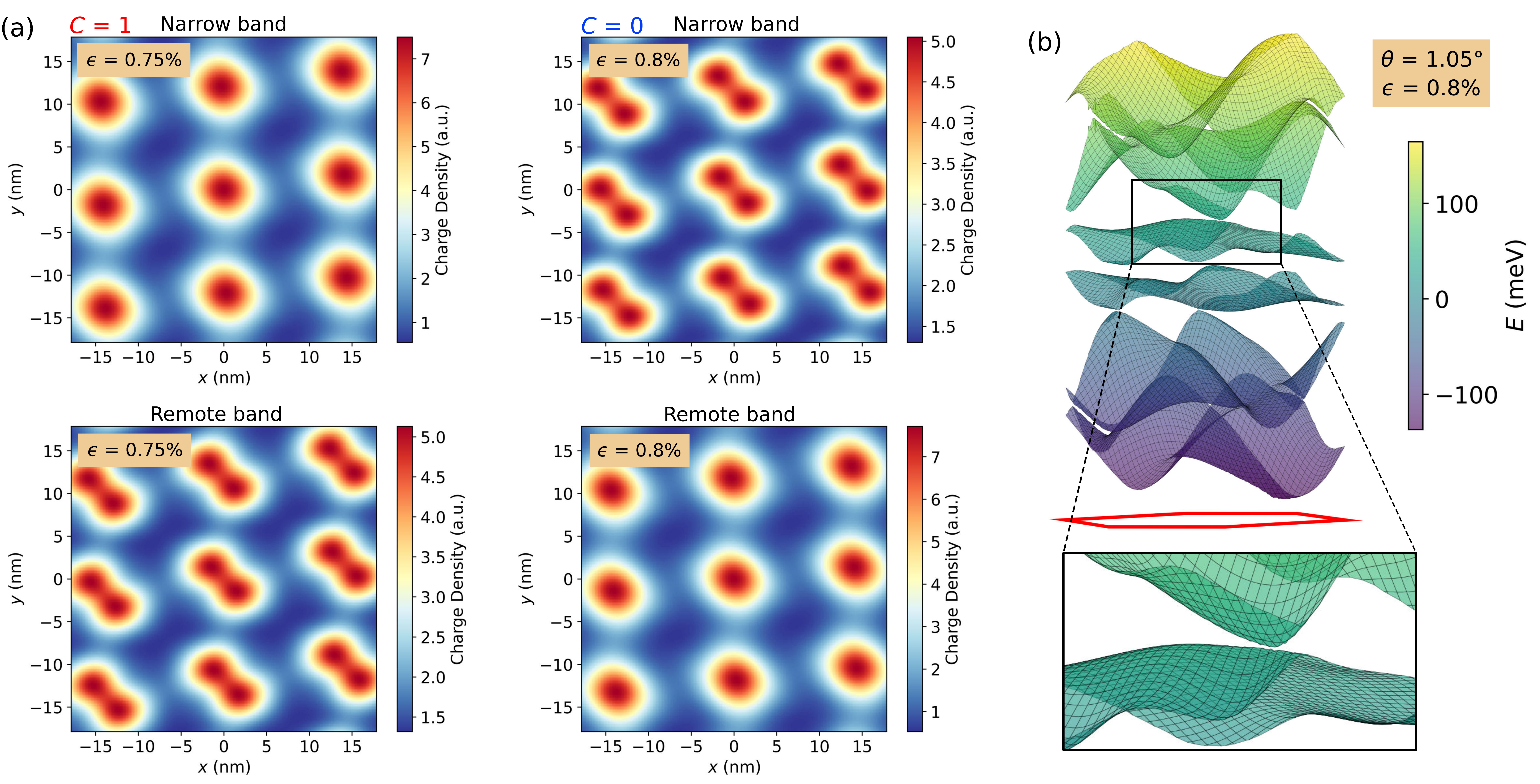}
    \caption{(a) Charge density $\rho_{\mathbf{k}}\left(\mathbf{r}\right)$ in the top narrow band and the closest remote band, at the momentum point $\mathbf{k}$ of minimum direct gap between them, for twist angle $\theta=1.05^{\circ}$ and uniaxial strain at $\phi=20^{\circ}$, with magnitudes $\epsilon=0.75\%$ and $\epsilon=0.8\%$. The transition from a topological ($C=1$) to trivial ($C=0$) narrow band takes place between the two strain magnitudes (cf. Figure \ref{fig:Chern_0}). When the transition occurs there is a charge density transfer from the top to the remote band. Panel (b) shows the noninteracting 3D band structure for the uniaxial strain magnitude $\epsilon=0.8\%$. The zoom inset highlights the minimum gap between the top narrow band and the remote band, at which the charge transfer takes place. }\label{fig:Charge_Density}
\end{figure*}

\subsection{Non-interacting case}

Figure~\ref{fig:Chern_0} shows the results for the non-interacting case at $\theta = 1.05^{\circ}$ under uniaxial and shear strain with different directions $\phi$ and magnitudes $\epsilon$. When interaction effects are neglected, the two narrow bands retain opposite Chern numbers of equal magnitude, so their total Chern number remains zero. In the absence of strain, the upper and lower narrow bands are topological with valley Chern numbers of +1 and -1, respectively. Upon introducing strain, a topological transition to trivial narrow bands ($C = 0$) occurs when the gap between the narrow and remote bands closes and reopens. This behavior arises because the strain increases the bandwidth of the narrow bands and simultaneously shifts the remote bands to lower energies (see Fig.~\ref{fig_relax_band}). The critical point at which the gap closes depends nontrivially on the strain magnitude and direction, and, in general, also on the twist angle. Interestingly, the non-interacting band topology is preserved up to relatively large strain magnitudes ($\epsilon\sim2\%$). The obtained topological transition could be realized with a recent strain technique that generates a position-dependent strain by bending the TBG across a nanoridge \cite{pan2025topological}. 

The touching between the narrow and the remote bands that triggers the topological transition is further accompanied by a transfer of charge density. Figure \ref{fig:Charge_Density} shows this transfer of charge density $\rho_{\mathbf{k}}\left(\mathbf{r}\right)$ for uniaxial strain (non-interacting case), at the momentum point $\mathbf{k}$ of minimum (direct) gap between the narrow and remote bands. As clearly seen, the charge density of the topological band ($C=1$) is directly transferred to the remote band after the gap closes and the narrow band becomes trivial ($C=0$). This charge transfer behavior occurs, in general, for any kind of twist and strain that induce a remote gap closing. The dependence on the specific type of strain is reflected in the profile of the charge density transferred (e.g, its symmetries), and the specific twist and strain parameters at which the topological transition takes place. For biaxial strain, for instance, the system retains the $C_{3}$ symmetry and the topological transition involves the transfer of ring-like to AA centered charge densities between the top and remote bands \cite{SM}. 

The strain-induced topological transitions are connected to changes in the Berry curvature \cite{xiao2010berry}, which reverses its momentum dependence before and after the transition \cite{SM}. With strain the Berry curvature exhibits, in general, three distinct peaks, but only one having a larger magnitude \cite{battilomo2019berry, moulsdale2020engineering, pantaleon2021tunable, pantaleon2021narrow, pantaleon2022interaction, cuypers2026evolution}. The largest peak in the Berry curvature occurs around the momentum point where the narrow and remote band close their gap. Thus, the peaks of the Berry curvature, and by extension of the Berry dipole, directly reflect the point where the narrow and remote band touch and the topology changes. Since the strain breaks the symmetries of the system, the peaks in the Berry curvature are in general distributed nonuniformly in the moiré Brillouin zone (their position depending on the twist and strain).

\subsection{Interacting case}
Figure~\ref{fig:Chern_Hartree} shows the topology evolution of the narrow bands after including electrostatic interactions. In contrast to the noninteracting case, we now see that by increasing the strain magnitude the topology of the top and bottom narrow bands becomes asymmetrical. That is, there are now strain configurations for which the sum of the top and bottom band Chern number is not zero. This is because for any nonzero filling the Hartree potential reshapes the top and bottom bands asymmetrically, cf. Fig.~\ref{fig:Hartree_bands}. In general, for positive fillings $\nu$ (electron-doped) the bottom narrow band is more strongly renormalized, and vice versa for negative fillings $\nu$. This leads to different strain parameters at which the gap with the remote bands closes and then reopens, and therefore, wider strain regimes in which only one band is topological. Compared to the noninteracting case of Fig.~\ref{fig:Chern_0}, we particularly see that the Hartree potential shrinks and increase, respectively, the regimes where the bottom and top bands are topological. Since the Hartree potential is practically symmetric with respect to charge neutrality~\cite{guinea2018electrostatic, cea2019electronic}, the regime in which only one band is topological is reversed when the system is hole-doped ($\nu<0$). 

\begin{table}[t]
\begin{ruledtabular}
\begin{tabular}{ccc}
\textbf{Perturbation} & \textbf{Topology Change} & \textbf{References}\tabularnewline
\hline 
Strain & $\checkmark$ & \cite{pantaleon2021tunable, pantaleon2022interaction}, This work\tabularnewline
Hartree & $\times$ & \cite{guinea2018electrostatic, cea2019electronic}\tabularnewline
Substrate & $\checkmark$ & \cite{cea2020band}\tabularnewline
Displacement field & $\times$ & \cite{gao2412double, dutta2025electric}\tabularnewline
\end{tabular}
\caption{Possible perturbations and their effect on being able to change the topology of the narrow bands in twisted bilayer graphene. The four perturbations are assumed to act on TBG at a fixed twist angle (e.g, the magic angle $\theta=1.05^{\circ}$), with a small mass term $\sim\sigma_{z}m$ that opens a gap at the Dirac points. The topological nature of the bands is taken with respect to transitions between nonzero and zero valley Chern numbers in the top and bottom narrow bands. The Hartree potential is assumed to be for fillings between $-4\leq\nu\leq4$. Non-listed perturbations involving two or more effects (e.g., strain and electrostatic interactions) can always change the topology of the narrow bands.}\label{table_top}
\end{ruledtabular}
\end{table}

It is interesting to compare the strain with other perturbations in their capacity to modify the topology of the narrow bands. Table \ref{table_top} list the topology effect of four common perturbations in TBG: Strain, Hartree (electrostatic interactions), substrate and displacement field. For a fixed twist angle (e.g., the magic angle), only a strain \cite{pantaleon2021tunable, pantaleon2022interaction} or a substrate \cite{cea2020band} can, by itself, modify the topology of the central narrow bands. In contrast, the topology of the narrow bands cannot be changed by means of solely electrostatic interactions \cite{guinea2018electrostatic, cea2019electronic} or a displacement field \cite{gao2412double, dutta2025electric}. Note that when two or more of the perturbations act in conjunction, the topology can always change (e.g., strain with Hartree, or Hartree with displacement field, and so on). 

Topology changes in the narrow bands, by means of any of the listed perturbations in Table \ref{table_top}, pose a restriction on topological heavy fermion models of the narrow bands in TSBG \cite{song2022magic,herzog2025topological,shi2022heavy}, as they rely on the topological nature of the bands.

\section{Conclusion}\label{sec:Conclusions}
In summary, we studied the combined effects of twist and strain in bilayer graphene using atomistic tight-binding and strain-extended continuum models. Strain reshapes the moiré geometry, broadens the narrow bands, splits the vHs, lifts valley degeneracy, and shifts the Dirac points within the mBZ. The shear strain introduces stronger distortion of both the geometrical and electronic properties. Specifically, under the same strain strength, the shear type induces a larger vHs separation than the uniaxial strain, in agreement with recent experimental results \cite{yu2024twist,carrasco2025twistraintronics}. Moreover, the strain direction is crucial: it controls both the bandwidth and the valley-resolved topology, and shifts the twist angle that minimizes the bandwidth.  

A continuum model with strain-induced scalar and gauge fields reproduces the atomistic spectra and the strain-driven topological transitions that occur when the gap to the remote bands closes and reopens. Including electrostatic (Hartree) interactions, we found nonrigid spectral shifts at low strain that evolve toward an almost rigid shift at higher strain; the interaction mainly reshapes the bands without reversing the strain-induced broadening. These results show that strain provides a practical knob to control band structure and valley topology in moiré graphene.

\section*{Acknowledgments}
We thank Christophe De Beule, Mikito Koshino and Eduardo V. Castro for fruitful discussions. Z.Z. thanks Wei Li for the discussion in the 2DSPM conference in San Sebastián, which inspired the initial idea of this work. IMDEA Nanociencia acknowledges support from the ``Severo Ochoa" Programme for Centres of Excellence in R\&D (Grant No. SEV-2016-0686), and from NOVMOMAT, Grant PID2022-142162NB-I00 funded by MCIN/AEI/ 10.13039/501100011033 and, by “ERDF A way of making Europe”. F.E. acknowledges support funding from the European Union's Horizon 2020 research and innovation programme under the Marie Skłodowska-Curie grant agreement No 101210351. Z.Z. acknowledges support funding from the European Union's Horizon 2020 research and innovation programme under the Marie Skłodowska-Curie grant agreement No 101034431 and from the ``Severo Ochoa" Programme for Centres of Excellence in R\&D (CEX2020-001039-S / AEI / 10.13039/501100011033). P.A.P acknowledges funding by Grant No.\ JSF-24-05-0002 of the Julian Schwinger Foundation for Physics Research.  S.Y. acknowledges funding from the National Natural Science Foundation of China (Grants No. 12425407, 12174291), the Natural Science Foundation of Hubei Province, China (Grant No. 2023BAA020). Numerical calculations presented in this paper have been performed in the Supercomputing Center of Wuhan University. 

~\\
\textbf{DATA AVAILABILITY}
~\\
All data needed to evaluate the conclusions in the paper are present in the paper and/or the Supplementary Materials. 

~\\
\textbf{CODE AVAILABILITY}
~\\
The codes that support the findings of this study are available from the corresponding
authors on reasonable request.

~\\
\textbf{AUTHOR CONTRIBUTIONS}
~\\
ZZ and FE supervised the project. DW performed the tight-binding calculations with the help of ZZ. FE performed the continuum calculations. All authors discussed the results. FE, DW and ZZ co-wrote the manuscript with inputs from all the authors. 

~\\
\textbf{COMPETING INTERESTS} 
~\\
The authors declare no competing interests. 

\bibliography{newcite}


\clearpage
\onecolumngrid

\onecolumngrid
\setcounter{section}{0}
\setcounter{equation}{0}
\setcounter{figure}{0}
\renewcommand\thesection{S\arabic{section}}
\renewcommand{\theequation}{S\arabic{equation}}
\renewcommand{\thefigure}{S\arabic{figure}}

\resumetoc 

\graphicspath{{figures_sup_comp/}}

\begin{center}
{\Large\emph{Supplemental Materials for}:\\
Straintronics and twistronics in bilayer graphene}{\Large\par}
\par\end{center}

\begin{center}
Federico Escudero, Dong Wang, Pierre A. Pantaleón, Shengjun Yuan, Francisco Guinea, and Zhen Zhan
\par\end{center}

\tableofcontents

\clearpage

\section{Algorithm to generate commensurate structures with twist and strain}
\label{algorithm_com}

As discussed in Sec. IIA.4 of the main text, one can obtain a commensurate superlattice structure with twist and uniaxial or shear strain by introducing an additional small biaxial strain in the system. The idea is, essentially, that given any twist and strain configuration  (which in generally gives a incommensurate structure), one can \textit{always} find the closest commensurate structure. It is important to note that the commensurate twist and strain parameters will end up being slightly different from the initial ones (the difference, however, is small and does not impact the electronic properties).

In this section we describe in more detail the \textit{algorithm} by which one can generate a commensurate structure with any twist and strain. Although we focus on relevant uniaxial and shear strain configurations, we emphasize that the described procedure is general and holds for any strain configuration (i.e., any strain tensor). The step-by-step procedure to generate the commensurate structure involves:

\begin{enumerate}
 \item Start with a set of four parameters ($\theta,\epsilon_{u/s},\phi,\epsilon_{b}$) that totally determine the twist and strain.
 \item Calculate the eight parameters by using Eqs. (9)--(12) of the main text. (Note that the direct solutions of the eight parameters are generally not integers.)
  \item Round the obtained values of the eight parameters to the nearest integers to find the closest commensurate case.
 \item Calculate ($a,b,c,d$) by constructing the Park-Madden transformation matrix according to Eq.~(13) of the main text.
 \item Determine the fitted geometrical parameters ($\theta^c,\epsilon_{u/s}^c,\phi^c,\epsilon_{b}^c$) according to Eq.~(14) of the main text.
 \item Finally, recalculate the strained lattice vectors and superlattice vectors in Eqs. (9) and (12) of the main text by using the commensurate twist and strain, respectively. 
 \end{enumerate}
This general procedure allows us to generate commensurate structures for arbitrary values of twist and different combinations of strain.

\clearpage
\section{Fitting Parameters for commensurate structures}
\label{sec:S1}
In this section, we list the geometrical parameters of the commensurate structures used in the tight-binding (TB) calculations. The method for generating a commensurate twisted and strained bilayer graphene (TSBG) with given twist angle and strain is in Sec. \Romannum{2}A of the main text (see also Sec. \ref{algorithm_com} above). As we can see from the tables below, for an initial proposed physical parameters, for instance, the twist angle and strain, the fitted values may be slightly different from the proposed ones. Moreover, in some cases we introduce a negligible biaxial strain or change the strain direction to get a commensurate structure. In all cases, the commensurate structures are described by a pair of eight integers $i,j,k,l,m,n,r,q$, which determine the superlattice vectors according to Eq. (12) in the main text. The position of the Dirac points in the moir\'e Brillouin zone (mBZ) are estimated by the eight integers with Eq. (17) of the main text. Note that in the shear strain fitting parameters, the strain direction list below is the real shear strain direction plus $\frac{\pi}{4}$, see the Eq. (8) of the main text. 

\begin{table*}[h]
    \centering
    \begin{tabular}{|c|c|c|c|}
    \hline
        ~ & proposed/fitted & proposed/fitted & proposed/fitted \\ \hline
        twist($\circ$) & 1.05/1.05012 & 1.05/1.05118 & 1.05/1.04671 \\ \hline
        uniaxial strain & 0/0 & 1e-03/1.16004e-03 & 2e-03/2.01835e-03 \\ \hline
        strain direction & 0/0 & 0/5.25590e-01 & 0/2.88086 \\ \hline
        biaxial strain & 0/0 & 0/-1.50613e-04 & 0/-1.40253e-05 \\ \hline
        $\begin{pmatrix} i & j \\ k & l\end{pmatrix}$/ $\begin{pmatrix} m & n \\ q & r\end{pmatrix}$ &
        $\begin{pmatrix} 32 & -63 \\ 63 &-31\end{pmatrix}$/$\begin{pmatrix} 31 & -63 \\ 63 &-32\end{pmatrix}$ & 
        $\begin{pmatrix} 33 & -63 \\ 65 &-34\end{pmatrix}$/$\begin{pmatrix} 32 & -63 \\ 65 &-35\end{pmatrix}$ & 
        $\begin{pmatrix} 33 & -63 \\ 67 &-37\end{pmatrix}$/$\begin{pmatrix} 32 & -63 \\ 67 &-38\end{pmatrix}$\\ \hline
    \end{tabular}
\end{table*}

\begin{table}[!ht]
    \centering
    \begin{tabular}{|c|c|c|c|}
    \hline
        ~ & proposed/fitted & proposed/fitted & proposed/fitted \\ \hline
        twist(deg) & 0.93/0.93180 & 0.93/0.93722 & 0.93/0.92942 \\ \hline
        uniaxial strain & 0/0 & 1e-03/1.16026e-03 & 2e-03/2.04312e-03 \\ \hline
        strain direction & 0/0 & 0/-2.82478 & 0/2.29995 \\ \hline
        biaxial strain & 0/0 & 0/-8.60503e-05 & 0/6.19560e-05 \\ \hline
        $\begin{pmatrix} i & j \\ k & l\end{pmatrix}$/ $\begin{pmatrix} m & n \\ q & r\end{pmatrix}$ &
        $\begin{pmatrix} 36 & -71 \\ 71 &-35\end{pmatrix}$/$\begin{pmatrix} 35 & -71 \\ 71 &-36\end{pmatrix}$ & 
        $\begin{pmatrix} 37 & -71 \\ 73 &-39\end{pmatrix}$/$\begin{pmatrix} 36 & -71 \\ 73 &-40\end{pmatrix}$ & 
        $\begin{pmatrix} 37 & -71 \\ 76 &-43\end{pmatrix}$/$\begin{pmatrix} 36 & -71 \\ 76 &-44\end{pmatrix}$\\ \hline
    \end{tabular}
\end{table}

\begin{table}[!ht]
    \centering
    \begin{tabular}{|c|c|c|c|}
    \hline
        ~ & proposed/fitted & proposed/fitted & proposed/fitted \\ \hline
        twist($\circ$) & 1.6/1.61354 & 1.6/1.60063 & 1.6/1.61739 \\ \hline
        uniaxial strain & 0/0 & 1e-03/7.77024e-04 & 2e-03/2.74423e-03 \\ \hline
        strain direction & 0/0 & 0/15.80032 & 0/0.80869 \\ \hline
        biaxial strain & 0/0 & 0/6.39745e-05 & 0/-3.55675e-04 \\ \hline
        $\begin{pmatrix} i & j \\ k & l\end{pmatrix}$/ $\begin{pmatrix} m & n \\ q & r\end{pmatrix}$ &
        $\begin{pmatrix} 21 & -41 \\ 41 &-20\end{pmatrix}$/$\begin{pmatrix} 20 & -41 \\ 41 &-21\end{pmatrix}$ & 
        $\begin{pmatrix} 21 & -41 \\ 42 &-21\end{pmatrix}$/$\begin{pmatrix} 20 & -41 \\ 42 &-22\end{pmatrix}$ & 
        $\begin{pmatrix} 22 & -41 \\ 43 &-23\end{pmatrix}$/$\begin{pmatrix} 21 & -41 \\ 43 &-24\end{pmatrix}$\\ \hline
    \end{tabular}
    \caption{The detailed geometrical parameters of TSBG with different twist angle and uniaxial heterostrain.}
\end{table}
\clearpage

\begin{table*}[h]
    \centering
    \begin{tabular}{|c|c|c|c|}
    \hline
        ~ & proposed/fitted & proposed/fitted & given/fit \\ \hline
        twist($\circ$) & 1.05/1.05012 & 1.05/1.05330 & 1.05/1.03850 \\ \hline
        shear strain & 0/0 & 1e-03/1.01078e-03 & 2e-03/1.79165e-03 \\ \hline
        strain direction & 0/0 & 0/5.26652e-01 & 0/2.02149 \\ \hline
        biaxial strain & 0/0 & 0/5.10839e-07 & 0/1.60502e-06 \\ \hline
        $\begin{pmatrix} i & j \\ k & l\end{pmatrix}$/ $\begin{pmatrix} m & n \\ q & r\end{pmatrix}$ &
        $\begin{pmatrix} 32 & -63 \\ 63 &-31\end{pmatrix}$/$\begin{pmatrix} 31 & -63 \\ 63 &-32\end{pmatrix}$ & 
        $\begin{pmatrix} 35 & -63 \\ 66 &-34\end{pmatrix}$/$\begin{pmatrix} 34 & -63 \\ 66 &-35\end{pmatrix}$ & 
        $\begin{pmatrix} 38 & -64 \\ 70 &-37\end{pmatrix}$/$\begin{pmatrix} 37 & -64 \\ 70 &-38\end{pmatrix}$\\ \hline
    \end{tabular}
\end{table*}

\begin{table}[!ht]
    \centering
    \begin{tabular}{|c|c|c|c|}
    \hline
        ~ & proposed/fitted & proposed/fitted & proposed/fitted \\ \hline
        twist($\circ$) & 0.93/0.93180 & 0.93/0.93576 & 0.93/0.92795 \\ \hline
        shear strain & 0/0 & 1e-03/1.06241e-03 & 2e-03/1.94041e-03 \\ \hline
        strain direction & 0/0 & 0/4.67881e-01 & 0/1.56608\\ \hline
        biaxial strain & 0/0 & 0/5.64364e-07 & 0/1.88262e-06 \\ \hline
        $\begin{pmatrix} i & j \\ k & l\end{pmatrix}$/ $\begin{pmatrix} m & n \\ q & r\end{pmatrix}$ &
        $\begin{pmatrix} 36 & -71 \\ 71 &-35\end{pmatrix}$/$\begin{pmatrix} 35 & -71 \\ 71 &-36\end{pmatrix}$ & 
        $\begin{pmatrix} 40 & -71 \\ 75 &-39\end{pmatrix}$/$\begin{pmatrix} 39 & -71 \\ 75 &-40\end{pmatrix}$ & 
        $\begin{pmatrix} 44 & -72 \\ 80 &-43\end{pmatrix}$/$\begin{pmatrix} 43 & -72 \\ 80 &-44\end{pmatrix}$\\ \hline
    \end{tabular}
\end{table}

\begin{table}[!ht]
    \centering
    \begin{tabular}{|c|c|c|c|}
    \hline
        ~ & proposed/fitted & proposed/fitted & proposed/fitted \\ \hline
        twist($\circ$) & 1.6/1.61354 & 1.6/1.61867 & 1.6/1.59548 \\ \hline
        shear strain & 0/0 & 1e-03/1.59108e-03 & 2e-03/1.94183e-03 \\ \hline
        strain direction & 0/0 & 0/8.09336e-01 & 0/-2.49564\\ \hline
        biaxial strain & 0/0 & 0/1.26578e-06 & 0/1.88535e-06 \\ \hline
        $\begin{pmatrix} i & j \\ k & l\end{pmatrix}$/ $\begin{pmatrix} m & n \\ q & r\end{pmatrix}$ &
        $\begin{pmatrix} 21 & -41 \\ 41 &-20\end{pmatrix}$/$\begin{pmatrix} 20 & -41 \\ 41 &-21\end{pmatrix}$ & 
        $\begin{pmatrix} 23 & -41 \\ 43 &-22\end{pmatrix}$/$\begin{pmatrix} 22 & -41 \\ 43 &-23\end{pmatrix}$ & 
        $\begin{pmatrix} 24 & -42 \\ 44 &-23\end{pmatrix}$/$\begin{pmatrix} 23 & -42 \\ 44 &-24\end{pmatrix}$\\ \hline
    \end{tabular}
    \caption{The detailed geometrical parameters of TSBG with different twist angle and shear heterostrain.}
\end{table}
\onecolumngrid\clearpage

\section{The tight-binding model}
We construct a TB model of TSBG consisting of only the $p_z$ orbital of the carbon atom \cite{trambly2012numerical}. The Hamiltonian of the graphene moir\'e system is:
\begin{equation}
H=\sum_i\varepsilon_i c^{\dagger}_{i} c_{i}+ \sum_{\langle i,j \rangle} t_{ij}c^{\dagger}_{i} c_{j}
\end{equation}
where $c_i$ is an annihilation operator for the $i$ state, $\varepsilon_i$ is the on-site potential, $\langle i,j \rangle$ is the sum over index with $i \neq j$, and $t_{ij}$ is the hopping integral between $i$ and $j$ orbitals. According to the Slater-Koster (SK) formalism, the hopping integral $t_{ij}$ between $p_z$ orbitals located at $\mathbf{r}_i$ and $\mathbf{r}_j$ has the form \cite{trambly2012numerical}:
\begin{equation}
    t_{ij}=n^2 V_{pp\sigma}(r_{ij})+(1-n^2)V_{pp\pi}(r_{ij}),
    \label{SK_t}
\end{equation}
where $r_{ij}=|\mathbf{r}_i-\mathbf{r}_j|$ is the distance between $i$ and $j$ orbitals and $n=z_{ij}/r_{ij}$ is the direction cosine of relative vector along $z$ axis. 
We use SK parameters $V_{pp\sigma}$ and $V_{pp\pi}$ as follow:
\begin{align}
    V_{pp\pi}(r_{ij})    & =-t_0e^{q_{\pi}(1-r_{ij}/d)}F_c(r_{ij}), \\
    V_{pp\sigma}(r_{ij}) & =t_1e^{q_{\sigma}(1-r_{ij}/h)}F_c(r_{ij}),
\end{align}
where $d=1.42\;\text{\AA}$ and $h=3.349\;\text{\AA}$ are the nearest in-plane distance and inter-layer spacing, respectively. $t_0$ and $t_1$ are commonly re-parameterized to fit different experimental results. In this work, we choose the TB intralayer and interlayer hopping parameters as $t_0=2.8$ eV and $t_1=0.44$ eV, respectively, which give a magic angle at $\theta \sim 1.05^\circ$. The parameters $q_{\sigma}$ and $q_{\pi}$ satisfy $q_{\sigma}/h=q_{\pi}/d=2.218\;\text{\AA}^{-1}$ and the smooth function is $F_c(r)=\left[1+e^{(r-r_c)/l_c}\right]^{-1}$, in which $l_c$ and the cutoff distance $r_c$ are chosen as $0.265\;\text{\AA}$ and $5.0\;\text{\AA}$. That is, for $r > r_c$, the hopping value is zero. All $p_z$ orbitals have the same on-site energy $\varepsilon_i$, which ensures that the energy $E_D$ of the Dirac point is zero.

We relax the moir\'e structure by using the classical simulation package LAMMPS \cite{plimpton1995fast}. The intralayer and interlayer interactions are simulated with the long-range carbon bond-order potential \cite{los2005improved} and Kolmogorov-Crespi potential \cite{kolmogorov2005registry}, respectively. We assume that the relaxed structure keeps the same period of the rigid case. In the relaxed system, the hopping terms will modified according to the Eq. \eqref{SK_t} with the relaxed structure.

We perform a numerical diagonalization of the tight-binding Hamiltonian using the FEAST eigen solver in the Intel math kernel library (MKL), and then calculate the band structure and density of states (DOS) around the Fermi energy \cite{hams_fast_2000,polizzi_density-matrix-based_2009}. All the TB calculations are performed in the TBPLaS simulator \cite{li2023tbplas}.

A valley operator is adopted to identify the valley flavor of a state of moir\'e graphene in the real space TB description \cite{ramires_electrically_2018,ramires_impurity-induced_2019}. 
The expectation of this operator is $\langle\mathcal{V}_z \rangle=+1$ for states in one valley and $\langle\mathcal{V}_z \rangle=-1$ for states in another valley.
The real-space operator in a honeycomb lattice can be expressed as \cite{ramires_electrically_2018}:
\begin{equation}
    \mathcal{V}_z = \frac{i}{3\sqrt{3}} \sum_{\langle\!\langle i,j \rangle\!\rangle} \eta_{ij} \sigma^{ij}_z c^{\dagger}_{i} c_{j},
\end{equation}
where $\langle\!\langle i,j \rangle\!\rangle$ denotes next-nearest-neighbor sites, $\eta_{ij}=\pm1$ for clockwise and counterclockwise hopping, and $\sigma_z^{ij}$ is a Pauli matrix associated with the degree of freedom of sublattice.

\onecolumngrid\clearpage
\section{The uniaxial and shear strains}
In this section, we check the strain effect on the electronic properties of twisted bilayer graphene (TBG) with different twist angle $\theta$. We first focus on the magic angle ($\sim 1.05^{\circ}$), and then angles below and above the magic angle. The strain direction is fixed to zero. 

\begin{figure*}[h]
    \centering
    \includegraphics[width=0.8\linewidth]{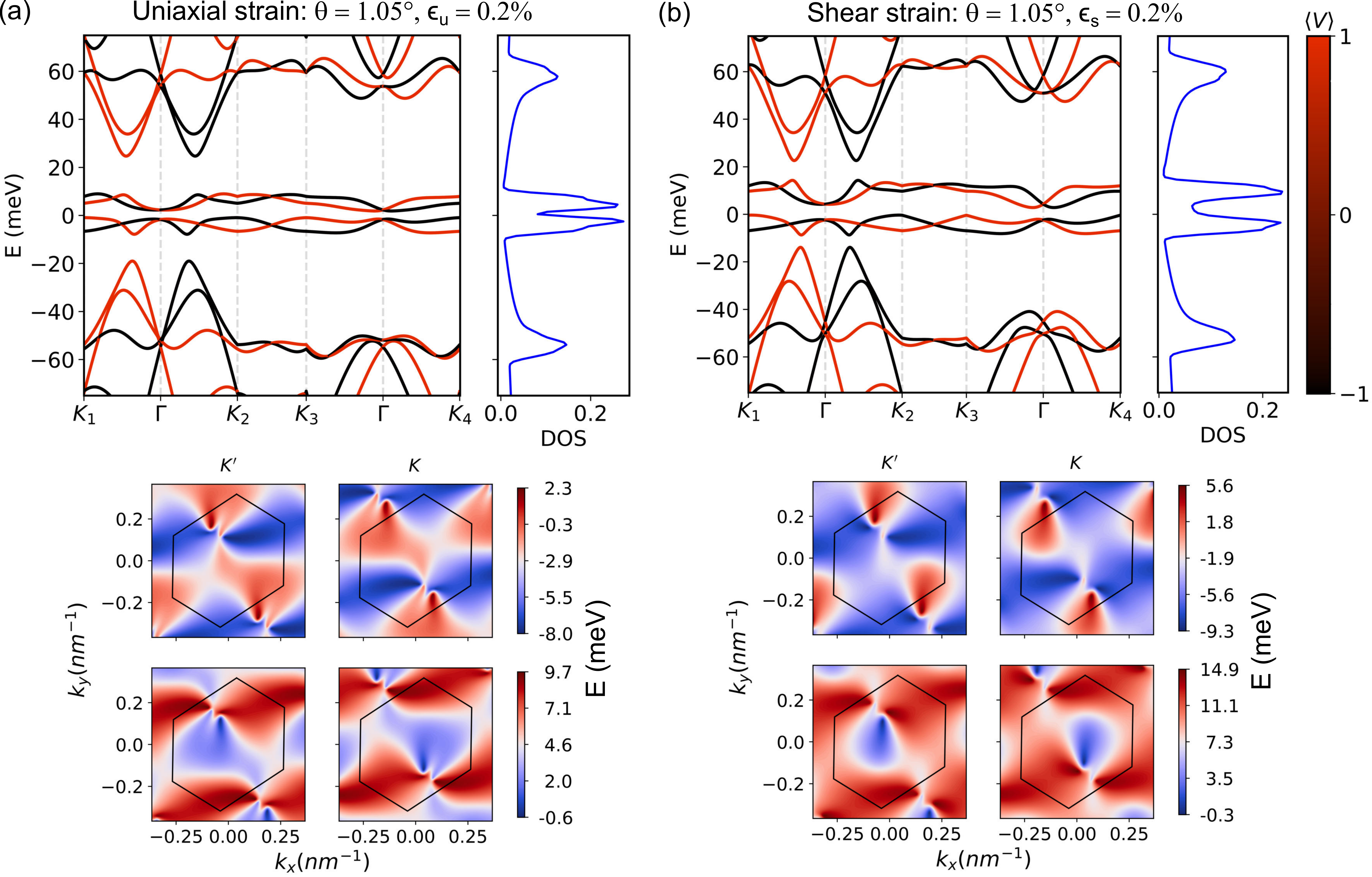}
    \caption{Tight-binding results of the (upper panel) band structure, density of states, and (lower panel) energy map of four middle narrow bands for the commensurate structure of rigid TSBG with $\theta=1.05\degree$, $\phi=0\degree$ and (a) uniaxial strain $\epsilon_{u}=0.2\%$, (b) shear strain $\epsilon_{s}=0.2\%$ (see Tables I and II in Sec. \ref{sec:S1} for the exact parameters). To compare with the continuum model results, we use the valley operator to distinguish the $K$ and $K'$ valleys. The color in the band structure represents the expectation of the valley operator with $\langle \hat{V}_z \rangle \approx 1$ if a state belongs to valley $K$ and $\langle \hat{V}_z \rangle \approx-1$ if a state belongs to valley $K'$.  In the energy map, the top panel are the valence narrow bands and the bottom panel are the conduction narrow bands, and the mBZ is illustrated with black line.}
    \label{Figs2_magic}
\end{figure*}

The uniaxial and shear strains have four generic features.  Let us focus on the magic angle case, shown in Fig. \ref{Figs2_magic} and Fig. 3 of the main text. First, the narrow bands are extremely sensitive to the strain. The strain divides one narrow peak into two peaks in the DOS results. The bandwidth and energy separation of the two peaks increase with the strain strength, which could be confirmed by the energy map of the narrow bands. Second, the strain breaks the valley degeneracy. The splitting of the two valleys increases with the strain strength. Third, the Dirac points are no longer only located at the corner of the mBZ, which are estimated by Eq. (17) of the main text with the parameters of eight integers. Fourth, both uniaxial and shear strain do not modify the peak ($\sim 60$ meV) at the remote bands. 

The uniaxial and shear strains have distinct effects. For the same magnitude of strain, the energy separation in the shear strain is larger than the uniaxial case. Moreover, the narrow bands in the shear strain are more dispersive than those of the uniaxial strain.    

The strain effects on TSBG with $\theta=1.6^{\circ}$ and $\theta=0.93^{\circ}$ are similar to the magic angle case. In the $\theta=1.6^{\circ}$ case, we find multiple peaks in the DOS of both conduction and valence bands, which means that the strain generate higher order van Hove singularities (vHs) in TBG.  

\onecolumngrid\clearpage

\begin{figure*}[h]
    \centering
    \includegraphics[width=0.8\linewidth]{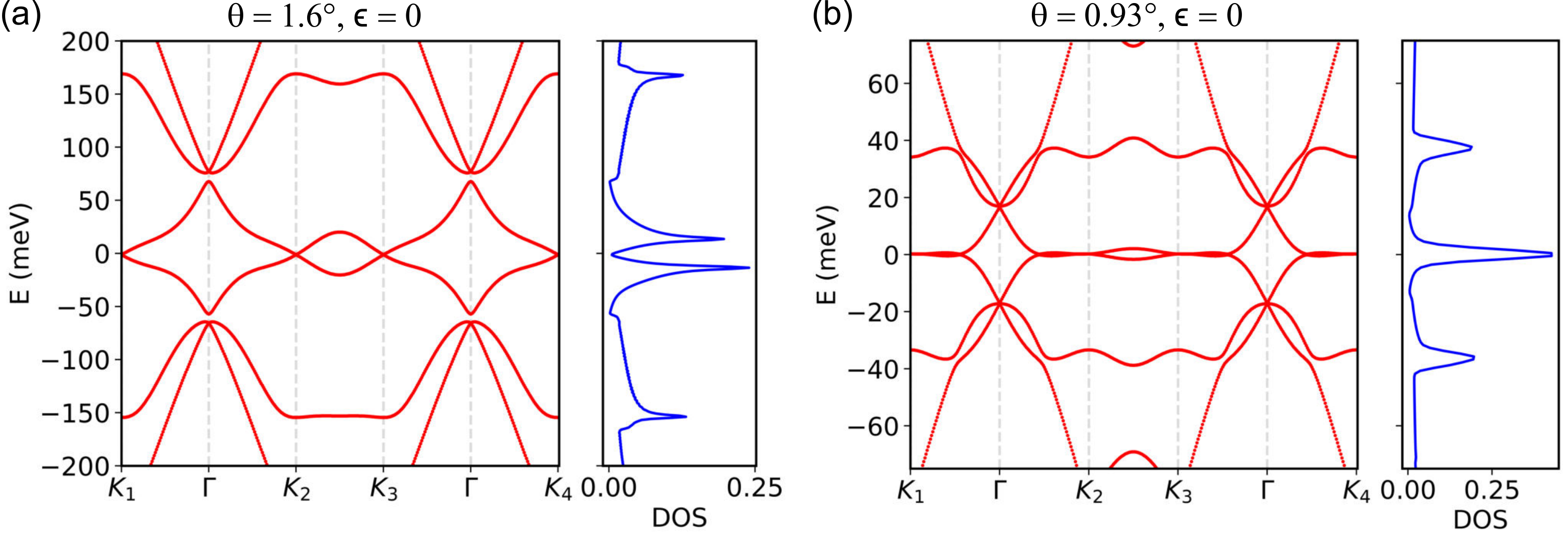}
    \caption{Tight-binding results of the band structure and density of states of rigid TBG with (a) $\theta=1.6\degree$ and (b) $\theta=0.93\degree$.}
\end{figure*}

\begin{figure*}[t]
    \centering
\includegraphics[width=0.8\linewidth]{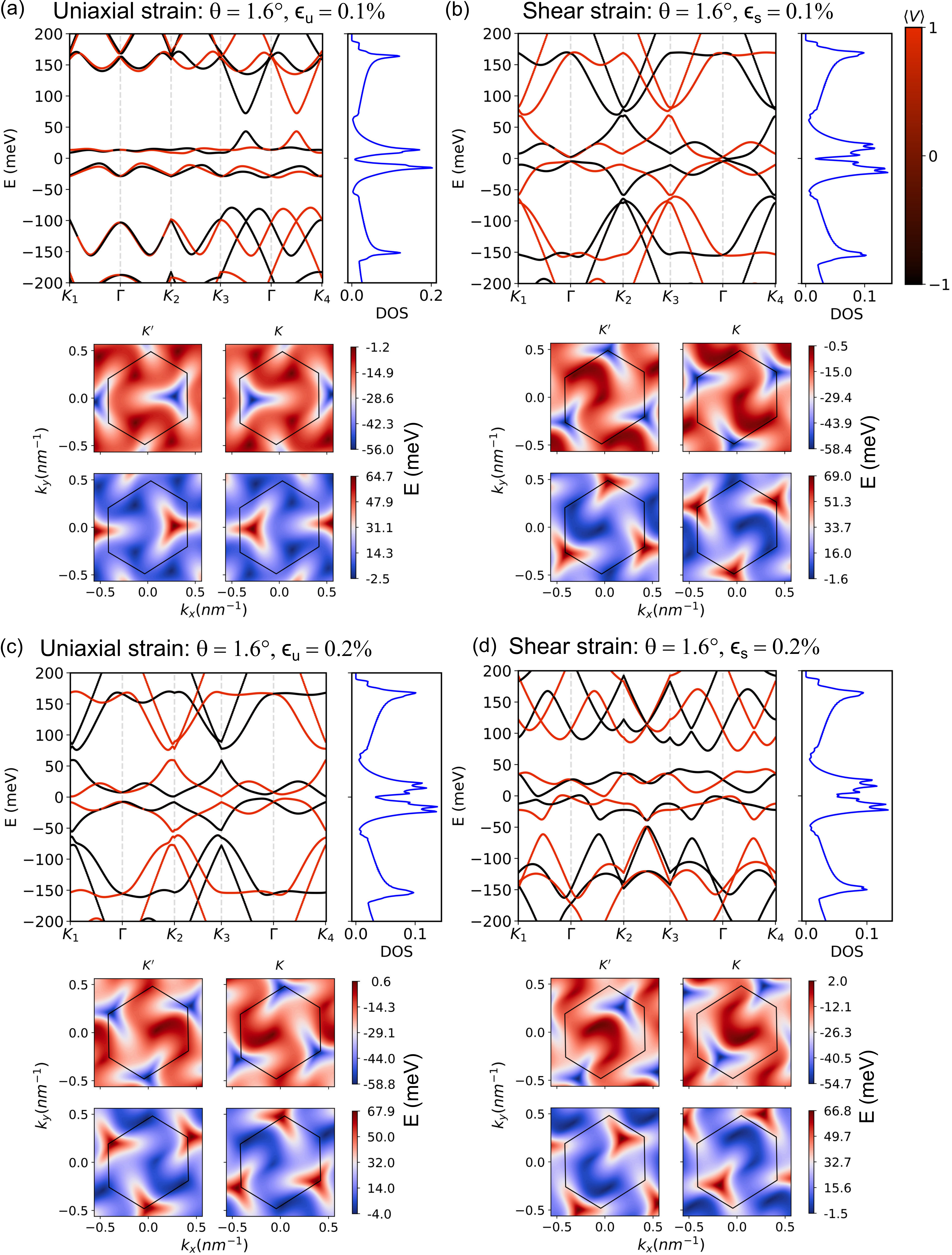}
    \caption{Tight-binding results of (upper panel) the band structure, density of states and (lower panel) energy map of four narrow bands for the commensurate structure of rigid TSBG with $\theta=1.6\degree$, $\phi=0\degree$ and (a) uniaxial strain $\epsilon_{u}=0.1\%$, (b) shear strain $\epsilon_{s}=0.1\%$, (c) uniaxial strain $\epsilon_{u}=0.2\%$, (d) shear strain $\epsilon_{s}=0.2\%$. In the energy map, the mBZ is illustrated with black line. The colors in the band structure are the same as in Fig. \ref{Figs2_magic}.}
\end{figure*} 

\begin{figure*}[h]
    \centering
\includegraphics[width=0.8\linewidth]{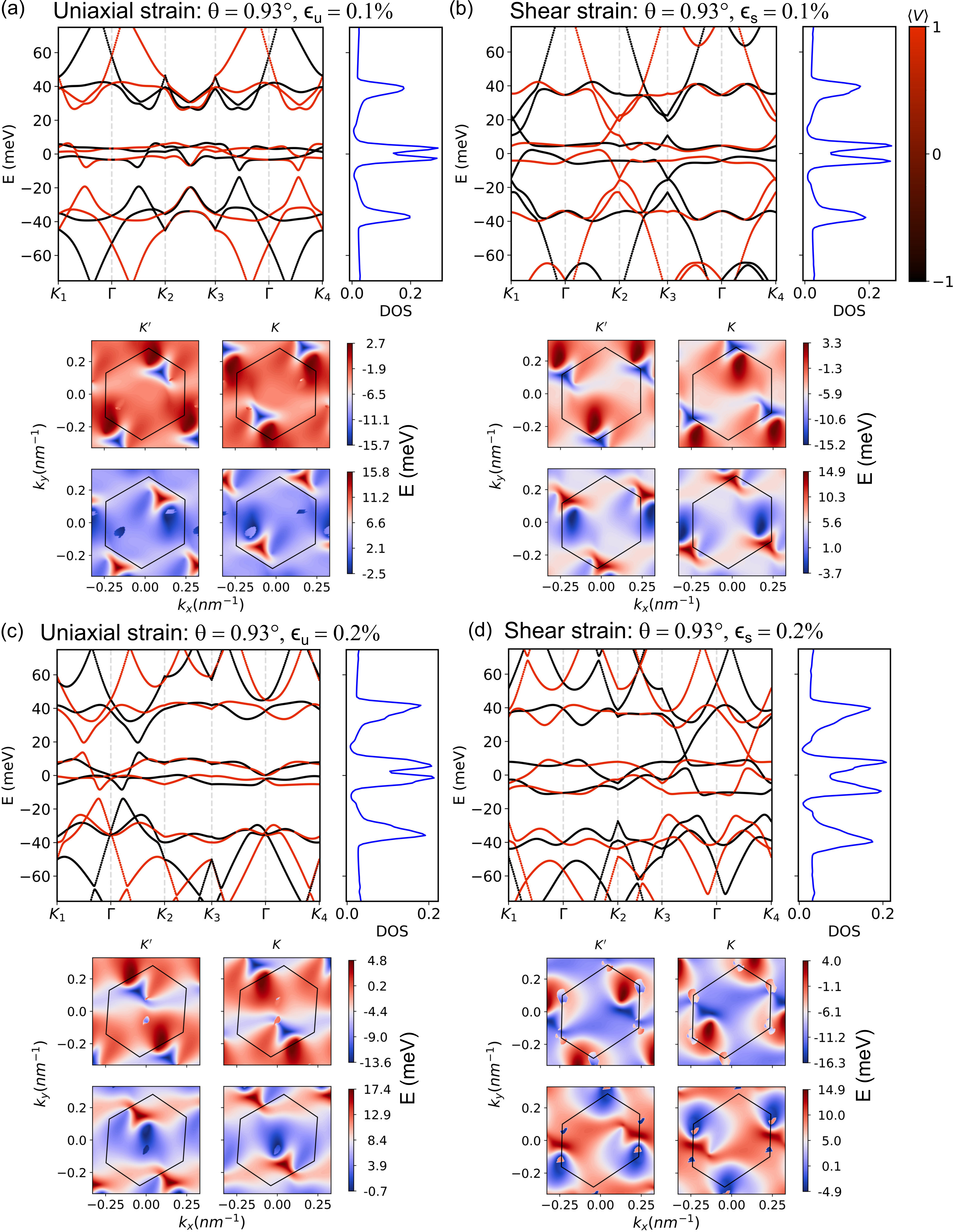}
    \caption{Tight-binding results of the band structure and density of states for the commensurate structure of rigid TSBG with $\theta=0.93\degree$, $\phi=0\degree$ and (a) uniaxial strain $\epsilon_{u}=0.1\%$, (b) shear strain $\epsilon_{s}=0.1\%$. In the energy map, the mBZ is illustrated with black line. The colors in the band structure are the same as in Fig. \ref{Figs2_magic}.}
\end{figure*}
\onecolumngrid\clearpage

\section{Flat band dome}
\begin{figure}[h]
    \centering
    \includegraphics[width=\linewidth]{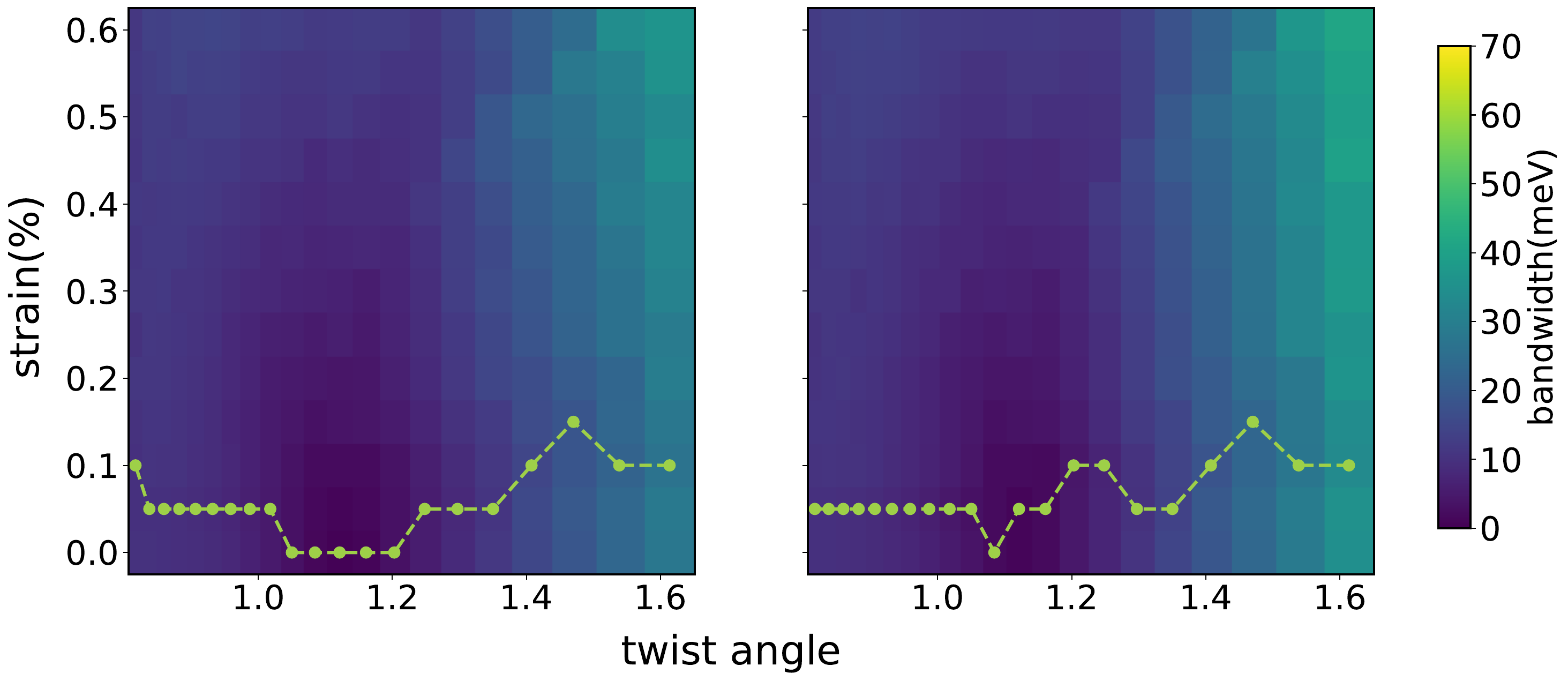}
    \caption{Color mesh of band width of valence (left) and conduction (right) bands for commensurate TSBG with twist angles from $0.8\degree$ to $1.6\degree$ and magnitudes of uniaxial strain from $0\%$ to $0.6\%$ with interval of $0.05\%$. The dots and dashed lines label the minimum band width for each twist angle. The strain direction is fixed to $\phi=0$.}
    \label{fig_mesh}
\end{figure}

We systematically study the change in the width of narrow bands with strain strength and twist angle via the TB calculations. The strain direction is fixed to $\phi=0$. The bandwidth is extracted from the band structure, for which a commensurate structure is required. The results are summarized in Fig. \ref{fig_mesh}. There are three interesting features in the results. First, in each twist angle, the minimum bandwidth appears around the small strain region. Second, for twist angle lower than $1.3^{\circ}$, the bandwidth is less than 20 meV, and is insensitive to the strain. There is a flat band dome around the magic angle ($1^{\circ} \sim 1.2^{\circ}$). Third, in the large angle region, with the strain magnitude increasing, the bandwidth first increases and then decreases.  

\onecolumngrid\clearpage

\section{The strain direction effect}

\begin{figure*}[h]
    \centering
    \includegraphics[width=\linewidth]{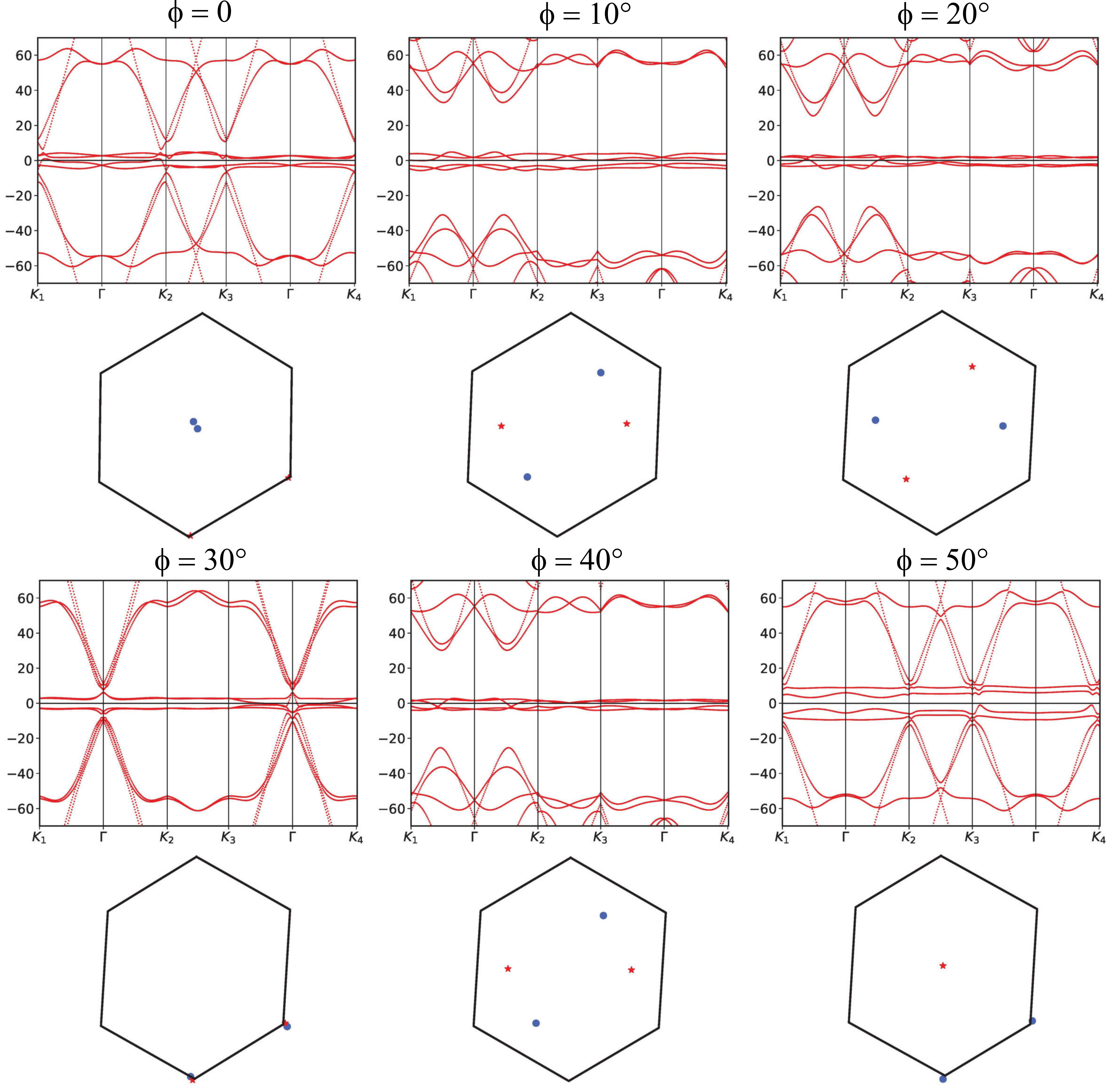}
    \caption{Tight-binding results of the band structure and the position of the Dirac point in the moir\'e Brillouin zone in rigid TSBG with different strain direction and fixed uniaxial strain. The strain direction changes from $0\degree$ to $50\degree$ with a step of $10^{\circ}$. The twist angle is $\theta=1.05\degree$ and strain magnitude is $\epsilon_{u}=0.1\%$. The blue dots corresponds to the projection of top two Dirac points and the red stars corresponds to bottom ones.}
    \label{Figs_direction}
\end{figure*}

In this section, we study the strain direction effect. We fix $\theta=1.05^{\circ}$ and $\epsilon_u=0.1\%$, and plot the band strauture and the Dirac positions for several values of strain direction $\phi$ as shown in Fig. \ref{Figs_direction}. Since the TBG has a $C_3$ symmetry, we can restrict our studies to $\phi \in[0,\pi/3)$. For different strain direction, the pair of eight integers to generate the commensurate structures are varied, leading to different positions of the Dirac points in mBZ. The narrow bands are also modified with the strain direction is changed. In the relative large twist case of $\theta=1.6^{\circ}$, the middle narrow bands are also sensitive to the strain direction, with a sharp peak appearing at $\phi=20^\circ$.  

\begin{figure*}[h]
    \centering
    \includegraphics[width=0.3\linewidth]{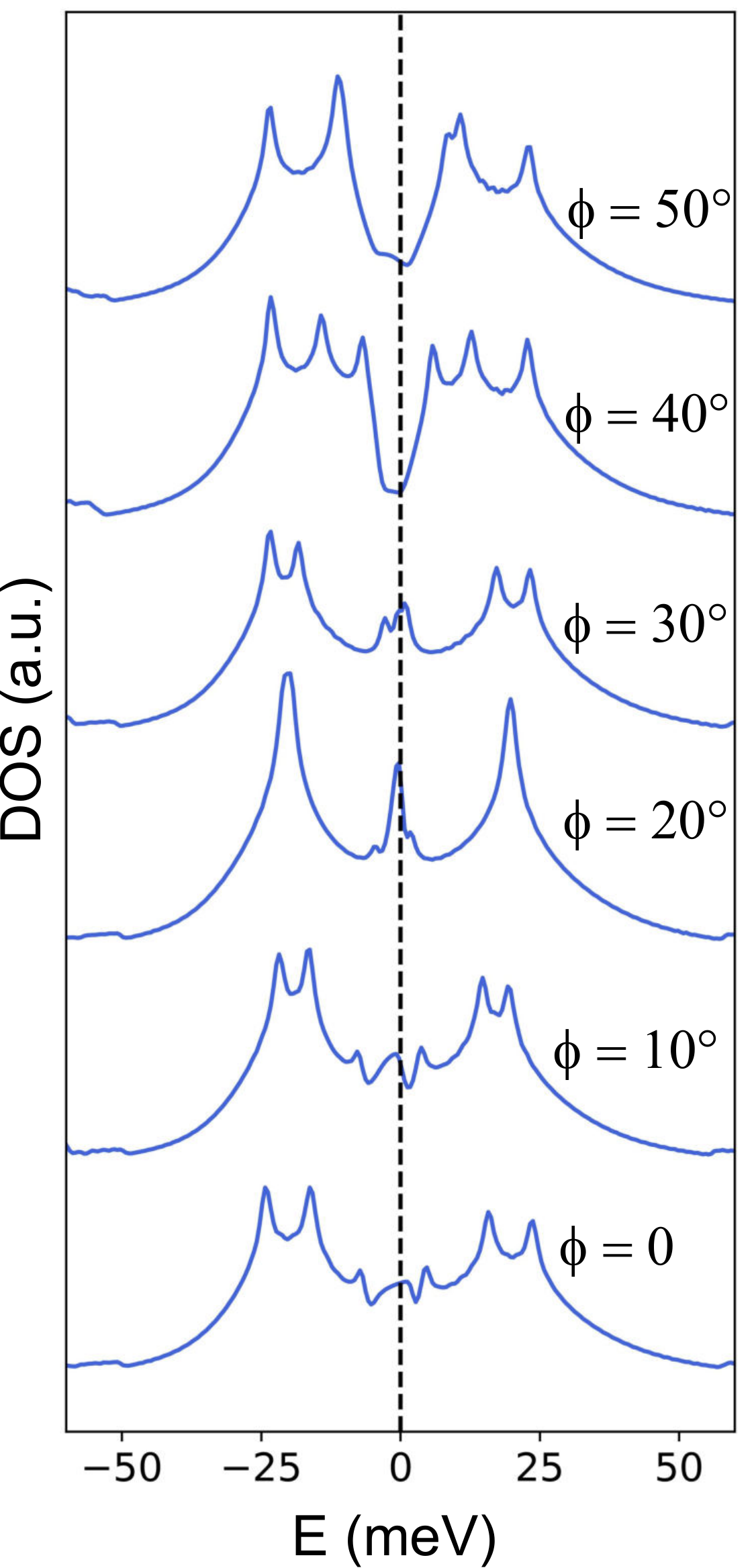}
    \caption{Evolution of DOS with uniaxial strain direction $\phi$ in TSBG with $\theta=1.6\degree$ and $\epsilon_u=0.3\%$, calculated by the TB model. We label $E=0\ \text{meV}$ with black vertical dashed lines. The curves are relatively shifted to make the plot clear.}
    \label{Figs_direction_dos}
\end{figure*}

\onecolumngrid\clearpage

\section{The lattice relaxation effect }

\begin{figure*}[h]
    \centering
    \includegraphics[width=\linewidth]{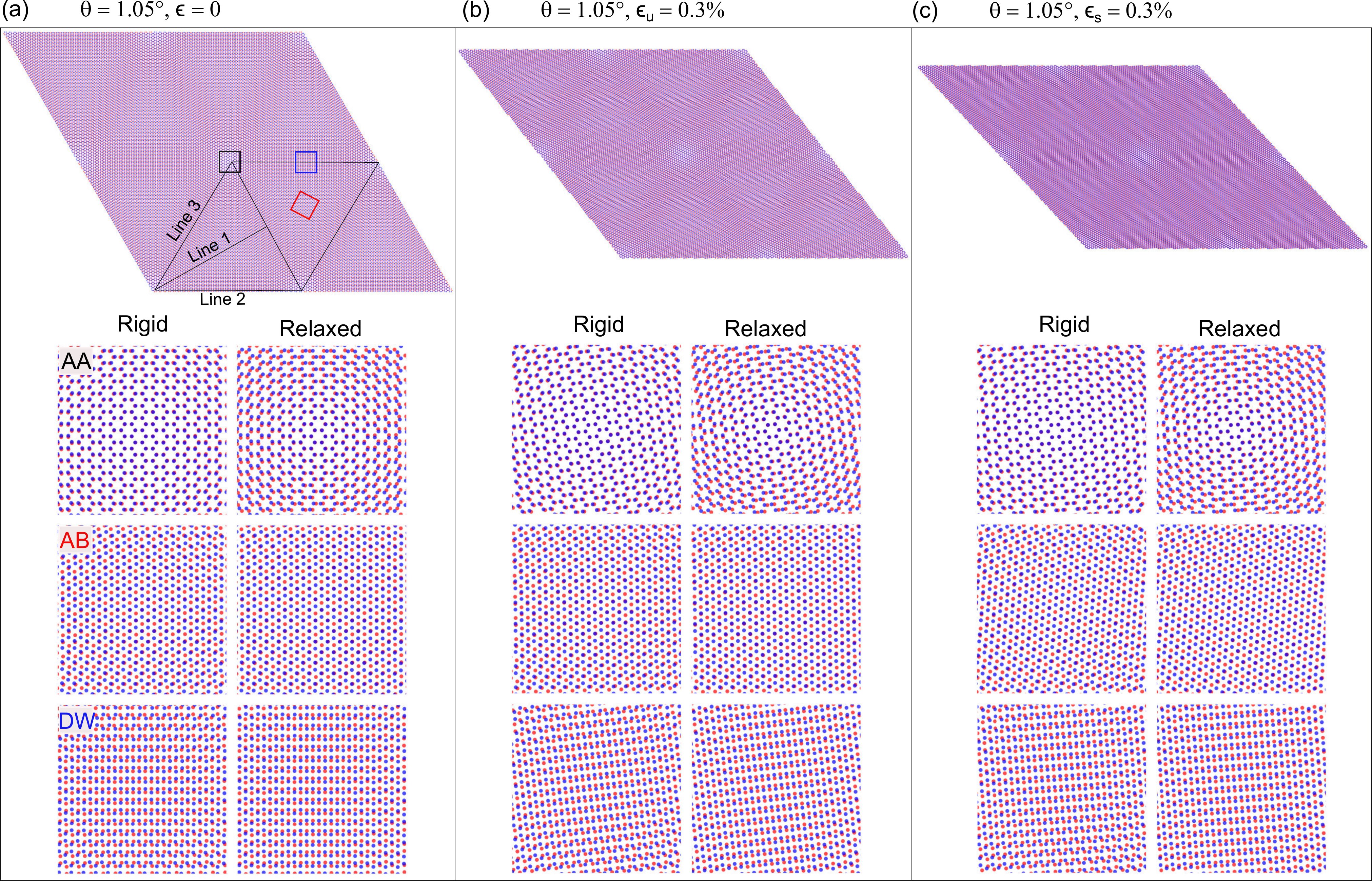}
    \caption{The lattice structure (top panel) and local atomic structure (bottom panel) near AA, AB, DW stackings before (left) and after (right) lattice relaxations for $\theta=1.05\degree$, $\phi=0\degree$ and (a) no strain, (b) uniaxial strain $\epsilon_{u}=0.3\%$, (c) shear strain $\epsilon_{u}=0.3\%$. The black parallelogram is the moir\'e unit cell. The local stackings is illustrated with colored square. Line 1, Line 2 and Line 3 are three paths that across different stackings.}
    \label{Figs_latt}
\end{figure*}

\begin{figure*}
    \centering
    \includegraphics[width=\linewidth]{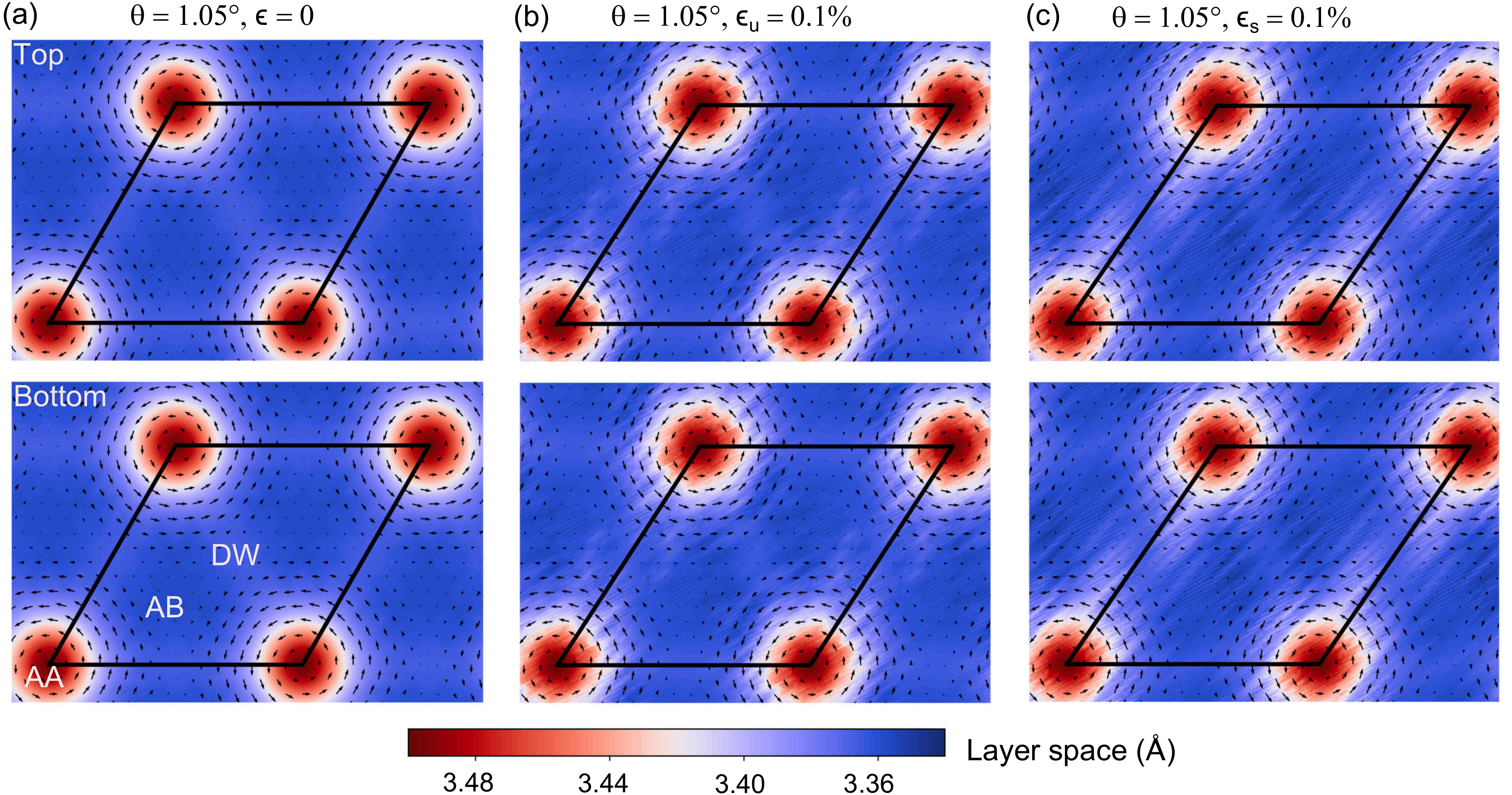}
    \caption{The position change of the atoms after relaxation  with $\theta=1.05\degree$ and (a) no strain, (b) uniaxial strain $\epsilon_{u}=0.1\%$, (c) shear strain $\epsilon_{s}=0.1\%$. The black arrows show in-plane displacement and the contours show layer spacing. The initial layer spacing is $0.3349$ nm. The unit cell is illustrated with black parallelograms.}
    \label{fig_relax_structure}
\end{figure*}

\begin{figure*}
    \centering
    \includegraphics[width=\linewidth]{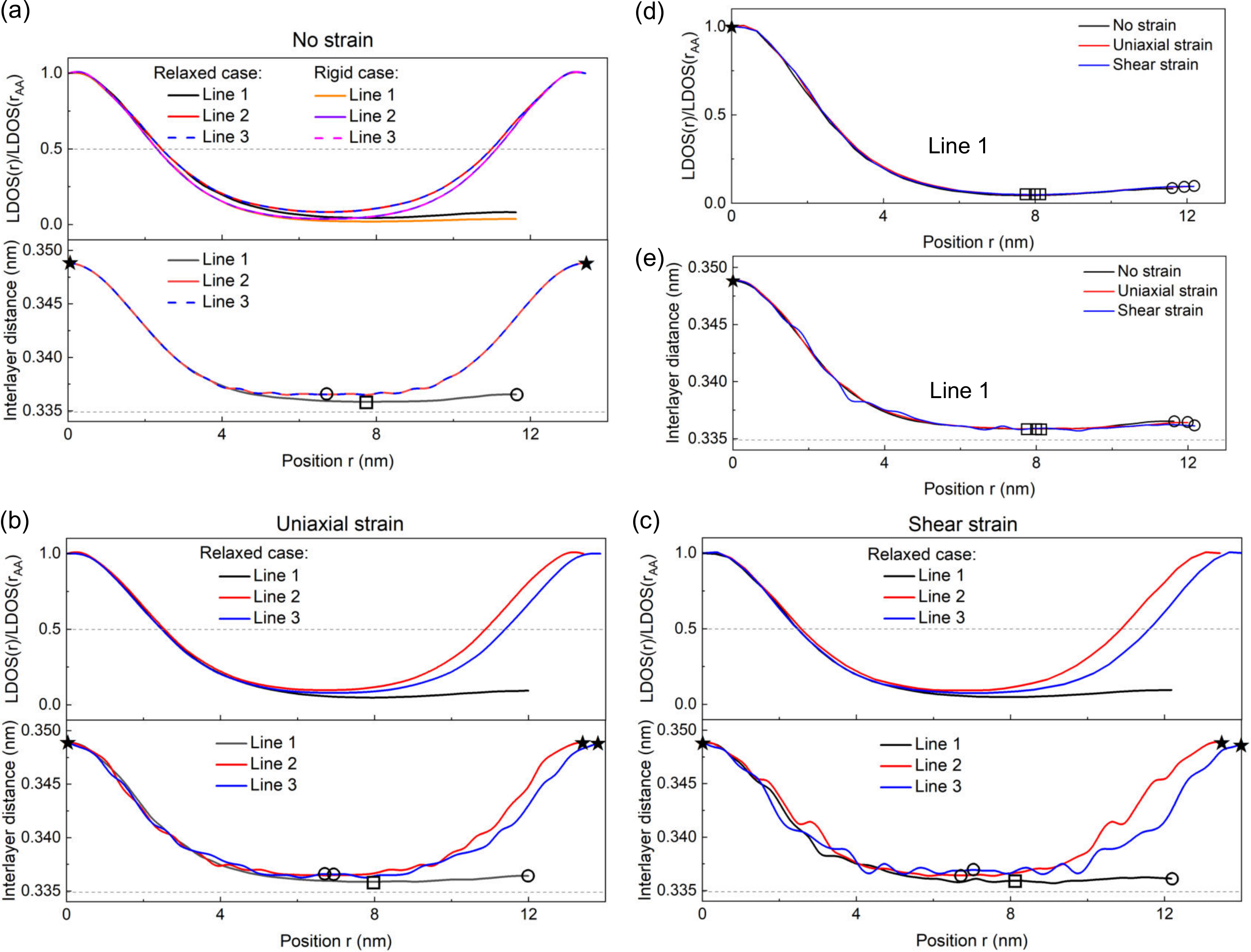}
    \caption{The in-plane and out-of-plane position changes of the atoms after lattice relaxation for TBG with $\theta=1.05\degree$ and different types of strain. (a) No strain case: (Top panel) The local DOS of the four narrow bands along the three lines. (Bottom panel) The interlayer distances along the three lines. The interlayer distance in the rigid case is 0.3349 nm (dashed horizontal line). The local DOS is normalized with LDOS(r$_{\mathrm{AA}}$), which is the DOS obtained in the AA point. Line 1, Line 2 and Line 3 are illustrated in Fig. \ref{Figs_latt}(a). Line 1 is from the AA stacking region (star) to the center of DW (circle) passing through AB domain (square). Line 2 and Line 3 are from AA to AA passing through the DW. The in-plane position changes are quantified by the changes of the local DOS. (b) The results for uniaxial strain $\epsilon_{u}=0.1\%$, (c) The results for shear strain $\epsilon_{s}=0.1\%$. (d) The local DOS and (e) interlayer distances along the Line 1 for no strain, uniaxial strain and shear strain cases. The fluctuations in the interlayer distances come from the variation in the atomic registry due to the twist and strain, and the convergence criteria \cite{guinea2019continuum}.}
    \label{fig_relax_structure1}
\end{figure*}

In this section, we study the lattice relaxation effect on both geometry and electronic structures of TBG under strain. First, let us focus on the geometric effect. Figures \ref{Figs_latt}, \ref{fig_relax_structure} and \ref{fig_relax_structure1} show respectively the local structure and atom displacements of magic angle in three cases: no strain, uniaxial and shear strains. The strain distorts the hexagonal structures, and makes the AA region elliptical. The elongated AA region can be quantified by the changes of the local DOS of the four narrow bands in different directions (for example, Line 1, Line 2 and Line 3) even in the case of 0.1$\%$ strain, shown in Fig. \ref{fig_relax_structure1}. This effect could be visible in local measurements such as scanning tunneling microscope. For the same strain magnitude, the shear strain makes the moir\'e more distorted. This may explain why the shear strain has more effect on the electronic structure of TBG than uniaxial strain with the same strength. 

There are two additional global features. First, the lattice relaxation effect on the geometry of TBG without or with strain are similar (in particular in small strain case in Fig. \ref{fig_relax_structure1}). Specially, the lattice relaxation significantly shrinks the AA region and expands the AB regions to form a triangular domain \cite{guinea2019continuum}. The interlayer distance in the AA region is larger than those in other stacking regions. This may explain why an unequal ration between the hopping energies of AA and Bernal stacking in the continuum model could capture the main features in the relaxed case. Second, compared to the nonstrain case, the strain causes a rotation of the local structures, as shown in Fig. \ref{Figs_latt}. Such a rotation introduces significant modulation of the domain wall (DW) region. In TBG without strain, due to the lattice relaxation, the system clearly exhibits a triangular domain pattern of AB and BA regions, and a shear domain boundary. In this shear domain boundary, the Burger vector is parallel to the DW. In the strained cases, the rotation changes the angle between the Burger vector and the DW boundary, modulating the DW from a shear type to mixed type of both shear and tensile \cite{lebedeva2020two,mesple2023giant}. Such modulation also affects the atom movements in the DW region. As shown in Fig. \ref{fig_relax_structure}, the atom movements in the DW region of TBG without and with strain are different. Consequently, due to the transition of the DW type and the distinct lattice relaxation, the electronic properties of the DW regions, which are in a high energy region, could be significantly different in the three strain cases \cite{nguyen2021electronic,timmel2020dirac}. 

\begin{figure*}[h]
    \centering
    \includegraphics[width=\linewidth]{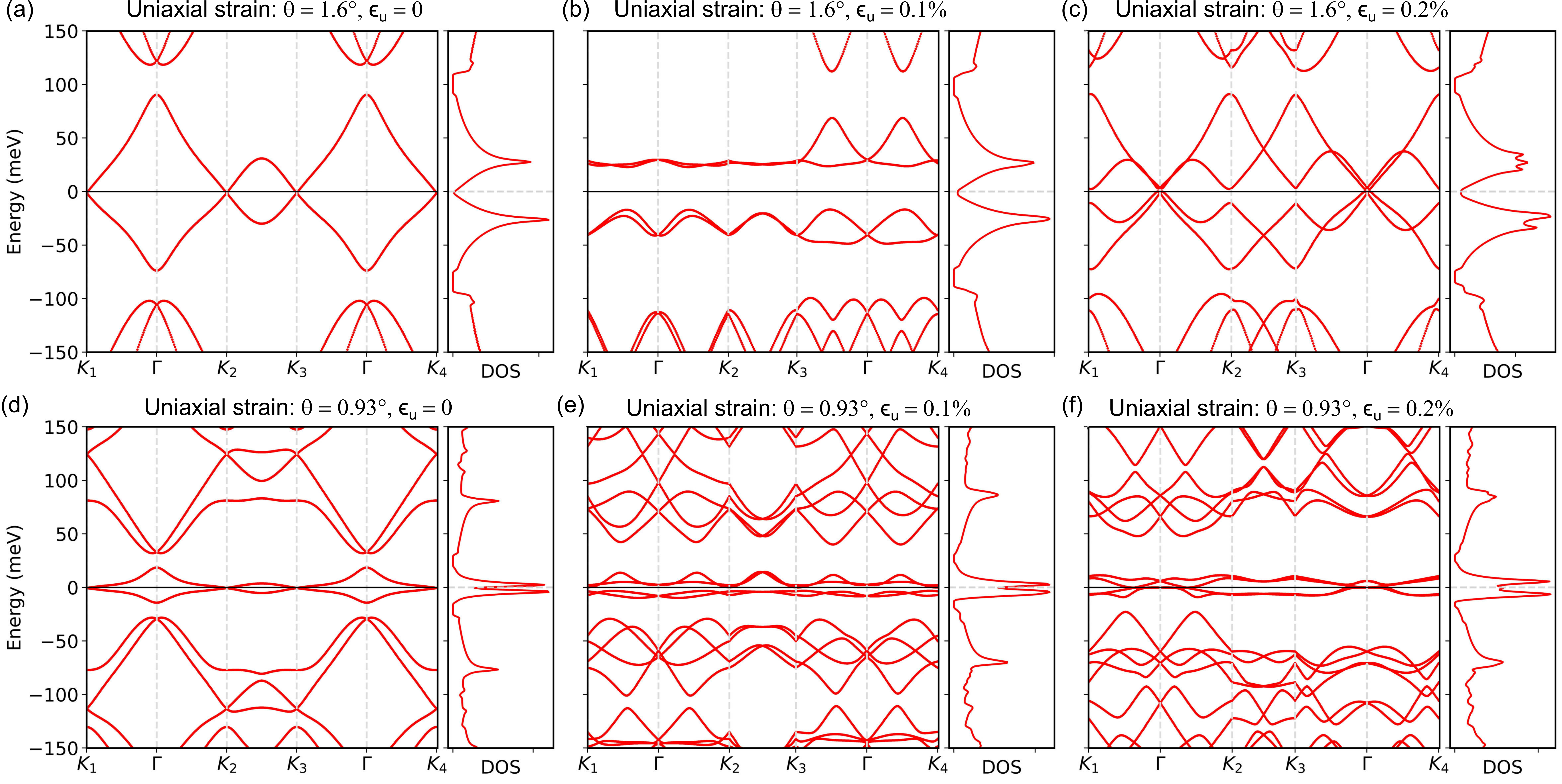}
    \caption{Tight-binding results of the band structure and density of states for the relaxed structure of moir\'e graphene with (upper panel) $\theta=0.93\degree$, $\phi=0\degree$ and (a) no strain, (b) uniaxial strain $\epsilon_{u}=0.1\%$, (c) uniaxial strain $\epsilon_{u}=0.2\%$, and (lower panel)  $\theta=1.6\degree$, $\phi=0$ with (d) no strain, (e) uniaxial strain $\epsilon_{u}=0.1\%$,(a) uniaxial strain $\epsilon_{u}=0.2\%$.}
    \label{Figs_bs_relaxed}
\end{figure*}

Figure \ref{Figs_bs_relaxed} show the band structure and DOS of relaxed TBG with $\theta_t=1.6^{\circ}$ and $\theta_t=0.93^{\circ}$ in the presence of an uniaxial strain. In general, the lattice relaxation opens a gap between the narrow and remote bands, increases the separation between the valence and conduction narrow bands, broadens the width of narrow bands. In the system with twist angle $\theta_t=0.93^{\circ}$, two peaks appear around the charge neutrality point (CNP) in the relaxed cases, whereas only one peak appears in the rigid cases. For $\theta_t=1.6^{\circ}$ and $\theta_t=0.93^{\circ}$, the DOS peaks from the conduction and valence bands have equal magnitudes.

\onecolumngrid\clearpage

\section{Continuum model band structures}

\begin{figure}[b]
    \centering
    \includegraphics[width=\linewidth]{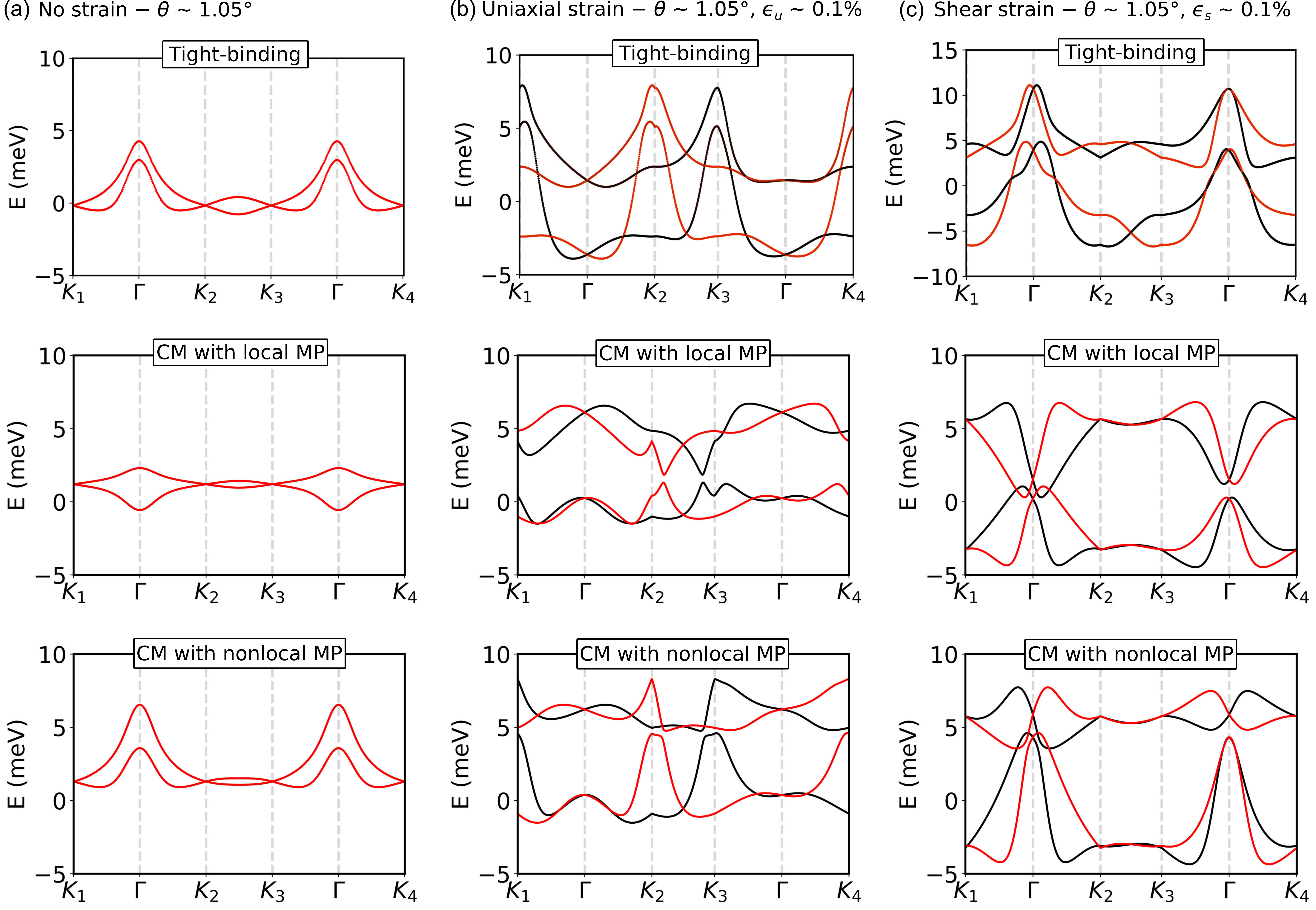}
    \caption{Narrow bands obtained with the tight-binding and the continuum model (with and without nonlocal moiré potential). The cases shown correspond the same commensurate twist and strain configurations considered in Figure 5 of the main text, with the same color profile.}\label{fig:BS_NarrowBands}
\end{figure}

Figure \ref{fig:BS_NarrowBands} shows a comparison between the narrow bands obtained by the tight-binding and continuum models, for the same commensurate twist and strain configurations considered in Figure 5 of the main text. The inclusion of the nonlocal potential in the continuum model clearly provides a better agreement with the TB results. 

\begin{figure}[t]
    \centering
    \includegraphics[width=\linewidth]{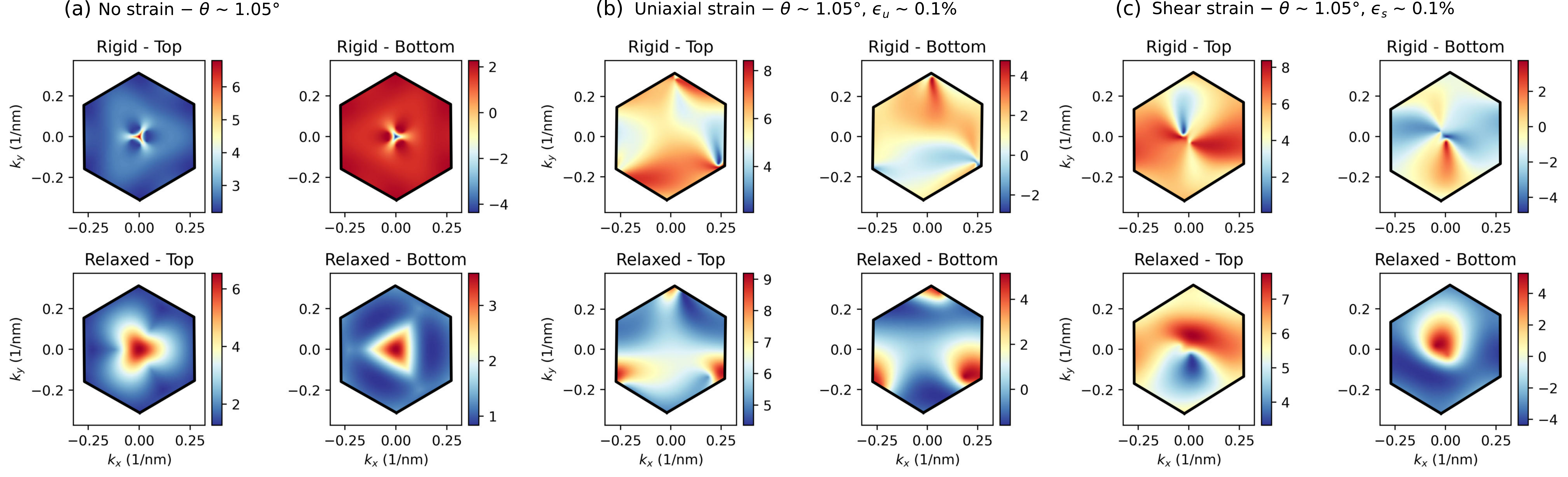}
    \caption{Continuum model density plots of the top and bottom narrow bands (K valley), for the same twist and strain configurations as in Figure 3 of the main text (commensurate solutions). The relaxed results take into account the nonlocal moiré potential. See also Figure 6 in the main text for the TB results (with strain cases).}\label{fig:Bands_Strain}
\end{figure}

Figure \ref{fig:Bands_Strain} shows density plots of the continuum model band structure at the magic angle, for the cases without strain, with uniaxial strain, and with shear strain. The twist and strain configurations are as in Figure 3 of the main text, corresponding to commensurate solutions. 

Compared to the TB results of Figure 6 in the main text, we observe that the continuum model results are in relatively good agreement. Importantly, the nonlocal moiré potential correctly captures the relaxation-induced particle-hole asymmetry, both in the case with and without strain. In general, we find a better agreement between TB and continuum in the rigid case. In part, this is because the local moiré potential only accounts for the relaxation of the AA, AB and DW regimes through an unequal ratio of the effective $u_0$ and $u_1$ hoppings. A more realistic treatment, which is expected to better capture the TB results, would be to include lattice relaxation fields within the continuum model \cite{kang2025analytical}.

\clearpage

\section{Continuum model DOS for relaxed TSBG with different direction}

\begin{figure}[h]
    \centering
    \includegraphics[width=1\linewidth]{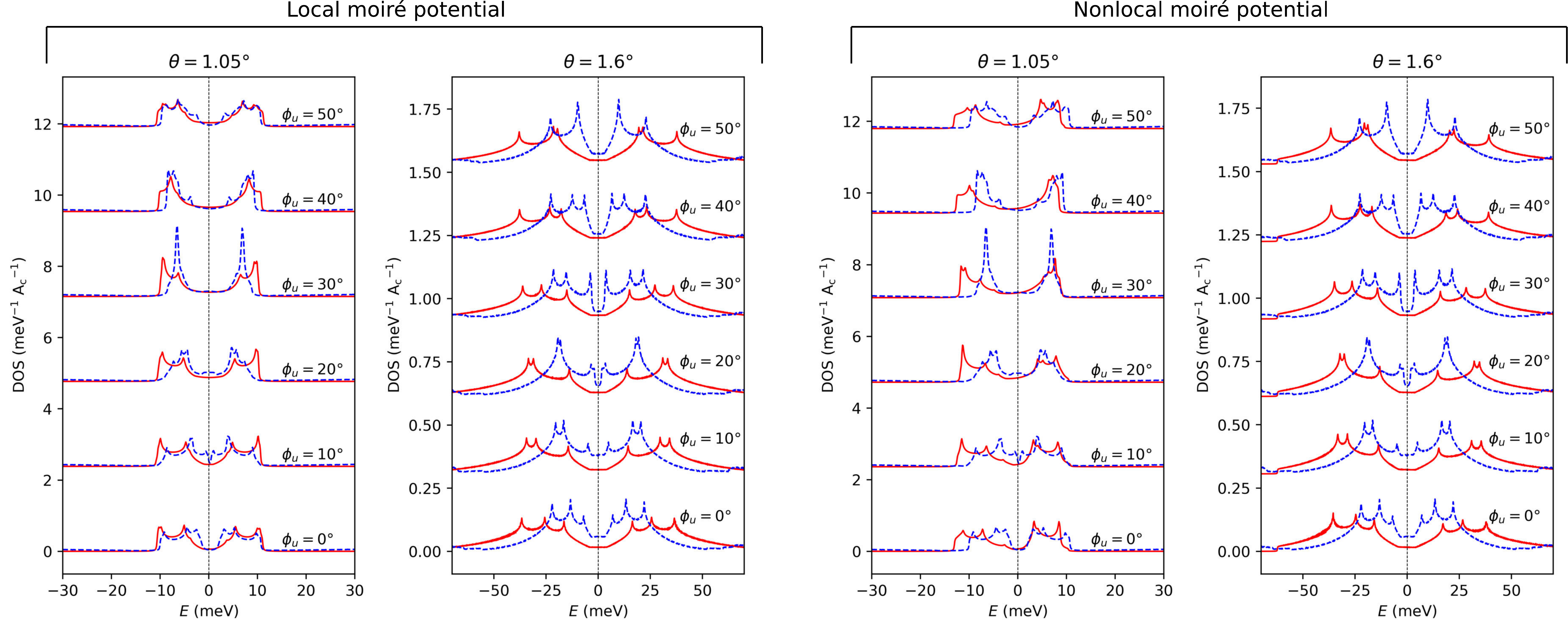}
    \caption{Continuum model total DOS for uniaxial heterostrain with magnitude $\epsilon_{u}=0.3\%$ along different directions $\phi$, and fixed twist angles (a) $\theta=1.05^{\circ}$ and (b) $\theta=1.6^{\circ}$. The solid red line correspond to the \textit{relaxed} results with $\hbar v/a=2.13\:\mathrm{eV}$, $w_{1}=0.096\,\mathrm{eV}$ and $w_{1}=0.05952\,\mathrm{eV}$, while the dashed blue line correspond to the \textit{rigid} results with $\hbar v/a=2.15\:\mathrm{eV}$, $w_0=w_1=0.1\,\mathrm{eV}$. Left and right panels show the relaxed results with and without the inclusion of the nonlocal moiré potential (see main text).}\label{fig:CM_dos_relax}
\end{figure}

In Fig. \ref{fig:CM_dos_relax} we show a comparison of the total density of states, for $\theta=1.05^{\circ}$ and uniaxial heterostrain with magnitude $\epsilon_{u}=0.3\%$, between the rigid and relaxed configurations, with and without the nonlocal moiré potential. We observe that the overall tendency of multiple VHs, highly sensitive to the strain, is preserved under relaxation. However, the particular location of the VHs is quite sensitive to relaxations. The relaxation-induced particle-hole asymmetry, accounted for by the nonlocal moiré potential, is reflected in unequal vHS with respect to charge neutrality. Overall, the relaxation tends to reduce and broaden the VHs, but this effect becomes more appreciable at larger twist angles. This confirms that the strain influence on the flat bands around the magic angle are less sensitive to relaxation effects.

\clearpage

\section{Role of the nonlocal moiré potential}

\begin{figure*}[h]
    \centering
         \includegraphics[width=\linewidth]{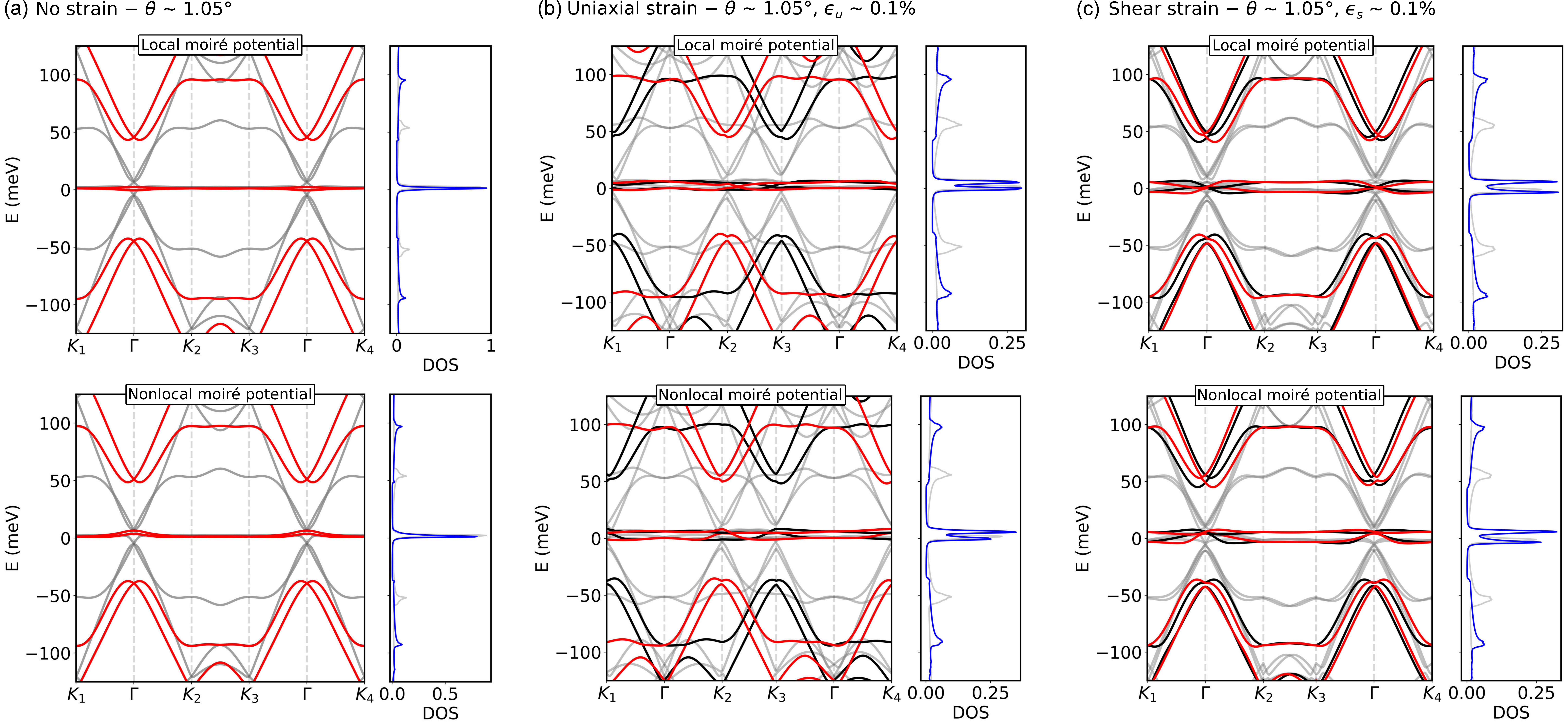}
     \caption{Comparison of the relaxed continuum model band structures, with and without the inclusion of the nonlocal moiré potential. The twist and strain configurations, and the continuum model parameters, are the same as in Figures 5 of the main text. }\label{fig:BS_comparison}
\end{figure*}

This can be seen in Figure \ref{fig:BS_comparison}, where we show a comparison of the continuum model with and without the inclusion of the nonlocal moiré potential. The twist and strain configurations considered correspond to those in Figures 3 and 5 of the main text.

As noted in the main text, the main effect of the nonlocal moiré potential is to capture the relaxation-induced particle-hole asymmetry of the band structure, both with and without strain. The nonlocal potential also shifts slightly the remote bands towards higher energies, in agreement with the relaxed TB band structure (see Figure 5 in the main text). 

However, besides the above effects, the role of the nonlocal moiré potential is in general small. For example, the shape and the gap between the narrow and remote bands is very similar to the case with only the local potential. Thus, the nonlocal moiré potential could be safely neglected when one is interested in general trends (particularly in situations when the inclusion of a nonlocal potential significantly increases the computational complexity, e.g., after interactions are included). 

\clearpage

\section{Continuum model LDOS at different stackings}

\begin{figure*}[h]
    \centering
         \includegraphics[width=\linewidth]{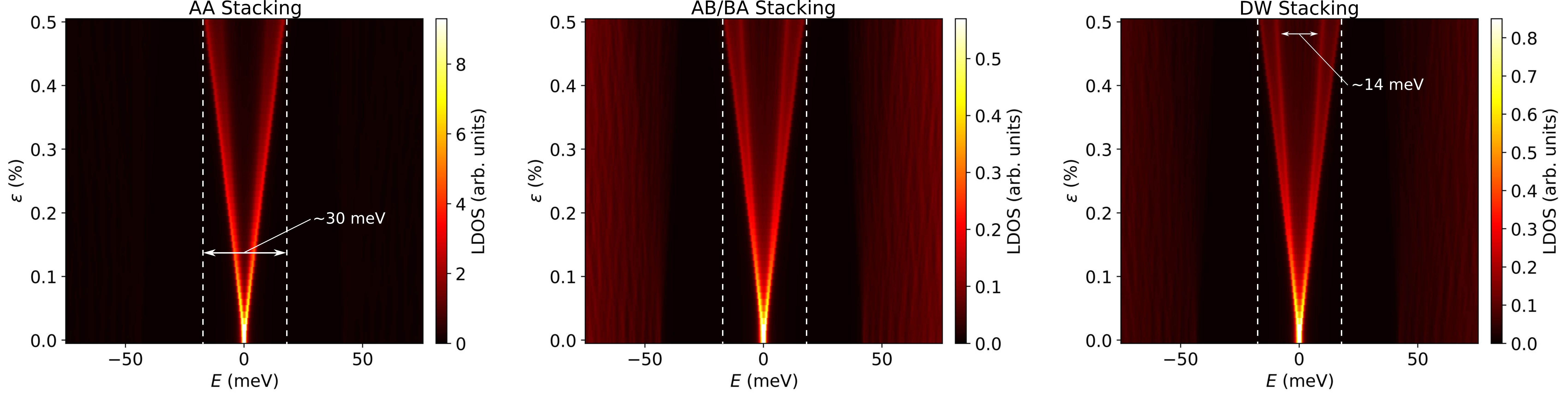}
     \caption{Continuum model LDOS at the AA, AB/BA and DW stacking regimes, as a function of the energy and uniaxial strain magnitude $\epsilon$, for fixed twist angle $\theta=1.05^{\circ}$ and strain direction $\phi=30^{\circ}$. The vertical dashed white line indicate the maximum splitting by $\sim30\,\mathrm{meV}$ of the van Hove singularities (vHs) at $\epsilon=0.5\%$. A smaller two fold splitting of each vHs takes place at larger strains, reducing the closest vHs to about $\sim14\,\mathrm{meV}$. All the results correspond to the relaxed configuration of the continuum model, but taking into account only the local moiré potential (see main text).}\label{fig:LDOS_strain}
\end{figure*}

The strain effect on the emergence of narrow bands naturally influences the local density of states (LDOS). This is seen in Figure \ref{fig:LDOS_strain}, which shows the LDOS as a function of the energy and uniaxial strain magnitude, for fixed a twist angle $\theta=1.05^{\circ}$and strain direction $\phi=30^{\circ}$, at the three stackings AA, AB/BA and DW. For the two central narrow bands, the largest LDOS is always at the AA stackings (about one order of magnitude larger than at the AB/BA and DW regimes). Within the narrow bands energy range, the effect of increasing the strain leads to a reduced magnitude of the LDOS and a splitting of the van Hove singularities (vHs). Both behaviors reflects the increase of the narrow bands bandwidth due to the strain (about $\sim30\,\mathrm{meV}$ at $\epsilon=0.5\%$; see Figure 7 of the main text). At relatively large strains we also see that the two separated vHs are further split by two, with almost equal magnitude at the AB/BA and DW stackings. Interestingly, the LDOS of the remote bands ($\left|E\right|<40\,\mathrm{meV}$) seem to be practically insensitive to the strain, with comparable magnitude for the three stackings regimes.

\section{Bandwidth with strain}

\begin{figure*}[t]
    \centering
\includegraphics[width=0.8\linewidth]{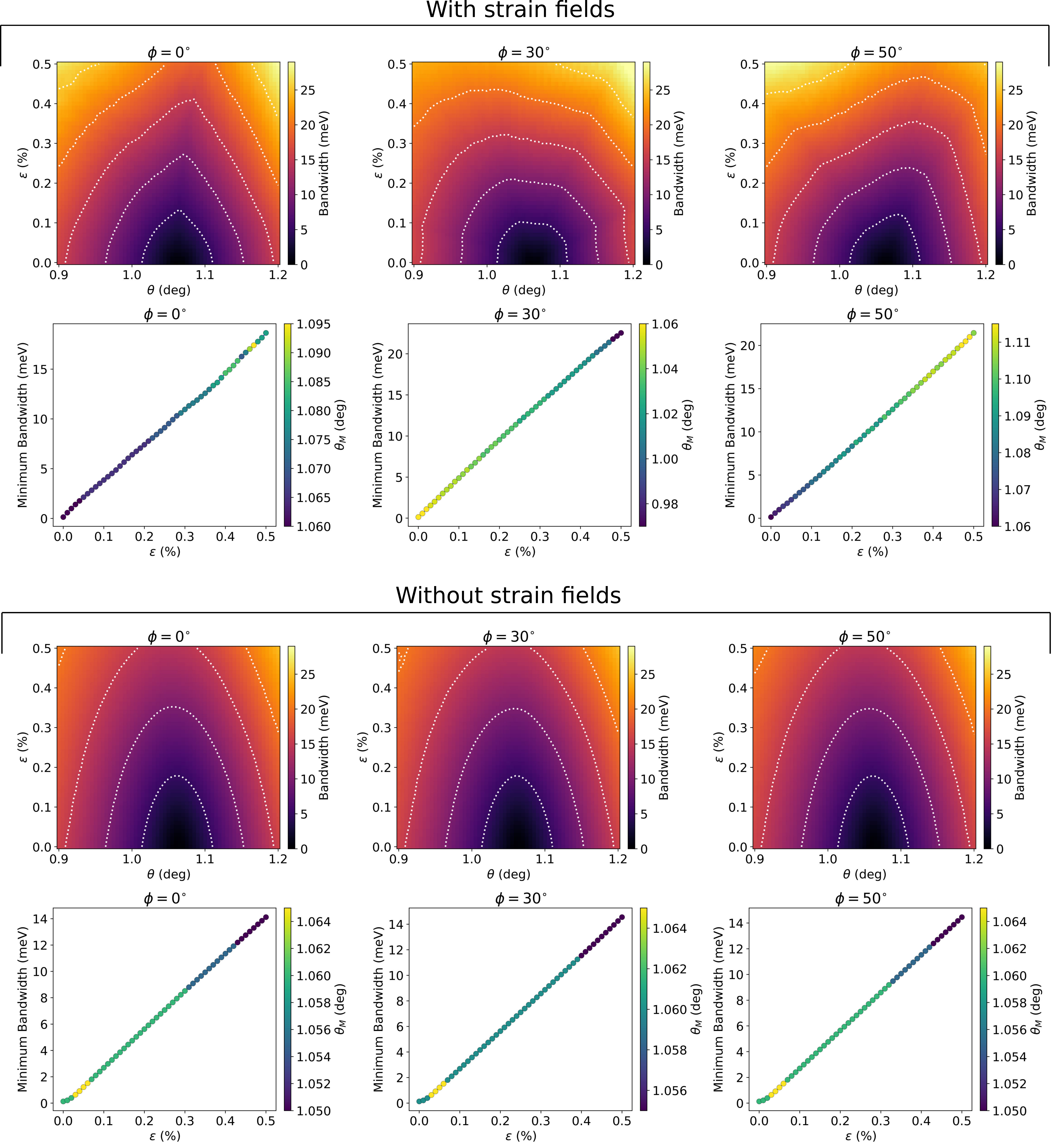}
     \caption{Numerical continuum model results for the bandwidth evolution in the top narrow band, for the same twist and strain configurations as in Figure 7 of the main text. The top panel shows the results with strain fields $V$ and $\mathbf{A}$ (top panels, same as in Figure 7 of the main text). The bottom panel shows results by turning off the strain fields $V$ and $\mathbf{A}$. }\label{fig:BW_strain_Nofields}
\end{figure*}

In Figure 7 of the main text we presented the evolution of the bandwidth as a function of the twist and the strain magnitude and direction. The main observation there is that the bandwidth is highly sensitive to the strain direction, a behavior which was already reflected in the DOS of Figure 4 in the main text. Within the continuum model, the bandwidth dependence on strain configuration comes from both the strain fields and the moiré potential (see Section III A.2 in the main text). 

To understand how the strain fields and the moiré potential contribute to the bandwidth evolution, here we repeated the calculation of Figure 7 in the main text, but turning off the strain-fields. The results are shown in Figure \ref{fig:BW_strain_Nofields}. Clearly, we observe that the strain direction dependence practically disappears without the strain fields. This result is again consistent with Figure 4 in the main text, whereby one sees that the variation of the DOS with the strain direction comes primarily from the strain fields. Another crucial difference we observe without the strain fields is that the twist angle of minimum bandwidth practically does not change as the strain increases. 

Notably, even after turning off the strain fields, we still observe that the minimum bandwidth scales linearly with the strain magnitude. This points out that is the moiré potential the main responsible for such linear scaling of the bandwidth with the strain. We further observe that without the strain fields the bandwidth at a given strain magnitude is always lower than with the strain fields. This is also consistent with Figure 4 of the main text, which shows a smaller bandwidth and larger VHs without the strain fields. 

From these observations we conclude that around the magic angle the strain fields tend to overall increase the bandwidth, but with a strength that depends on both the strain magnitude and direction.

\clearpage

\section{Electrostatic interactions with strain}

Here we briefly review the treatment of electrostatic interactions within the continuum model \cite{guinea2018electrostatic, cea2019electronic, goodwin2020hartree}. The Hartree interaction is the direct (classical) interaction of an electron with the surrounding charge density. Due to the moiré potential, the charge density is not homogeneously distributed in real space \cite{trambly2010localization, rademaker2018charge}. Within a jellium model, the net charge density $\delta\rho\left(\mathbf{r}\right)$
is given by
\begin{equation}
\delta\rho\left(\mathbf{r}\right)=\sum_{\mathbf{k}}\sum_{n,\eta,i}^{\prime}\left|\psi_{n,\mathbf{k},\eta,i}\left(\mathbf{r}\right)\right|^{2},
\end{equation}
where the prime indicates summation only over occupied (or unoccupied) states from CNP. The plane-wave expansion of the Bloch states in TSBG reads
\begin{align}
\psi_{n,\mathbf{k},\eta,i}\left(\mathbf{r}\right) & =\frac{1}{\sqrt{A_{c}}}\sum_{\mathbf{G}}u_{n,\mathbf{k},\eta,i}\left(\mathbf{G}\right)e^{i\left(\mathbf{k}+\mathbf{G}\right)\cdot\mathbf{r}},
\end{align}
where $A_{c}$ is the moiré unit cell area, and $n,\eta,i$ are the band, valley/spin and layer/sublattice indices, respectively. The Fourier coefficients $u_{n,\mathbf{k},\eta,i}\left(\mathbf{G}\right)$ are normalized as \cite{cea2019electronic} $\sum_{\mathbf{G},i}u_{n,\mathbf{k},\eta,i}^{*}\left(\mathbf{G}\right)u_{m,\mathbf{k},\eta,i}\left(\mathbf{G}\right)=\delta_{n,m}$, which ensures that the Bloch wave function is normalized within the moiré unit cell: $\sum_{i}\int_{\mathrm{unit\,cell}}d\mathbf{r}\left|\psi_{n,\mathbf{k},\eta,i}\left(\mathbf{r}\right)\right|^{2}=1$. The Fourier expansion of the charge density then reads
\begin{align}
\delta\rho\left(\mathbf{r}\right) & =\sum_{\mathbf{G}\neq0}\delta\rho\left(\mathbf{G}\right)e^{-i\mathbf{G}\cdot\mathbf{r}},\\
\delta\rho\left(\mathbf{G}\right) & =A_{c}^{-1}\sum_{\mathbf{k},\mathbf{G}'}\sum_{n,\eta,i}^{\prime}u_{n,\mathbf{k},\eta,i}^{*}\left(\mathbf{G}'+\mathbf{G}\right)u_{n,\mathbf{k},\eta,i}\left(\mathbf{G}'\right).
\end{align}
The Hartree potential felt by an electron at position $\mathbf{r}$ then corresponds to the classical interaction
\begin{equation}
V_{H}\left(\mathbf{r}\right)=\int d\mathbf{r}'v_{C}\left(\mathbf{r}-\mathbf{r}'\right)\delta\rho\left(\mathbf{r}'\right),
\end{equation}
where $v_{C}\left(\mathbf{r}-\mathbf{r}'\right)$ is the bare Coulomb potential. Replacing the Fourier expansion of the charge density leads to Eqs. (32) and (31) in the main text:
\begin{align}
V_{H}\left(\mathbf{r}\right) & =\sum_{\mathbf{G}\neq0}V_{H}\left(\mathbf{G}\right)e^{-i\mathbf{g}\cdot\mathbf{r}},\\
V_{H}\left(\mathbf{G}\right) & =\frac{v_{C}\left(\mathbf{G}\right)}{A_{c}}\sum_{\mathbf{k},\mathbf{G}'}\sum_{n,\eta,i}^{\prime}u_{n,\mathbf{k},\eta,i}^{*}\left(\mathbf{G}'+\mathbf{G}\right)u_{n,\mathbf{k},\eta,i}\left(\mathbf{G}'\right),
\end{align}
where $v_{C}\left(\mathbf{G}\right)$ is the Fourier transform of the bare Coulomb potential. In the main text we consider a gated configuration of two metallic plates \cite{bernevig2021twisted}, for which $v_{C}\left(\mathbf{G}\right)=e^{2}\tanh\left(d\left|\mathbf{G}\right|\right)/2\varepsilon_{0}\varepsilon_{r}\left|\mathbf{G}\right|$, where $d$ is the distance between the two metallic plates, and $\varepsilon_{r}$ is the relative primitivity of the system. The results obtained, and our conclusions in the main text, qualitatively do not differ if we instead consider a nongated potential $v_{C}\left(\mathbf{G}\right)\propto1/\left|\mathbf{G}\right|$). 

\clearpage

\section{Berry curvature}

\begin{figure*}[t]
    \centering
\includegraphics[width=0.8\linewidth]{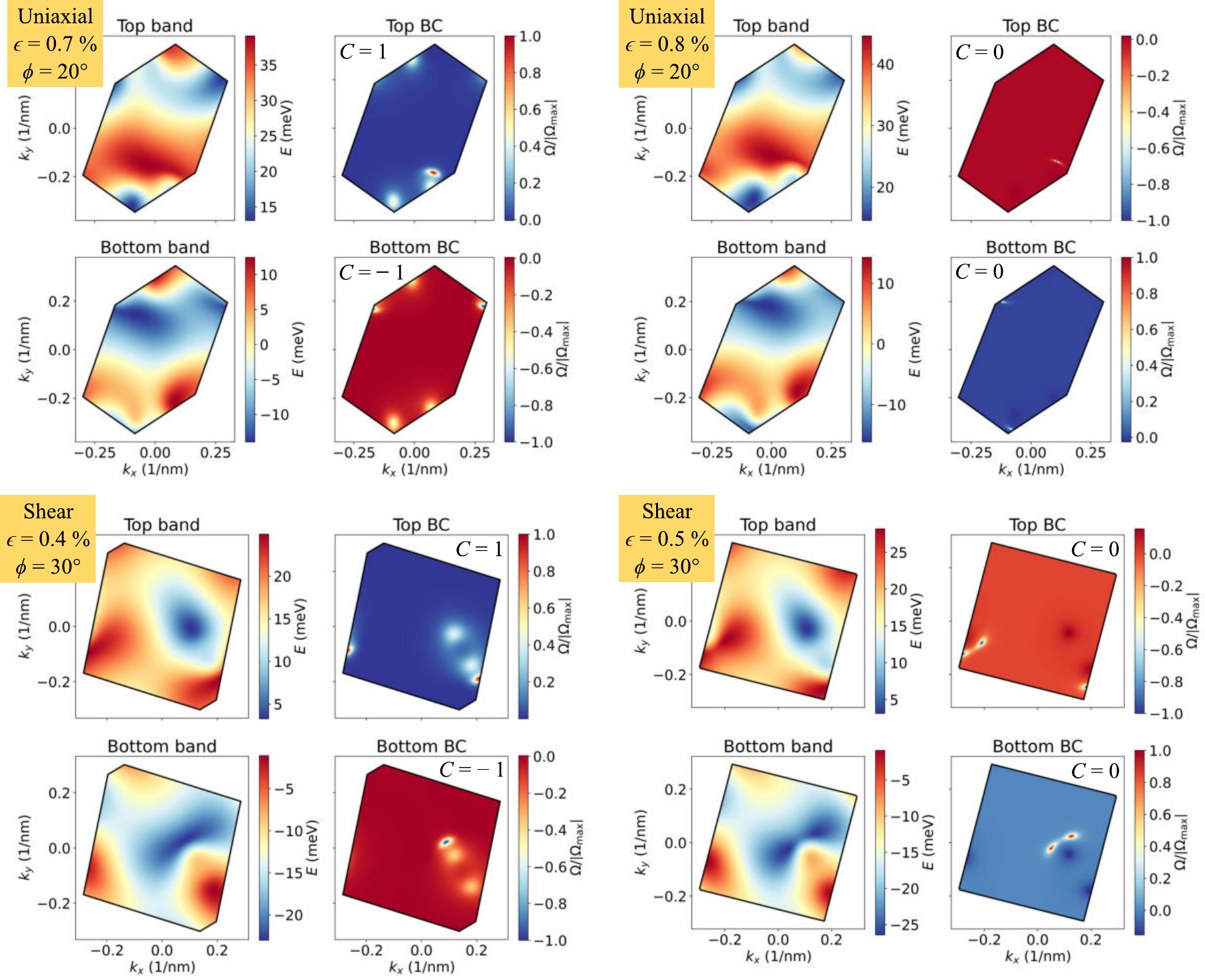}
     \caption{Density plots of the band structure and Berry curvature $\Omega$ of the top and bottom narrow bands in TSBG, for uniaxial and shear strain. The Berry curvature is normalized to its maximum value $\left|\Omega_{\mathrm{max}}\right|$. The cases displayed correspond to the situations before and after the topological transition, from valley-Chern numbers $C=\pm1$ to $C=0$ (trivial). This transition reverses the behavior of the Berry curvature due to the touching of the narrow bands with the closest remote bands. The Berry curvature peaks around the momentum point where this touching takes place. All results correspond to twist angle $\theta=1.05^{\circ}$, with a small small mass term $m=10\,\mathrm{meV}$ that opens a gap at the Dirac point. All other parameters as in Figure 10 of the main text. 
}\label{fig:Berry_strain}
\end{figure*}

The strain-induced topological transitions, from valley-topological $C=\pm1$ to trivial $C=0$, come from changes in the Berry curvature \cite{xiao2010berry}
\begin{equation}
\boldsymbol{\Omega}_{n}\left(\mathbf{k}\right)=i\left\langle \partial_{\mathbf{k}}\psi_{n\mathbf{k}}\right|\times\left|\partial_{\mathbf{k}}\psi_{n\mathbf{k}}\right\rangle .
\end{equation}
Strain-induced changes in the Berry curvature and the Berry curvature dipole and been extensively studied and reported in previous works \cite{pantaleon2021narrow, pantaleon2022interaction, moulsdale2020engineering, cuypers2026evolution}. The Berry curvature dipole, in particular, is determined by the momentum derivative of the Berry curvature and is directly related to the nonlinear Hall current in the system \cite{sodemann2015quantum,low2015topological, Sinha2022Berry}. 

Figure \ref{fig:Berry_strain} shows the changes in the Berry curvature of the top and bottom narrow bands, before and after the valley-resolved topological transition, for the cases of uniaxial and shear strain. The results correspond to the same configuration as in Figure 10 of the main text, obtained by introducing a small mass $m=10\,\mathrm{meV}$ in the continuum model that opens a gap at the Dirac point. The topological transition clearly inverts the behavior of the Berry curvature, from positive/negative to mostly negative/positive. This reverse is connected to the band touching that triggers the topology change of the bands, which is also accompanied by a transfer of charge density between the narrow and remote bands (see Figure 12 in the main text). 

In line with previous work \cite{pantaleon2021narrow, pantaleon2022interaction, moulsdale2020engineering, cuypers2026evolution}, we further see that with strain the Berry curvature exhibits, in general, three distinct peaks, but only one having a larger magnitude. This behavior becomes more pronounced when close to the topological transition, where the Berry curvature peaks strongly around the point where the narrow and remote bands close their gap. Thus, the peaks of the Berry curvature, and by extension of the Berry dipole, directly reflect the point where the narrow and remote band touch and the topology changes. Note that these peaks are in general distributed nonuniformly in the moiré Brillouin zone (their position depending on the twist and strain), due to the strain-induced broken symmetries. 


\section{Band topology with biaxial strain}

\begin{figure*}[h]
    \centering
         \includegraphics[width=\linewidth]{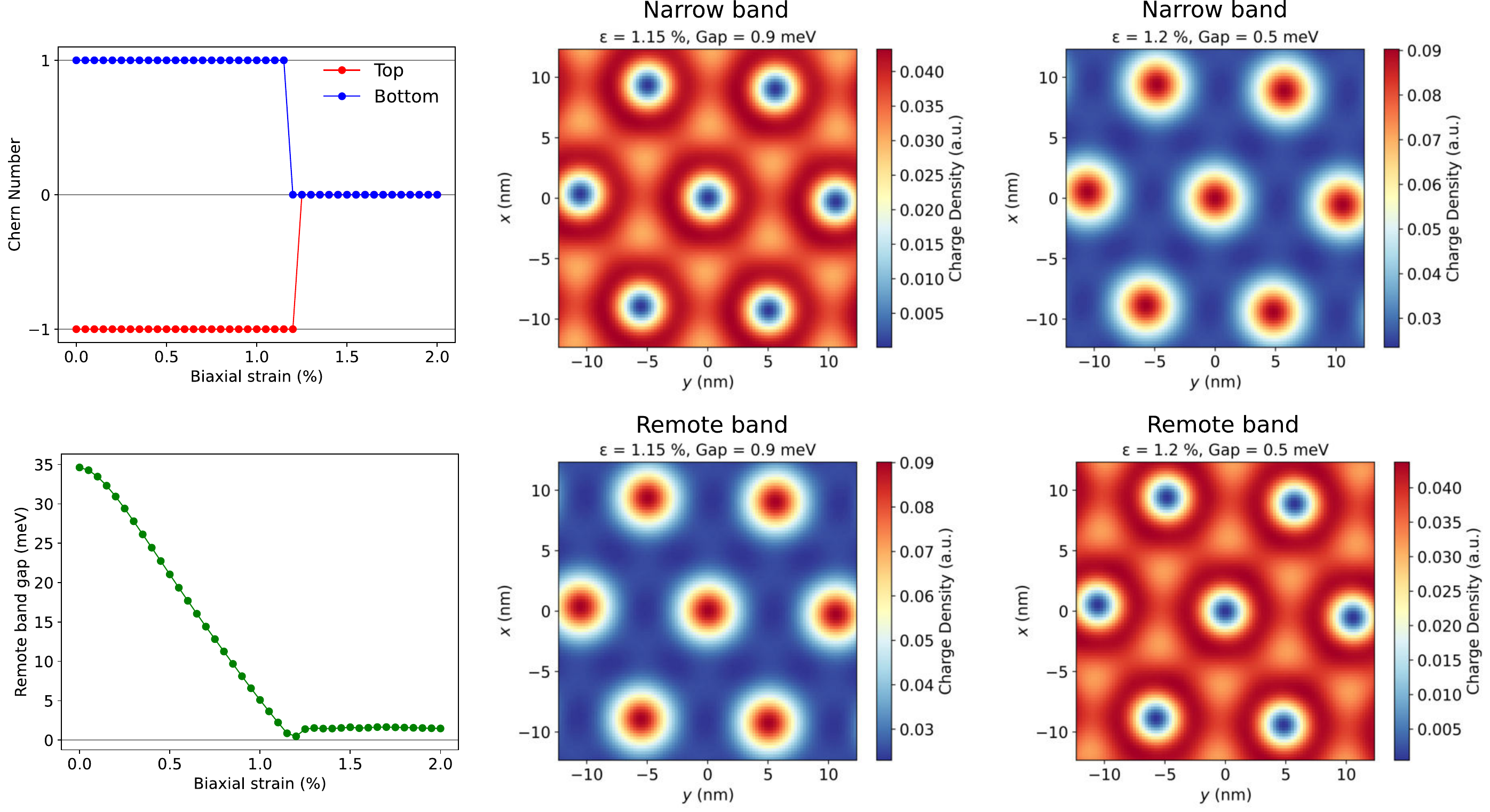}
     \caption{(Top left) The valley Chern numbers of the top and bottom narrow bands, as a function of the biaxial strain strength. (Bottom left) The gap between the narrow and remote bands, as a function of strain strength. The charge density of the narrow (top) and remote (bottom) bands before (middle panel) and after (right panel) the topological transition.}
     \label{fig:Topology_Biaxial}
\end{figure*}

Figure \ref{fig:Topology_Biaxial} shows a topological transitions driven solely by increasing biaxial strain. The results correspond to the twist angle $\theta=1.05^{\circ}$ and different biaxial strain magnitudes, for the same twist and strain configuration considered in the main text (see Sec. V in the main text). The panels on the left show the Chern number (top panel) in the top and bottom narrow bands and the gap between the top and narrow band (bottom panel), as a function of the biaxial strain magnitude. As in the uniaxial strain case (Figures 10 and 11 in the main text), we also observe a transition from topological to trivial as the strain increases, with the transition occurring when the gap between the narrow and remote band close and then reopens. The last four panels in \ref{fig:Topology_Biaxial} show the charge densities in the top and remote bands, before and after the topological transition, reflecting again the charge transfer that take place. Note that in this case of purely biaxial strain, the charge retain the $C_3$ symmetry, and the touching of the bands takes place around the $\Gamma$ point. 

\end{document}